\documentclass[%
 reprint,
superscriptaddress,
 amsmath,amssymb,
 aps,
]{revtex4-2}

\usepackage{subcaption} 
\usepackage{graphicx}
\usepackage{dcolumn}
\usepackage{bm}
\usepackage{hyperref}
\usepackage{float}

\begin{document}

\preprint{APS/123-QED}

\title{First Search for Neutral Current Coherent Single-Photon Production in MicroBooNE}

\newcommand{\ANL}{Argonne National Laboratory (ANL), Lemont, IL, 60439, USA}
\newcommand{\Bern}{Universit{\"a}t Bern, Bern CH-3012, Switzerland}
\newcommand{\BNL}{Brookhaven National Laboratory (BNL), Upton, NY, 11973, USA}
\newcommand{\UCSB}{University of California, Santa Barbara, CA, 93106, USA}
\newcommand{\Cambridge}{University of Cambridge, Cambridge CB3 0HE, United Kingdom}
\newcommand{\CIEMAT}{Centro de Investigaciones Energ\'{e}ticas, Medioambientales y Tecnol\'{o}gicas (CIEMAT), Madrid E-28040, Spain}
\newcommand{\Chicago}{University of Chicago, Chicago, IL, 60637, USA}
\newcommand{\Cincinnati}{University of Cincinnati, Cincinnati, OH, 45221, USA}
\newcommand{\CSU}{Colorado State University, Fort Collins, CO, 80523, USA}
\newcommand{\Columbia}{Columbia University, New York, NY, 10027, USA}
\newcommand{\Edinburgh}{University of Edinburgh, Edinburgh EH9 3FD, United Kingdom}
\newcommand{\FNAL}{Fermi National Accelerator Laboratory (FNAL), Batavia, IL 60510, USA}
\newcommand{\Granada}{Universidad de Granada, Granada E-18071, Spain}
\newcommand{\IIT}{Illinois Institute of Technology (IIT), Chicago, IL 60616, USA}
\newcommand{\ICL}{Imperial College London, London SW7 2AZ, United Kingdom}
\newcommand{\Indiana}{Indiana University, Bloomington, IN 47405, USA}
\newcommand{\KSU}{Kansas State University (KSU), Manhattan, KS, 66506, USA}
\newcommand{\Lancaster}{Lancaster University, Lancaster LA1 4YW, United Kingdom}
\newcommand{\LANL}{Los Alamos National Laboratory (LANL), Los Alamos, NM, 87545, USA}
\newcommand{\Louisiana}{Louisiana State University, Baton Rouge, LA, 70803, USA}
\newcommand{\Manchester}{The University of Manchester, Manchester M13 9PL, United Kingdom}
\newcommand{\MIT}{Massachusetts Institute of Technology (MIT), Cambridge, MA, 02139, USA}
\newcommand{\Michigan}{University of Michigan, Ann Arbor, MI, 48109, USA}
\newcommand{\MSU}{Michigan State University, East Lansing, MI 48824, USA}
\newcommand{\Minnesota}{University of Minnesota, Minneapolis, MN, 55455, USA}
\newcommand{\Nankai}{Nankai University, Nankai District, Tianjin 300071, China}
\newcommand{\NMSU}{New Mexico State University (NMSU), Las Cruces, NM, 88003, USA}
\newcommand{\Oxford}{University of Oxford, Oxford OX1 3RH, United Kingdom}
\newcommand{\Pitt}{University of Pittsburgh, Pittsburgh, PA, 15260, USA}
\newcommand{\QMUL}{Queen Mary University of London, London E1 4NS, United Kingdom}
\newcommand{\Rutgers}{Rutgers University, Piscataway, NJ, 08854, USA}
\newcommand{\SLAC}{SLAC National Accelerator Laboratory, Menlo Park, CA, 94025, USA}
\newcommand{\SDSMT}{South Dakota School of Mines and Technology (SDSMT), Rapid City, SD, 57701, USA}
\newcommand{\Maine}{University of Southern Maine, Portland, ME, 04104, USA}
\newcommand{\Syracuse}{Syracuse University, Syracuse, NY, 13244, USA}
\newcommand{\TelAviv}{Tel Aviv University, Tel Aviv, Israel, 69978}
\newcommand{\UTA}{University of Texas, Arlington, TX, 76019, USA}
\newcommand{\Tufts}{Tufts University, Medford, MA, 02155, USA}
\newcommand{\VTech}{Center for Neutrino Physics, Virginia Tech, Blacksburg, VA, 24061, USA}
\newcommand{\Warwick}{University of Warwick, Coventry CV4 7AL, United Kingdom}

\affiliation{\ANL}
\affiliation{\Bern}
\affiliation{\BNL}
\affiliation{\UCSB}
\affiliation{\Cambridge}
\affiliation{\CIEMAT}
\affiliation{\Chicago}
\affiliation{\Cincinnati}
\affiliation{\CSU}
\affiliation{\Columbia}
\affiliation{\Edinburgh}
\affiliation{\FNAL}
\affiliation{\Granada}
\affiliation{\IIT}
\affiliation{\ICL}
\affiliation{\Indiana}
\affiliation{\KSU}
\affiliation{\Lancaster}
\affiliation{\LANL}
\affiliation{\Louisiana}
\affiliation{\Manchester}
\affiliation{\MIT}
\affiliation{\Michigan}
\affiliation{\MSU}
\affiliation{\Minnesota}
\affiliation{\Nankai}
\affiliation{\NMSU}
\affiliation{\Oxford}
\affiliation{\Pitt}
\affiliation{\QMUL}
\affiliation{\Rutgers}
\affiliation{\SLAC}
\affiliation{\SDSMT}
\affiliation{\Maine}
\affiliation{\Syracuse}
\affiliation{\TelAviv}
\affiliation{\UTA}
\affiliation{\Tufts}
\affiliation{\VTech}
\affiliation{\Warwick}

\author{P.~Abratenko} \affiliation{\Tufts}
\author{D.~Andrade~Aldana} \affiliation{\IIT}
\author{L.~Arellano} \affiliation{\Manchester}
\author{J.~Asaadi} \affiliation{\UTA}
\author{A.~Ashkenazi}\affiliation{\TelAviv}
\author{S.~Balasubramanian}\affiliation{\FNAL}
\author{B.~Baller} \affiliation{\FNAL}
\author{A.~Barnard} \affiliation{\Oxford}
\author{G.~Barr} \affiliation{\Oxford}
\author{D.~Barrow} \affiliation{\Oxford}
% ...
\collaboration{The MicroBooNE Collaboration}
\thanks{microboone\_info@fnal.gov}\noaffiliation

\date{\today}

\begin{abstract}
This article presents the first search for neutrino-induced neutral current coherent single-photon production (NC coherent 1$\gamma$). The search makes use of data from the MicroBooNE 85-tonne active volume liquid argon time projection chamber detector, situated in the Fermilab Booster Neutrino Beam (BNB), with an average neutrino energy of $\langle E_{\nu}\rangle \sim 0.8$~GeV. A targeted selection of candidate neutrino interactions with a single photon-like electromagnetic shower in the final state and no visible vertex activity was developed to search for the NC coherent 1$\gamma$ process,  along with two auxiliary selections used to constrain the dominant background from NC$\pi^0$ production. With an integrated exposure of $6.87 \times 10^{20}$ protons on target delivered by the BNB, we set the world’s first limit for this rare process, corresponding to an upper limit on the flux-averaged cross section of $\sigma<1.49 \times 10^{-41}\text{cm}^2$ at 90\% C.L.
\end{abstract}

\maketitle


\section{\label{sec:intro}Introduction}

Neutrino-nucleus cross sections have been the subject of intense study in recent years~\cite{pdg_review,PhysRevD.106.L051102,PhysRevD.107.012004,PhysRevD.104.052002,PhysRevD.105.L051102,PhysRevLett.125.201803,PhysRevD.102.112013,PhysRevLett.123.131801} due to their role in interpreting neutrino oscillation measurements and searches for other rare processes in neutrino experiments. Inclusive and exclusive neutral current (NC) neutrino interactions, in particular, are important to understand in their own right and are a contributing background to neutrino oscillation measurements in current and next-generation experiments.
NC neutrino interactions are particularly important for $\nu_e$ and $\overline{\nu}_e$ measurements in the energy range of less than 1~GeV, especially for detectors that lack robust methods to differentiate between photon- and electron-induced electromagnetic showers. In such detectors, any NC process involving one or more photons in the final state could potentially contribute as a background to $\nu_e$ or $\overline{\nu}_e$ charged current (CC) scattering. This necessitates experimental measurements of NC single-photon production to validate or improve existing theoretical predictions for such processes. Misidentification of photons as electrons also complicates the interpretation of $\nu_e$ appearance measurements aiming to measure subtle oscillation effects. These include, for example, past short-baseline appearance searches~\cite{MiniBooNE_combine_osc,MicroBooNE:2022sdp}, upcoming sterile neutrino oscillation searches with the Short Baseline Neutrino (SBN) program~\cite{SBN}, and planned CP violation measurements and mass hierarchy determination with the future Deep Underground Neutrino Experiment (DUNE)~\cite{DUNE:2021mtg}. 

Neutrino-induced NC coherent single-photon production is predicted to be a sub-dominant source of single photons in neutrino-argon scattering below 1~GeV~\cite{coh_gamma_model}. The dominant single-photon production process in this energy range is that of NC resonance production of $\Delta$ baryons followed by $\Delta$ radiative decay~\cite{coh_gamma_pred}. The NC $\Delta$ radiative decay signal is expected to be roughly an order of magnitude larger than NC coherent single-photon production on argon. Although this NC coherent single-photon (NC coherent $1\gamma$) process is predicted in the Standard Model (SM), it has never been directly searched for or observed in neutrino scattering. 

In this article, we present the first experimental search for neutrino-induced NC coherent 1$\gamma$ production on argon. This measurement is performed at a mean neutrino energy of $\langle E_{\nu}\rangle \sim 0.8$~GeV making it especially relevant to the physics programs of experiments operating in a similar energy regime, or with argon as a target material. Additionally, much like MicroBooNE's previous search for NC $\Delta$ baryon production and radiative decay~\cite{glee_delta}, this measurement serves as another test for photon~\cite{Gninenko:2009ks,Fischer:2019fbw} and photon-like $e^+e^-$~\cite{Bertuzzo:2018itn,Ballett:2018ynz} final state interpretations of the MiniBooNE ``low energy excess''~\cite{MiniBooNE_combine_osc}. This analysis and the techniques involved also serve as a blueprint for future higher-sensitivity searches for this process that will be enabled, for example, by the upcoming Short Baseline Near Detector (SBND) due to its much higher anticipated event rate~\cite{SBND}. This result is one of three complementary next-generation single-photon searches released simultaneously by MicroBooNE, alongside an updated analysis of the previous NC $\Delta$ radiative decay search~\cite{uboone_enhanced_nc_delta} and an inclusive search for anomalous single-photon events~\cite{uboone_inclusive_gamma}.

The interaction final state targeted in this analysis is defined as:\\
\begin{equation}
    \nu (\overline{\nu}) + \text{Ar}_{gs} \rightarrow \nu (\overline{\nu}) + \text{Ar}_{gs} + \gamma,
\end{equation}
where $\text{Ar}_{gs}$ represents the struck and residual (argon) nucleus, which remains in the ground state after scattering. Due to the coherent (low momentum transfer) nature of the interaction, the outgoing photon almost always has a forward direction relative to the beam. Thus, we aim to identify this process in the MicroBooNE detector by searching for a single photon shower in the forward direction. No other observable activity is expected in the final state, and this informs the signal selection strategy. This result makes extensive use of the capability of liquid argon time projection chamber~\cite{Nygren:1974nfi,Rubbia:117852} (LArTPC) detectors to identify MeV-scale energy deposits (often called ``blips'') in the detector from small, unresolved protons in order to study the coherent nature of selected events in more detail. While there has been much development and demonstration~\cite{PhysRevD.99.012002,Andringa:2023aax} of the potential and capability of low-energy blips to help with neutron identification, supernova and solar neutrino energy reconstruction~\cite{Castiglioni:2020tsu,Abratenko_2022}, and ambient background radiation~\cite{PhysRevD.109.052007}, this result represents the first development and application of MeV-scale blip reconstruction in a high-level MicroBooNE physics analysis. 

In the remainder of this article we introduce the MicroBooNE experiment in more detail, followed by an overview of the analysis, detailed descriptions of the event selection, systematic uncertainty evaluation, and the sample utilized to constrain the NC $\pi^0$ backgrounds in the signal region. We then present the results of our search and conclude with a summary and future outlook.

\section{\label{sec:uB}The MicroBooNE Experiment}
\subsection{The Detector and Neutrino Beam}
The MicroBooNE experiment is a LArTPC neutrino experiment located at Fermilab. The MicroBooNE LArTPC detector~\cite{ub_TPC_design} has a rectangular prism shape with dimensions of 2.3 m in height, 2.6 m in width, and 10.4 m in length, containing $\sim$ 85 metric tons of liquid argon in its active volume. The cathode plane and anode planes are oriented vertically and parallel to the neutrino beam direction, with $-70$~kV high voltage applied to the cathode plane during normal operation, creating a uniform electric field of 273~V/cm inside the detector. Charged particles traversing the detector ionize the argon atoms, generating ionization electrons along their path. Under the effect of the electric field, these ionization electrons drift toward the anode plane and induce electronic signals. Three sensing wire planes are placed 3~mm apart at the anode; two wire planes record induction signals and are referred to as ``plane 0'' and ``plane 1'' in the rest of the paper, and a third wire plane collects the drifting electron charge and is referred to as the collection plane (``plane 2''). The wires within each wire plane are spaced 3~mm apart, and each plane’s wires are oriented at an angle relative to vertical: $0^{\circ}$ (``plane 2''), and $\pm 60^{\circ}$ (``plane 1'' and ``plane 0'', respectively). Together, these produce three fine-grained two-dimensional (2D) views of the neutrino interaction.

A light collection system consisting of 32 photomultiplier tubes (PMTs) is distributed behind the anode plane outside of the drift volume. The PMTs collect liquid argon scintillation light produced by the interactions with a time resolution of a few ns. This provides precise timing information for the neutrino interaction and an estimate of the interaction position within the detector to enable cosmic-ray rejection. During data collection, a coincidence  requirement between the light collected and the neutrino beam time is applied to reduce cosmic ray backgrounds in the recorded data. In addition, at the reconstruction stage, the magnitude and time of the light signal recorded by each PMT allows time-matching of the light pulses to individual particle candidates in the TPC, resulting in further rejection of cosmic rays. Both the LArTPC and light collection system are immersed in a cylindrical, foam-insulated cryostat vessel holding a total of 170 tonnes of liquid argon and serving as a Faraday cage.

The MicroBooNE detector is exposed to an on-axis neutrino flux from the Booster Neutrino Beam (BNB)~\cite{miniboone_BNB_flux}. The BNB uses protons accelerated to a momentum of 8.9 GeV/$c$ in the Booster synchrotron and directed onto a beryllium target producing secondary hadrons, such as pions and kaons. These are focused by an electromagnetic horn before decaying in the decay pipe, producing a directed beam of neutrinos. The simulation of the BNB neutrino flux is based on the work by the MiniBooNE Collaboration~\cite{miniboone_BNB_flux} and adjusted for the MicroBooNE location 72~m upstream of MiniBooNE. The resulting neutrino beam has an average energy of $\langle E_{\nu} \rangle \sim 0.8$ GeV, consisting predominantly of $\nu_{\mu}$s (93.6\%) with a small contamination of  $\overline{\nu}_{\mu}$ (5.8\%) and $\nu_{e}/\overline{\nu}_{e}$ ($<$0.6\%)~\cite{miniboone_BNB_flux}.

The MicroBooNE detector ran for five years between 2015-2020 and collected approximately $1.3\times 10^{21}$ protons-on-target (POT) of BNB data. This analysis makes use of data collected during the first three years of operation (Runs 1-3) from February 2016 to July 2018, corresponding to a total of $6.87\times 10^{20}$ POT. 

\subsection{Event Simulation and Reconstruction}\label{sec:sim_reco}
MicroBooNE utilizes the \textsc{genie}~\cite{GENIE} neutrino event generator to simulate the neutrino-argon interactions inside and outside the detector. More specifically, \textsc{genie} \texttt{\detokenize{v3.0.6 G18_10a_02_11a}} is adopted as the default event generator with a customized tune~\cite{ub_genie_tune} of CC interaction models based on fits to T2K data~\cite{T2K_cc_data}. As the focus of this analysis is aimed at neutral current interactions, the simulation of background NC $\pi^0$ events is of utmost importance. In our \textsc{genie} simulation, both coherent and resonant pion production use the Berger-Sehgal model~\cite{PhysRevD.79.053003}, with axial form factors derived from a fit of MiniBooNE data and vector form factors derived from electron scattering fits of BBBA07~\cite{Bodek:2007ym}. After the neutrino interaction, typically the struck nucleus is left in a highly excited state and will de-excite by emitting nuclear fission fragments, nucleons, or photons. The simulation of emitted nucleons is via the \textsc{intranuke}~\cite{PhysRevD.45.743} intranuclear cascade simulation framework with an empirical, data-driven hA2018 final-state interaction model. Current versions of \textsc{genie} only simulate deexcitation photon emission for oxygen, and so these are not modeled in this analysis.  

The outgoing final-state particles from \textsc{genie} are passed to the LArSoft~\cite{larsoft} framework for a full event simulation. LArSoft is a software toolkit that supports event simulation, reconstruction, and analysis for LArTPC experiments. \textsc{geant4}~\cite{geant4} \texttt{\detokenize{v4_10_3_03c}} simulates the propagation of particles in the detector, including the energy deposition and production of optical light and ionization electrons along the particle trajectory. The simulation of the propagation of light and ionization charge in the detector is done in LArSoft. Dedicated detector simulation tools~\cite{ub_signal_process_1, ub_signal_process_2} are developed to simulate the response of the sensing wires and TPC readout electronics while taking into account different detector effects~\cite{electron_recomb,ub_SCE_effect,uB_Efield_measurement}. 

As a surface detector, MicroBooNE is exposed to a constant cosmic-ray rate of approximately 4.5~\text{kHz}~\cite{cosmic_rate}. To model the cosmic-ray activity in the detector during the few-millisecond readout window, MicroBooNE takes a data-driven approach. Simulated neutrino event TPC waveforms are overlaid onto data events collected without a PMT optical trigger requirement when the BNB beam was off, to account for cosmic activity during the TPC readout window. The resulting simulated samples are referred to as ``overlay'' samples. Cosmic-ray backgrounds (``cosmics'') can additionally generate light and charge mimicking a neutrino interaction even when no neutrino interaction happens in the detector during the beam-spill. To evaluate this background, data triggered by the PMTs during beam-off periods is recorded. 

The event reconstruction chain in MicroBooNE starts with signal processing. The TPC wire signals are processed through an offline noise filter to remove excessive coherent and incoherent noise from the electronics~\cite{ub_signal_process_1}, and then run through 2D wire-time deconvolution to deconvolve the detector response and reconstruct the actual  ionization electron distribution~\cite{ub_signal_process_2}. Regions of interest are identified on the deconvolved signals, and Gaussian pulses are extracted by a hit-finding algorithm; the reconstructed Gaussian hits are fed into a reconstruction framework to form high-level reconstruction objects. 

This analysis utilizes the Pandora Software Development Kit~\cite{pandora_reco} for pattern recognition and particle trajectory reconstruction. Pandora takes reconstructed Gaussian hits with 2D (time and wire) information as inputs and then uses the temporal and spatial features to perform three-dimensional (3D) reconstruction of the particles. First, Gaussian hits are clustered to form 2D clusters on each plane. A three-dimensional (3D) interaction vertex is then identified from pairs of 2D clusters from different planes while taking into account the distribution of hits of the clusters. Each 2D cluster is assigned a score from 0 to 1 based on the linearity, hit dispersion, extent, and orientation of the clusters with respect to the reconstructed vertex and assumed beam direction. Next, clusters are classified as ``shower-like'' or ``track-like''. The 3D shower (track) objects are formed by matching 2D shower (track) clusters from the three planes. 

Analysis-specific high-level reconstruction is performed on the outputs of the Pandora pattern recognition and particle trajectory reconstruction. Due to the single-shower topology of the targeted signal, we limit the discussion here to the high-level reconstruction only of the reconstructed shower. 
The $dE/dx$ of the shower start is evaluated by identifying the main trunk of the shower start through a Kalman filter based procedure~\cite{kalman_filter} and taking the median of the $dE/dx$ calculated over the first 4 cm of the start of shower. The energy of the shower is calculated by summing up the energy of all hits associated with the shower on a plane-by-plane basis and taking the maximum of all three planes. The resulting calorimetric energy of the shower is found to be systematically about 20\% lower than the true shower energy in simulated events~\cite{MicroBooNE:2017kvv}. This is due to lossy effects such as thresholding, where small energy depositions during the shower cascade are too low to be reconstructed, and misclustering, where reconstructed hits from the shower activity are missed during clustering. We account for this by correcting the reconstructed shower energy following the procedure in Ref.~\cite{gLEE_pi0} where an approximate 20\% increase is applied to ensure that the $\pi^0$ mass peak reconstructed from our Monte-Carlo (MC) simulations agrees with the measured value. 

\section{Analysis Overview}
\begin{figure} 
    \centering
    \includegraphics[width=\linewidth]{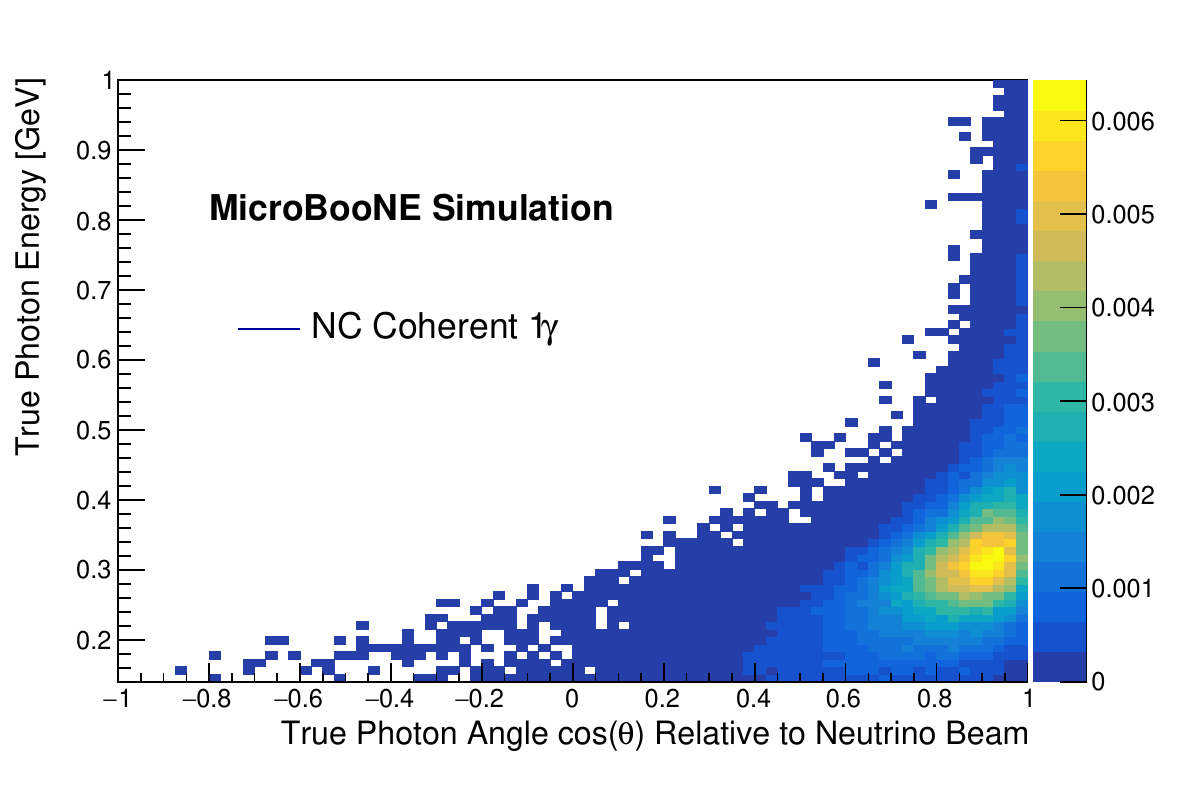}
    \caption{The area-normalized 2D distribution of the true energy and angle with respect to the beam of outgoing photons in NC coherent $1\gamma$ events. This highlights the phase space of the outgoing photon, which populates the forward region and has a peak energy at $\sim0.3$~GeV.}
    \label{fig:signal_2D}
\end{figure}

This analysis searches for the NC coherent $1\gamma$ production on argon. 
The theoretical model is proposed and described in detail in~\cite{coh_gamma_model}, with a predicted total cross section on argon of $\frac{\sigma}{A}\sim \mathcal{O}(10^{-43} \text{~cm}^2)$ where $A$ is the atomic mass number of the target nucleus. This model is not implemented in the \textsc{genie} \texttt{\detokenize{v3.0.6}} generator used by the default MicroBooNE software; instead it has been implemented and validated 
in an updated \textsc{genie} version v.3.2 ~\cite{genie_v3.2} on the development branch tag \textit{NCGammaFix}. Thus, unlike backgrounds that are simulated using the MicroBooNE software suite as described in Sec.~\ref{sec:sim_reco}, the NC coherent 1$\gamma$ is simulated in a standalone \textsc{genie} simulation, and the outputs from \textsc{genie} simulation are fed into the MicroBooNE software suite for a full simulation of the final-state particle propagation and detector response. These events are then added to the nominal MicroBooNE neutrino event rate prediction used for this analysis. 
Figure~\ref{fig:signal_2D} shows the distribution of the simulated outgoing photons as a function of the true photon energy and its true angle with respect to the neutrino beam direction, highlighting the forward nature of the outgoing photon from the coherent interaction. 

The analysis was performed using a blinding procedure. It was developed primarily using MC overlay samples, while a 10\% subset of the data was also accessible during the analysis development stage for validation purposes. A sideband sample was additionally defined (see Sec.~\ref{sec:sideband}), and the analysis was fully frozen and validated on the sideband data sample before the signal region was unblinded. When the statistical significance of the observation was evaluated, two independent high-statistics $\pi^0$ measurements were used to constrain the predicted background rate and systematic uncertainties in the signal region. Section~\ref{sec:selection} details the event selection, Sec.~\ref{sec:sys_error} describes the systematic uncertainties involved and how they are incorporated in this analysis, and Sec.~\ref{sec:pi0_constraint} shows the power of the $\pi^0$ constraint. The results obtained with the Run 1-3 data are discussed in Sec.~\ref{sec:sideband} and~\ref{sec:final_results} for the sideband and signal region, respectively.

\section{Event Selection}\label{sec:selection}
Driven by the rarity of the targeted signal, a series of selections, including traditional and gradient-boosted decision-tree (BDT) based selections, were developed with the goal of maximizing signal efficiency and minimizing background efficiency. The following sections present the detailed selections and tools developed for this analysis. 

\subsection{Event Breakdown}
MC simulations are broken down into different categories based on truth information, in a manner similar to the analyses in Refs.~\cite{glee_delta} and~\cite{gLEE_pi0}, and shown in Fig.~\ref{fig:precut}. For ease of reading, we reproduce the definitions of different categories here: 
\begin{itemize}
    \item NC coherent $1\gamma$: This is the targeted signal. It includes all NC coherent single-photon production on argon inside the active TPC volume, regardless of incoming neutrino flavor. 
    \item NC 1$\pi^0$: All NC interactions in the active volume that produce one exiting $\pi^0$ regardless of incoming neutrino flavor. This category is further split into two sub-categories, ``NC 1$\pi^0$ Coherent'' and ``NC 1$\pi^0$ Non-Coherent'' contributions, based on their interaction types with coherent production making up $2.3\%$ of all NC $1\pi^0$ interactions in GENIE. Non-Coherent scattering occurs when a neutrino interacts with a nucleon inside the argon nucleus, potentially knocking out one or more nucleons. NC 1$\pi^0$ Coherent, like the NC coherent $1\gamma$ signal, is produced from coherent scattering of the neutrino with the nucleus, and the resulting $\pi^0$ tends to be very forward relative to the incoming neutrino beam. While  NC 1$\pi^0$ coherent is the closest in appearance to our signal, it is expected to be $\mathcal{O}(3\%)$ of all NC 1$\pi^0$ events and the non-coherent contribution remains our largest background.
    \item NC $\Delta \rightarrow N\gamma$: NC single-photon production from the radiative decay of the $\Delta$(1232) baryon in the active volume. This is the largest contribution of NC single-photon production relevant to the BNB beam. In this analysis, this is further broken down into two exclusive subsets, NC $\Delta \rightarrow N\gamma$ (1+p) and NC $\Delta \rightarrow N\gamma$ (0p), depending on the outgoing nucleon and its kinetic energy (KE). NC $\Delta \rightarrow N\gamma$ (1+p) refers to NC $\Delta$ radiative decay events that have at least one proton exiting the nucleus with KE larger than 50~MeV. $\Delta \rightarrow N\gamma$ (0p) is the complementary subset and includes NC $\Delta$ radiative decay events with either no protons exiting the nucleus, or protons exiting the nucleus but with KE $<50$~MeV. In all cases neutrons of any energy are allowed to exit the nucleus. 
    \item CC $\nu_\mu\;1\pi^0$: All $\nu_\mu$ CC interactions in the active volume that have exactly one true exiting $\pi^0$. CC $\overline{\nu}_\mu\;1\pi^0$ is not included in this category but in ``BNB Other,'' defined below.
    \item CC $\nu_e/\overline{\nu}_{e}$ Intrinsic: All CC $\nu_e$ or $\overline{\nu}_{e}$ interactions in the active volume. 
    \item BNB Other: All remaining BNB neutrino interactions that take place in the active TPC volume and are not covered by the above five categories, such as multiple $\pi^0$ events and $\eta$ meson decay.
    \item Dirt (Outside TPC): All BNB neutrino interactions that take place outside the MicroBooNE active TPC but have final states that enter and deposit energy inside the active TPC detector. This can originate from scattering off liquid argon in the cryostat vessel outside the active TPC volume or from interactions in the concrete and ``dirt'' surrounding the cryostat itself. 
    \item Cosmic Data: Coincident cosmic ray interactions that take place during a BNB beam-spill but without any true neutrino interaction present.
\end{itemize}

\subsection{Topological Selection and Preselection}
\begin{figure}[h!]
\centering 
    \begin{subfigure}{0.48\textwidth}
        \includegraphics[trim=0 50 0 0,clip, width =\textwidth]{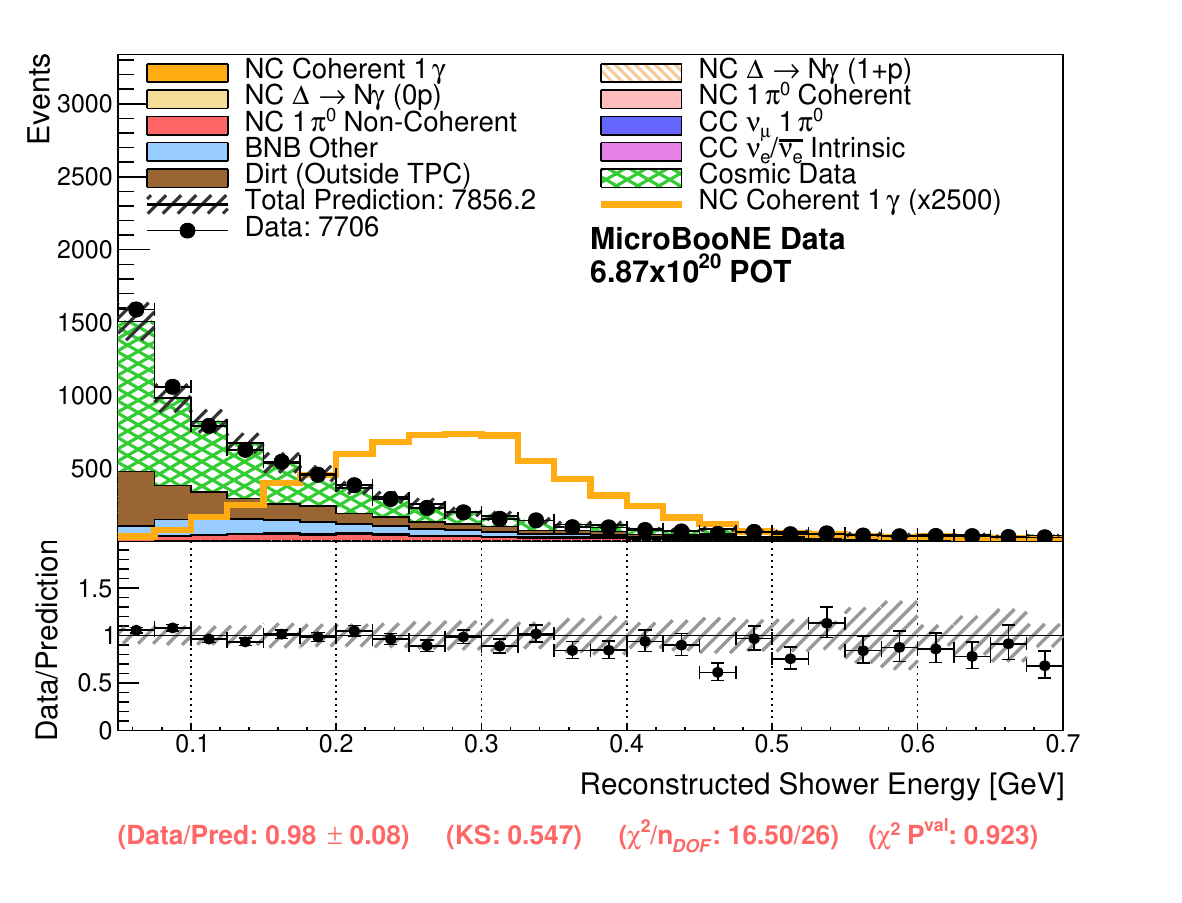}
        \caption{}
        \label{fig:cosmic_bdt}
  \end{subfigure} 
  \begin{subfigure}{0.48\textwidth}
        \includegraphics[trim=0 50 0 0,clip, width =\textwidth]{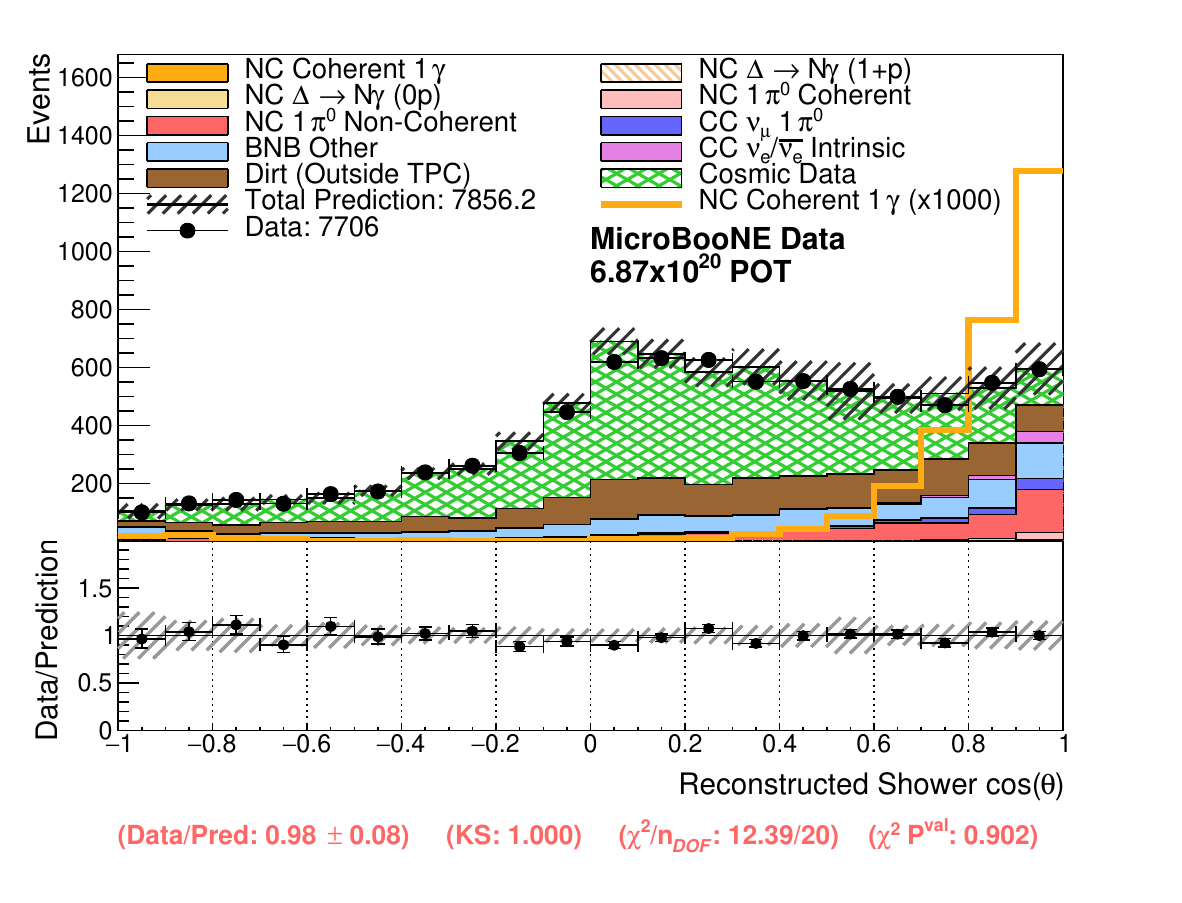}
        \caption{}
        \end{subfigure}
    \caption{Predicted background distributions and observed data as a function of (a) the reconstructed shower energy and (b) the reconstructed shower angle with respect to the neutrino beam, at preselection. At this early stage many categories contain too few events to observe, including the coherent signal. As such, the orange histogram overlaid (not stacked) shows the distribution of NC coherent $1\gamma$ signal scaled by 2500 in (a) and 1000 in (b) for it to be visible. At this stage, the signal and backgrounds populate different regions of phase space in reconstructed shower energy, with the $\sim300$~MeV peak in signal being dwarfed by low energy shower activity. In the angular phase space one can clearly see the forward nature of the signal compared to the downward nature of cosmic-originating showers.}
    \label{fig:precut}
\end{figure}

The event selection starts with a topological selection on outputs from Pandora, specifically requiring one reconstructed shower and zero reconstructed tracks ($1\gamma0p$), which is the expected topology for NC coherent $1\gamma$ signal in the detector. This topological requirement selects 28.1\% of simulated NC coherent 1$\gamma$ events; the remainder of the events do not pass the selection cut because $32.7$\% of the true signal events are reconstructed with no track or no shower by Pandora (i.e no candidate neutrino interaction), while $39.2$\% are reconstructed with either $\ge 1$ tracks or $\ge 2$ showers. Then, preselection cuts are applied in order to remove obvious backgrounds.  
The preselection cuts require that the reconstructed shower energy be larger than 50 MeV to remove Michel electrons from muon decays. Additionally a fiducial volume cut of at least 2 cm away from the space charge boundary~\cite{spacecharge} is made on the reconstructed shower start point, in order to reduce the number of partially contained showers. Relative to the topological stage, the preselection cuts further remove 33.9\% of the overall background---mostly dirt events and cosmic ray background---while preserving 98.6\% of the signal. At the preselection stage, 27.7\% of the NC coherent $1\gamma$ signal remains out of all simulated signal events in the TPC, and the corresponding signal to background ratio is about $1:2700$. Figure~\ref{fig:precut} shows the predicted NC coherent $1\gamma$ signal scaled by a factor of $2500$ and the nominal predicted background at preselection, highlighting the rarity of the signal and the different regions of phase space in reconstructed shower energy populated by the signal and backgrounds.

\subsection{Event-level Boosted Decision Tree Based Selection}\label{subsec:event_BDT}
At preselection, BDTs are developed and optimized to further differentiate the signal from background events. There are in total six (6) tailored BDTs developed using the XGBoost library~\cite{xgboost}, each trained with variables derived from reconstructed objects. Every BDT targets a different background and thus is trained with a different category of background events. There are four BDTs trained with event-level variables, which are variables characterizing the event as a whole, and two BDTs trained with cluster-level variables associated with clusters of reconstructed hits in each event. Simulated MC events are separated into statistically exclusive sets; test sets independent from the BDT training sets are used to obtain the MC prediction, which later is used to choose the cut values and perform data-MC comparisons. In this section we discuss the event-level BDTs; the formation of clusters and cluster-level BDTs are described in Sec.~\ref{subsec:clusterBDT}. 
\begin{figure*}[hbt!]
    \centering 
    \begin{subfigure}{0.48\textwidth}
        \includegraphics[trim=0 45 0 0,clip, width =\textwidth]{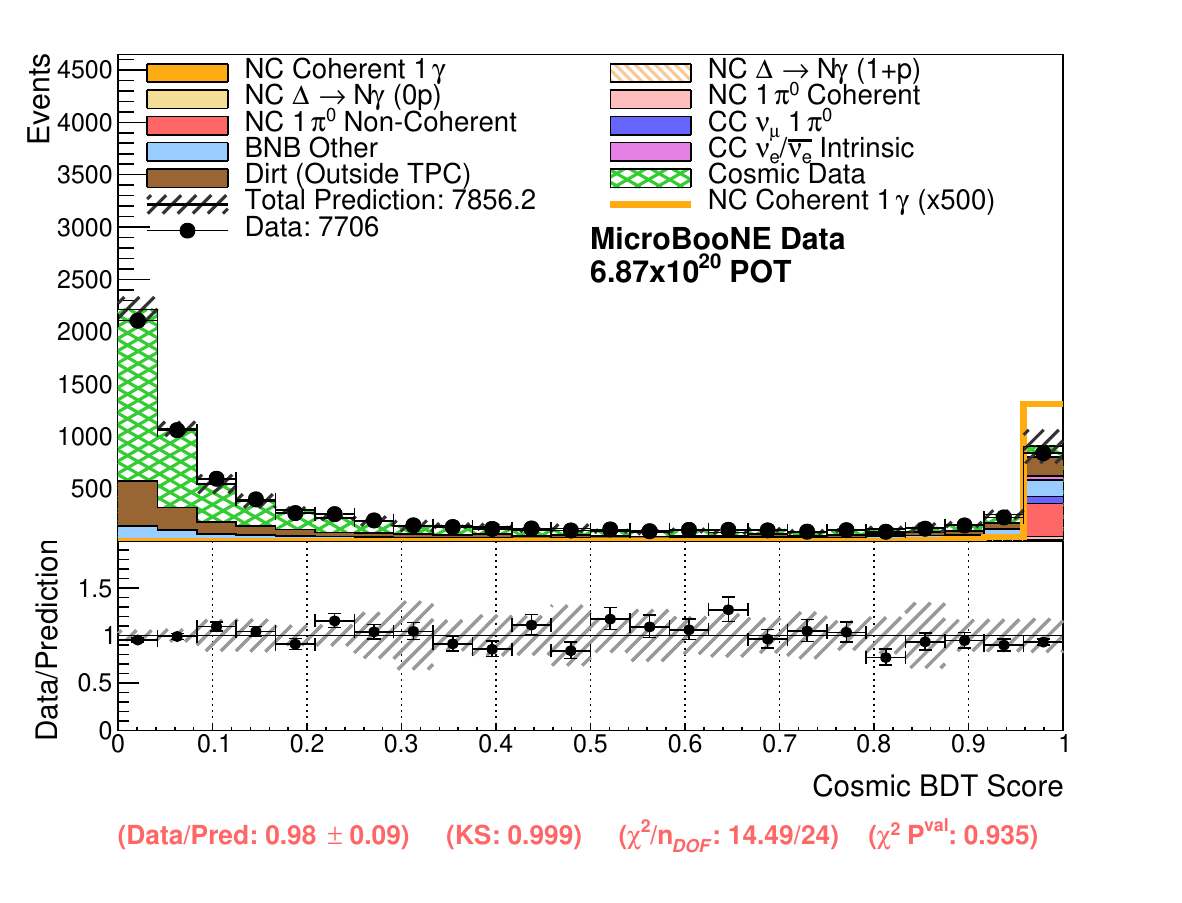}
        \caption{Cosmic Rejection BDT }
        \label{fig:cosmic_bdt}
  \end{subfigure} 
  \begin{subfigure}{0.48\textwidth}
        \includegraphics[trim=0 45 0 0,clip, width =\textwidth]{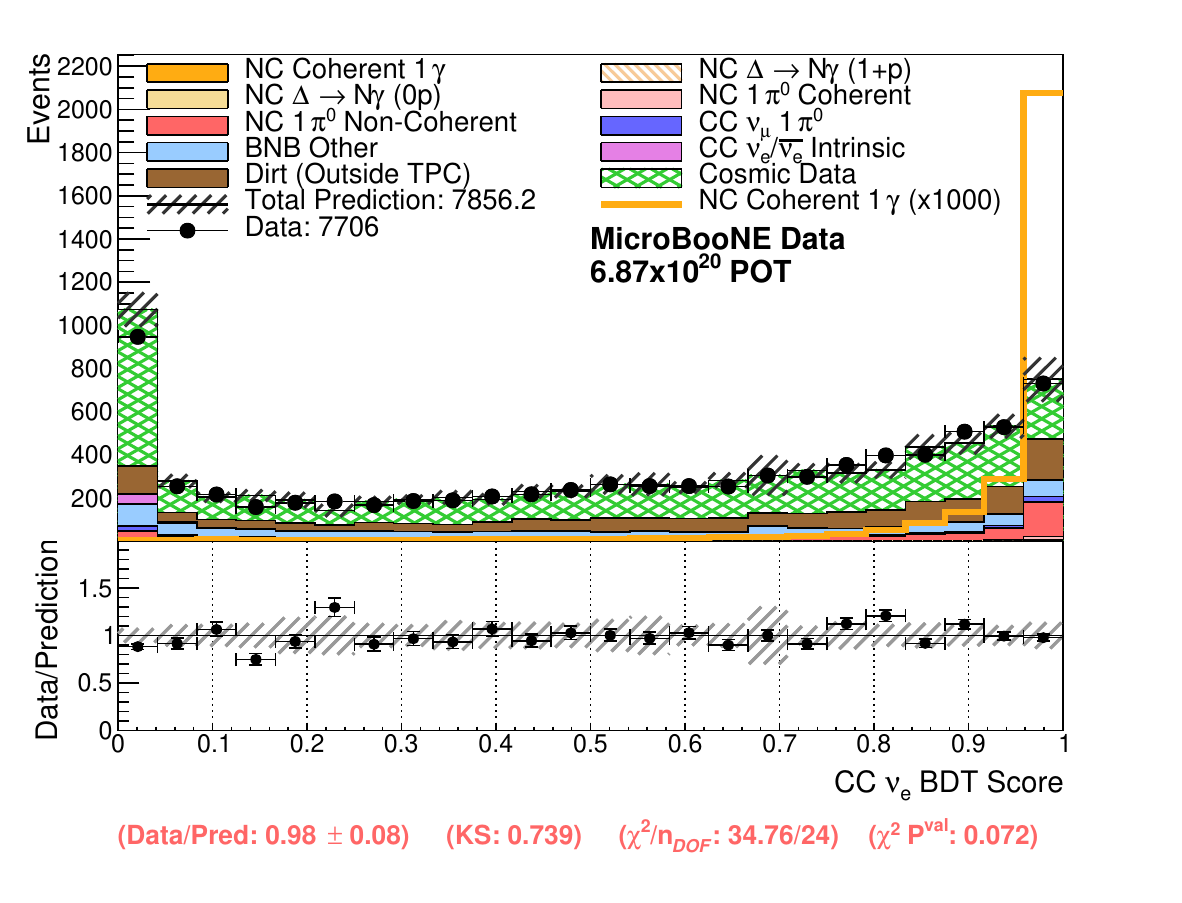}
        \caption{CC $\nu_{e}$ Rejection BDT}
  \end{subfigure} 
  \begin{subfigure}{0.48\textwidth}
        \includegraphics[trim=0 45 0 0,clip, width =\textwidth]{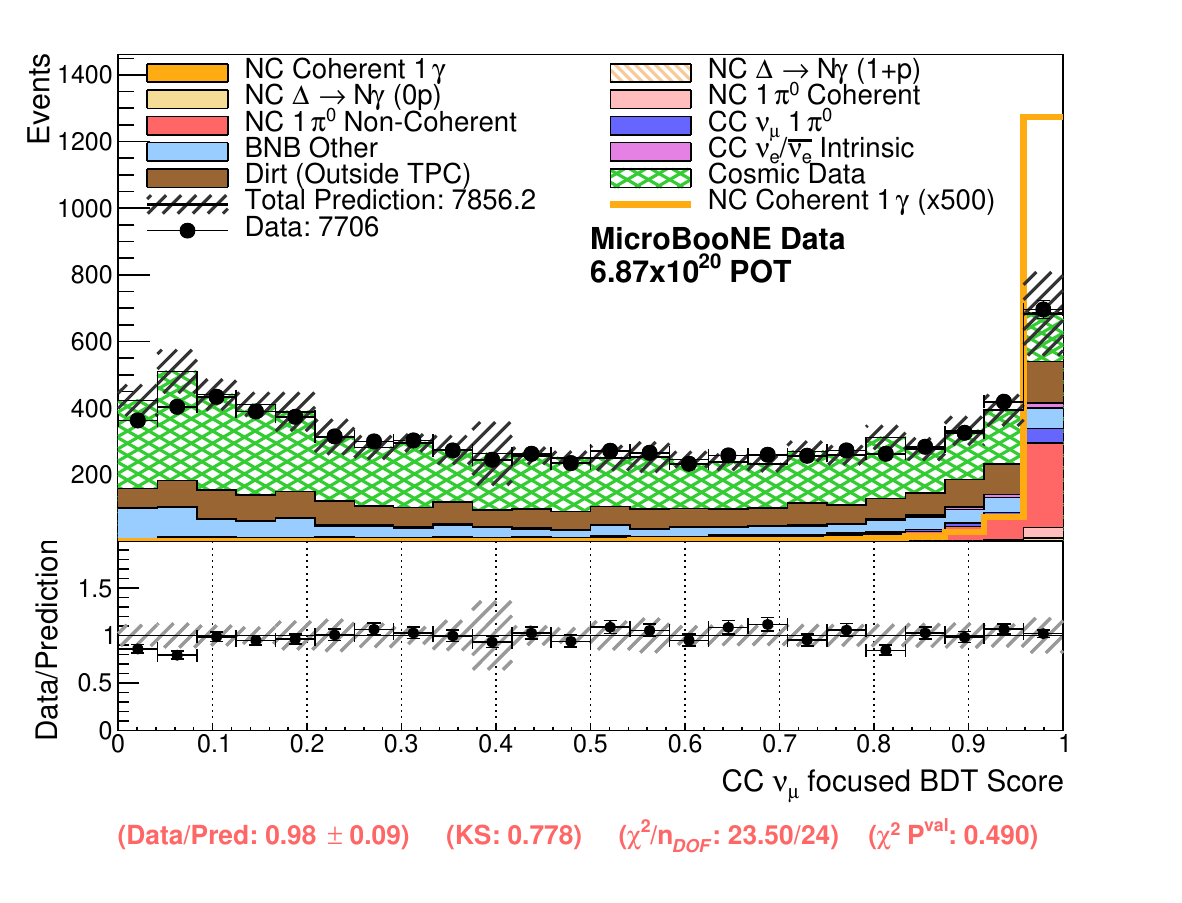}
        \caption{CC $\nu_{\mu}$ Rejection BDT}
  \end{subfigure} 
  \begin{subfigure}{0.48\textwidth}
        \includegraphics[trim=0 45 0 0,clip,width = \textwidth]{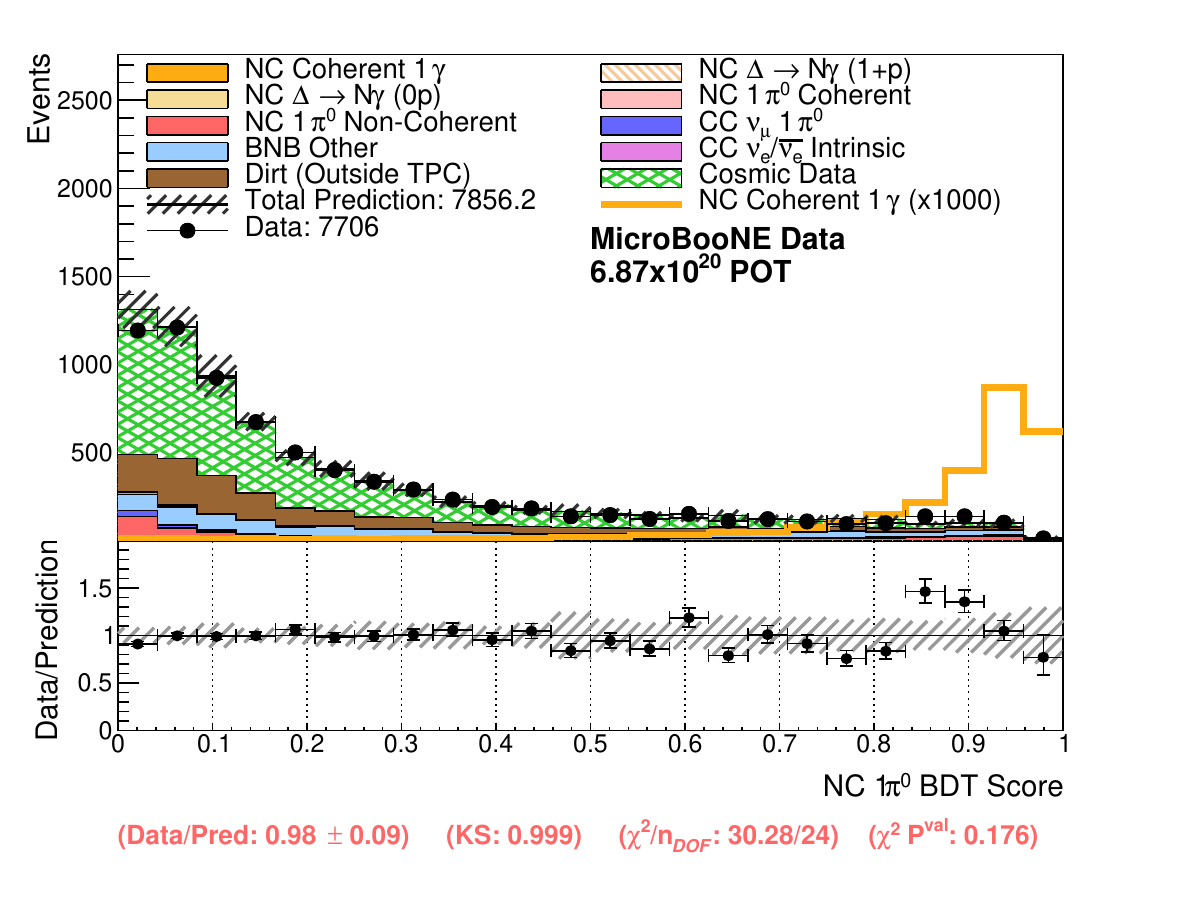}
        \caption{NC $1\pi^0$ Rejection BDT}
  \end{subfigure} 
  \caption{The BDT classifier scores for (a) the Cosmic BDT; (b) CC $\nu_{e}$ BDT; (c) CC $\nu_{\mu}$ focused BDT; and (d) NC $1\pi^0$ BDT. Higher scores indicate more NC coherent $1\gamma$ signal-like events. The stacked histograms show the predicted background distribution, while the orange histogram (not stacked) represents the distribution for the NC coherent $1\gamma$ signal, both at preselection stage. The signal histogram is scaled for visual purposes by a factor of 1000 in (c) and (d) and 500 in (a) and (b).}
  \label{fig:bdt_responses}
\end{figure*}

The largest background contributions remaining at preselection are cosmic-ray background, dirt events ($\sim50\%$ of which are neutrino interactions containing a final state $\pi^0$), BNB Other, and NC non-coherent $1\pi^0$, listed in decreasing order of the size of the predicted contribution. Three event-level BDTs are employed to target these backgrounds, and another BDT is designed to specifically target the electron background from CC interactions induced by the intrinsic $\nu_{e}/\overline{\nu}_{e}$ in the beam. To enhance the ability of BDTs to learn the differences between NC coherent $1\gamma$ events and backgrounds, all four BDTs are trained with selections of events passing the preselection requirements. The same NC coherent $1\gamma$ 
overlay sample is used as the ``signal'' during training of the four BDTs with, in addition to the preselection, the requirements that the reconstructed shower originates from photon activity in truth and $>50$\% of the reconstructed hits in the shower result from energy deposition of the simulated activity. The background definitions used and the key features in training the four BDTs are described below:
\paragraph{Cosmic BDT:} The goal of the cosmic BDT is to differentiate the NC coherent $1\gamma$ signal from misidentified cosmic backgrounds. The cosmic BDT trains on BNB external data as background, and makes use of the fact that cosmic rays usually travel from the top to the bottom of the detector, resulting in reconstructed showers that are oriented more vertically and have less extent in the beam direction when compared to that of the signal.
\paragraph{CC $\nu_{e}$ BDT:} The CC $\nu_{e}$ BDT aims to remove reconstructed showers originating from electrons instead of photons, and trains on simulated, preselected CC $\nu_{e}/\overline{\nu}_{e}$ events. One key handle for photon/electron separation is the shower $dE/dx$. Most photons in the energy range relevant to this analysis lose energy through $e^{+}e^{-}$ pairproduction; these then further radiate, producing an electromagnetic shower. For electron showers, $dE/dx$ at the shower start is similar to that of a minimum ionizing particle (MIP), around 2 MeV/cm, whereas the photon shower start typically has a $dE/dx \sim 4$~MeV/cm. 
\paragraph{CC $\nu_{\mu}$ focused BDT:} The CC $\nu_\mu$ focused BDT aims to remove any backgrounds other than the cosmic, NC $\Delta$ radiative decay, CC $\nu_{e}/\overline{\nu}_{e}$ and NC 1$\pi^{0}$. Among these background events, about 76\% are CC $\nu_{\mu}/\overline{\nu}_{\mu}$ events, 60\% of which are CC events without a $\pi^{0}$ exiting the nucleus, and for which the muon is misreconstructed as a shower. Since muons are minimum ionizing, variables such as shower $dE/dx$, average energy per hit in the reconstructed shower, and the Pandora shower score are important for the separation of misidentified muons from true coherent single photons. The CC $\nu_{\mu}$ focused BDT is trained with simulated $\nu/\overline{\nu}$-interactions other than NC $\Delta$ radiative decay, CC $\nu_{e}/\overline{\nu}_{e}$ and NC 1$\pi^{0}$ as background events. 
\paragraph{NC $1\pi^0$ BDT:} The NC 1$\pi^{0}$ background is harder to remove compared to cosmic and other $\nu$-induced backgrounds because in most cases the reconstructed shower in misidentified NC 1$\pi^{0}$ events is indeed from a true photon (most likely the leading photon from $\pi^{0}$ decay). There are a variety of reasons that can lead to single shower reconstruction in the $\pi^{0}$ sample: 1) The second shower from $\pi^{0}$ decay is not visible in the detector; this could happen when the photon is absorbed by the medium, leaves the detector before pair-producing, or is not of sufficient energy to be detected. 2) The second shower deposits energy in the detector but the 3D reconstruction fails; this could happen when the cluster-matching across planes in Pandora fails. 3) The hits from the second photon are reconstructed in 3D but not associated with the neutrino interaction; this happens when cosmic-rays interfere with the energy deposition on one or more planes. 
To mitigate failure case (2), a cluster-level second-shower veto (SSV) BDT is developed following the exact same approach as in Ref.~\cite{glee_delta} to identify the possible presence of the second shower that is not reconstructed. The resulting SSV BDT score is not used as an independent variable to directly decide whether to retain or reject an event. Instead, it serves as a training variable input for the NC $1\pi^0$ BDT to improve its $\pi^0$ rejection efficiency.  The SSV BDT is discussed in more detail in the next section, and a selection of the SSV BDT outputs are used to train the NC 1$\pi^{0}$ BDT. In addition to variables involving the missing second shower, variables associated with the primary reconstructed shower that contribute the most in separating the signal from NC $1\pi^0$ backgrounds are the reconstructed shower energy, angle, and projected momentum on the vertical plane perpendicular to the neutrino beam direction.

The outputs of the event-level BDTs are scores in the range of [0,1]; a higher event-level BDT score indicates the event is more NC coherent $1\gamma$ signal-like. Figure~\ref{fig:bdt_responses} shows the predicted BDT score distributions at preselection. The NC coherent $1\gamma$ signal is scaled up in order to highlight the separation between the signal and the targeted background in the BDT responses. 

Selections for these four event-level BDTs are chosen by placing cuts on each resulting BDT score in order to maximize the statistical significance of the NC coherent $1\gamma$ signal, i.e.~$\frac{N_{\text{sig}}}{\sqrt{N_{\text{bkg}}}}$, where $N_\text{sig}$ is the number of signal events and $N_\text{bkg}$ is the number of background eventshe possible selection cut values are explored simultaneously on all four BDTs and the signal significance is evaluated for every choice of BDT cut values. Significant degeneracy is found in the four-dimensional space of possible selection cut values. The selection cuts chosen to maximize signal selection efficiency within the degenerate region are shown in Tab.~\ref{tab:BDT_cuts}. 
The set of events for which their BDT scores are higher than the corresponding selection values constitutes the  ``Single-Photon'' rich selection. 

\begin{table}[h!]
    \centering
    \begin{tabular}{|c|c|}
    \hline 
      Event-level BDTs   & Optimized BDT Score Value\\
      \hline 
       Cosmic  & 0.990 \\
        CC $\nu_{e}$ & 0.885 \\
       CC $\nu_{\mu}$  & 0.992 \\
       NC $1\pi^0$  & 0.891 \\
       \hline 
    \end{tabular}
    \caption{Single-Photon selection values on the four primary event-level BDTs, optimized based on the signal statistical significance ($\frac{N_{\text{sig}}}{\sqrt{N_{\text{bkg}}}}$).}
    \label{tab:BDT_cuts}
\end{table}

\subsection{Cluster-level Boosted Decision Tree Based Selection}\label{subsec:clusterBDT}
Cluster-level BDTs are designed to identify activity of interest that is missed by Pandora pattern recognition. There are two cluster-level BDTs employed in this analysis. First, a second-shower veto BDT targets events that are missing the second shower from $\pi^0$ decay. Second, a proton-stub veto (PSV) BDT aims to identify proton activity near the shower vertex, in order to remove non-coherent backgrounds. Both BDTs build on hits that are not reconstructed in 3D by Pandora; individual scattered hits are clustered together to form candidate clusters of interest which then get classified by the BDTs. We will discuss the clustering mechanism first before describing the specifics of each BDT. 
\subsubsection{Hit Clustering}
Each reconstructed hit consists of a calibrated charge integral and associated wire and timing information. Nearby neighboring hits are grouped together on a plane-by-plane basis; this is referred to as ``hit clustering''. Hits with lower than 25 charge integral units are considered ``noise'' hits and ignored during clustering formation. The separation of two hits on the same plane is defined by the following metric: 
\begin{equation}
    D = \sqrt{((w_1-w_2)*0.3\text{~cm})^2 + (\frac{t_1-t_2}{25\text{~cm}^{-1}})^2},
\end{equation}
where $w_1, w_2$ are the wire number of two hits, $t_1, t_2$ are the associated time tick (1 tick = 0.5~$\mu$s) information and where the normalization factors are chosen to make the wire spacing and time interval difference comparable in physical extent. 

The clustering is performed using the density-based spatial clustering of applications with noise (DBSCAN) algorithm~\cite{dbscan} with two required parameters: 
\begin{itemize}
    \item $N_{\text{minPts}}$: the minimum number of hits required to form a cluster.
    \item $\epsilon$: the maximum distance between two hits. This distance requirement is used in both initial cluster formation, which requires at least $N_{\text{minPts}}$ hits all within a distance of $\epsilon$ from each other, and cluster expansion, where an additional hit is absorbed as long as it is within a distance of $\epsilon$ from at least one hit in an existing cluster. 
\end{itemize}
The result of DBSCAN is a group of clusters on three planes that are robust to spatially outlying hits and arbitrarily shaped, complying with the restriction from $\epsilon$.

\subsubsection{Second Shower Veto BDT}
The SSV BDT, as mentioned in Sec.~\ref{subsec:event_BDT}, targets the second shower from $\pi^0$ decay in the detector, which has an expected topology of a sparse cascade of hits near the primary reconstructed shower. Thus, the candidate clusters for the second shower are formed with $N_{\text{minPts}} = 8$ and $\epsilon = 4$ cm. Undesirable clusters can result from electronic noise along a single wire during continuous time intervals and from correlated noise across several wires at the same time. To avoid clustering on such noise, the candidate clusters are required to span at least four time ticks and three different wires. Further, the spatial variance of hits within a cluster along any single axis cannot exceed 99.9\% of the total spatial variance of the cluster. This is evaluated by performing principal component analysis (PCA). 
This further reduces the misclustering of noise and yields one or more second shower candidate clusters per plane per event.

\begin{figure}[h!]
    \centering
    \includegraphics[trim= 0 50 0 0, clip, width=1\linewidth]{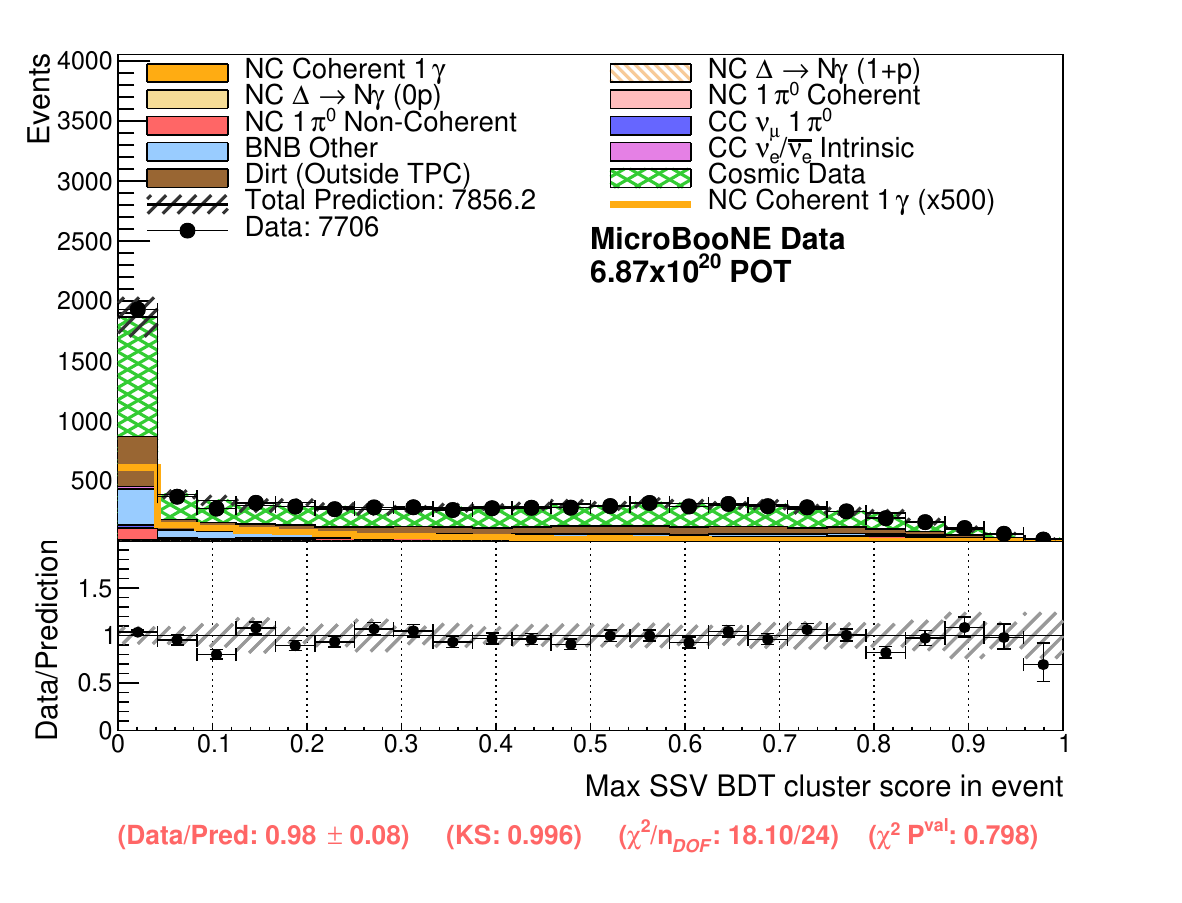}
    \caption{Distribution of the maximum SSV BDT score of all clusters formed in events at preselection. Nominal predictions for backgrounds are stacked in the colored histograms, while the signal prediction is scaled by a factor of 1000 and plotted separately in the orange histogram (not stacked). A higher score means a higher probability of a second shower being present in the event. Note that, if no second shower candidate cluster is formed in an event, the maximum SSV BDT score for this event is set to zero. }
    \label{fig:maxSSV}
\end{figure}

The reconstructed calorimetric and spatial variables associated with second shower candidate clusters are the inputs to the SSV BDT. More specifically, second shower candidate clusters that are truth-matched to a different photon from the primary reconstructed shower in the NC $1\pi^0$ sample serve as the training signal, while second shower candidate clusters formed in simulated NC coherent $1\gamma$ events are used as training background. 
The output of the BDT is a score assigned to each cluster from 0 to 1. The higher the score, the higher the probability that the cluster originates from a photon from $\pi^0$ decay. The output of the SSV BDT is not itself directly cut on in event selection; instead, the maximum SSV BDT score of all individual clusters on each of the three planes (i.e.~the most likely second-shower candidate per plane) is used as input variables to the NC $1\pi^0$ BDT. Figure~\ref{fig:maxSSV} shows the distribution of the maximum SSV BDT scores of all clusters formed in events.

\subsubsection{Proton Stub Veto BDT}\label{subsec:psv_bdt}
Besides the SSV score, another handle is leveraged to aid in the rejection of background: small traces of protons. Pandora can consistently perform particle identification on proton tracks starting at around 40 MeV of kinetic energy, however, hand scanning of NC $\Delta$ radiative decay events shows visible true low momentum proton-like activity on one or more planes near the reconstructed shower in some $1\gamma0p$ events. %
This proton-like activity usually sits very close to the backward projection of the shower direction. The PSV BDT is designed to identify such indications of protons missed by reconstruction, through which we can remove non-coherent backgrounds, specifically misidentified NC backgrounds with protons exiting the nucleus. 

The building block of the PSV BDT is proton candidate clusters, similarly to the SSV BDT. Since most protons that are missed by the reconstruction have low energy, and proton tracks are expected to be very straight with dense energy deposition, proton candidate clusters are defined by requiring that they are dense, specifically by setting $\epsilon = 1$~cm during the DBSCAN clustering. Driven by handscan observations, we also set $N_{\text{minPts}}$ to 1 to allow the capture of tiny clusters originating from proton activities. This allows small noise hits to leak in, leading to $\mathcal{O}(100)$ proton candidate 2D clusters per event, approximately evenly distributed per plane. Thus the PSV BDT must be efficient in correctly identifying true proton clusters from a substantial amount of background clusters arising from other activity. Two handles help solve this problem: the geometric relation between true proton candidate clusters and the reconstructed shower, and the distinct calorimetric profile of a proton track. 

\begin{figure}[h!]
    \centering
    \includegraphics[width=\linewidth, trim = 3cm 4cm 4cm 1cm, clip]{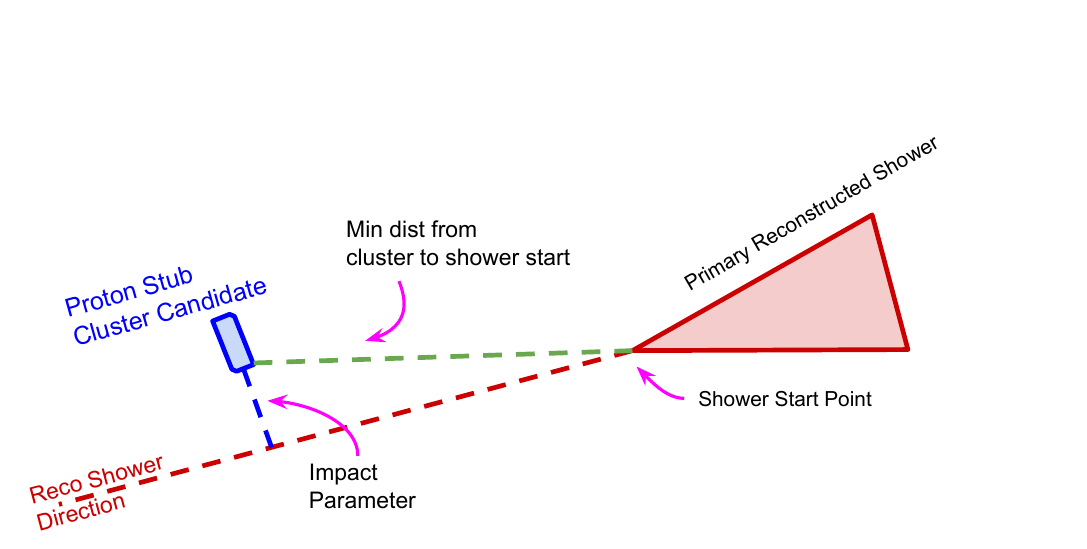}\\
    \vspace{0.5cm}
    \includegraphics[width=\linewidth]{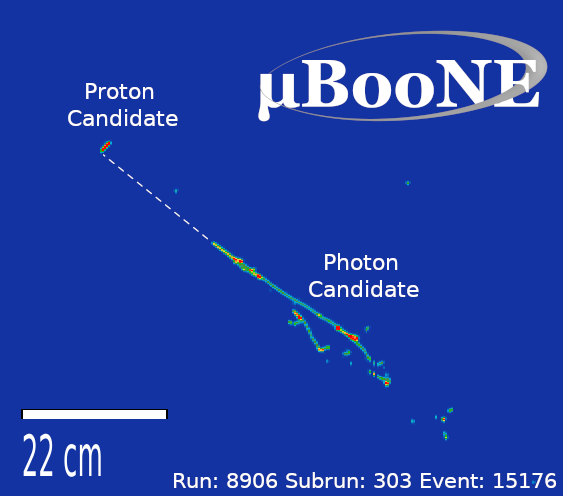}

    \caption{\emph{Top:} Cartoon showing the relative position of a proton candidate cluster to the primary reconstructed shower on a plane. Two important variables for the PSV BDT are also highlighted: the minimum distance between the proton cluster and the reconstructed shower start point, and the impact parameter of the candidate cluster to the back-projection of the reconstructed shower direction. If the proton stub candidate is indeed from a proton that comes from the same vertex as the reconstructed shower, the impact parameter is expected to be close to zero, assuming the reconstructed shower direction is accurate. \emph{Bottom:} An example data event display showing a clear candidate proton stub in the back-projection of the reconstructed shower on plane 2, with a high PSV BDT score of 0.94 and a reconstructed energy of 13.5 MeV.}
    \label{fig:psv_cartoon}
\end{figure}

The cartoon in Fig.~\ref{fig:psv_cartoon} shows a potential geometric relation between a proton candidate cluster and the reconstructed shower on any given plane. In backgrounds that have true protons exiting the nucleus and a reconstructed shower from a true photon, the proton cluster is expected to intersect the line of backward projection of the reconstructed shower. This means the minimum perpendicular distance from the candidate cluster to the direction of the primary reconstructed shower, i.e. the impact parameter of the cluster, should be zero for the proton candidate clusters. Additionally, the PSV BDT takes advantage of the straightness and the Bragg peak signature expected for a proton track to identify true proton clusters.

A sample of simulated NC $\Delta$ radiative decay events is used to train the PSV BDT: proton candidate clusters in the NC $\Delta$ sample that are truly from protons are used as training signals, while all remaining candidate clusters, be it from noise, cosmogenic origin or non-proton neutrino induced activity,  are the training background. Unlike the SSV BDT, which is not used as a selection cut, 
a cut is applied directly on a derived PSV BDT output to further improve the signal-to-background ratio after the single-photon selection.

The derived PSV BDT output used is the maximum PSV BDT score of all proton candidate clusters on plane 0 and plane 2 (abbreviated as ``maximum PSV score on planes 0 and 2''). The reason for not including clusters formed on plane 1 is due to the fact that an excess amount of proton candidate clusters is found in data in the open dataset. The excess is localized in specific regions of plane 1 and found to be due to a mismatch of run periods used in the MC overlay sample and the open data. This was resolved by requiring the overlay sample in simulation and the collected data to come from overlapping run periods. However, to be conservative, the PSV BDT score information on plane 1 is not used. 

\begin{figure}[h!]
    \centering
    \includegraphics[trim= 0 50 0 0, clip,width=1\linewidth]{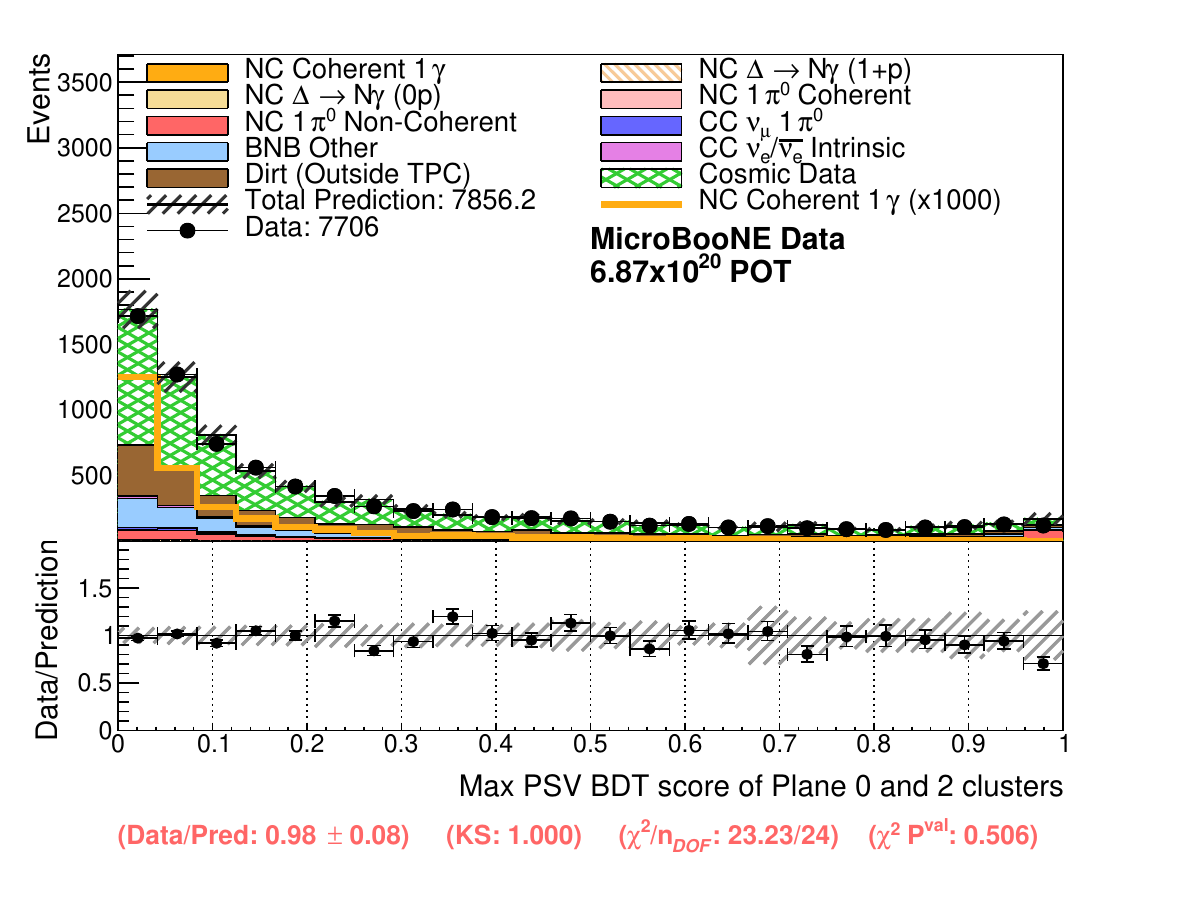}
    \caption{The maximum PSV score among all clusters on plane 0 and plane 2 for events at preselection. Higher score indicates increasing confidence of a proton exiting the nucleus that is missed by Pandora. Nominal predictions for backgrounds are stacked in the colored histograms, while the signal prediction is scaled by a factor of 1000 and plotted separately in the orange histogram (not stacked). }
    \label{fig:maxPSV}
\end{figure}

\begin{figure}[h!]
    \centering
    \includegraphics[trim= 0 50 0 0, clip, width=1\linewidth]{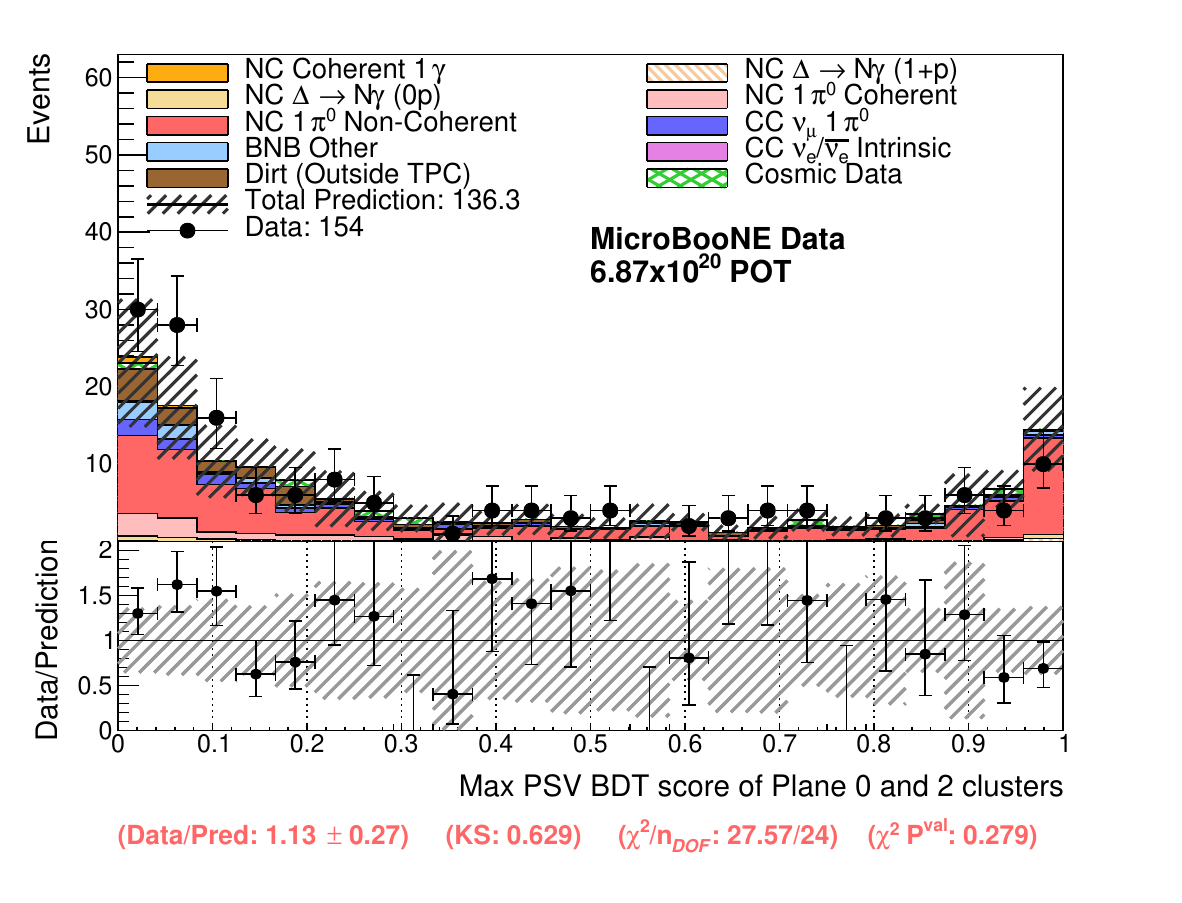}
    \caption{The maximum PSV score among all clusters on plane 0 and plane 2 for events passing the three event-level BDTs (cosmic, CC $\nu_{e}$ and CC $\nu_{\mu}$ focused BDTs). While there can be many clusters per event, this plot is showing just the maximum so there is one entry per whole event. The nominal prediction for the signal is shown on top of the background stacked histograms as an orange filled histogram, which can be seen in the first bin on the left.}
    \label{fig:maxPSV_photonRich}
\end{figure}

Figure~\ref{fig:maxPSV} shows the distribution of the maximum PSV score on planes 0 and 2 for the predicted background at preselection. As expected, the NC coherent $1\gamma$ signal and most of the background pile up on the left, while the right corner is most populated by the NC non-coherent $1\pi^0$. The distribution of the same variable with the cosmic, CC $\nu_{e}$ and CC $\nu_{\mu}$-focused BDT requirements applied is shown in Fig.~\ref{fig:maxPSV_photonRich}. The resulting sample is a photon-rich sample dominated by the NC non-coherent $1\pi^0$s. The peak near 1 is mostly populated by NC non-coherent $1\pi^0$ background, highlighting the strong separation power of the PSV BDT. The NC coherent $1\gamma$ signal and majority of the NC coherent $1\pi^0$ background cluster at low BDT score regions due to their coherent nature, as expected.

The requirement on the maximum PSV score on planes 0 and 2 is applied to events in the single-photon selection, and the cut value is optimized to maximize the signal significance with the neutrino flux and neutrino interaction uncertainties (discussed in Sec.~\ref{sec:sys_error}) included and constrained from the $2\gamma$ samples (discussed in Sec.~\ref{sec:pi0_constraint}). This optimized selection value is found to be 0.2. Thus, events that have a maximum PSV score on planes 0 and 2 less than 0.2 comprise the the ``Coherent-Rich Subset'' of the single-photon selection. 

\begin{figure}
    \centering
    \includegraphics[width=1\linewidth]{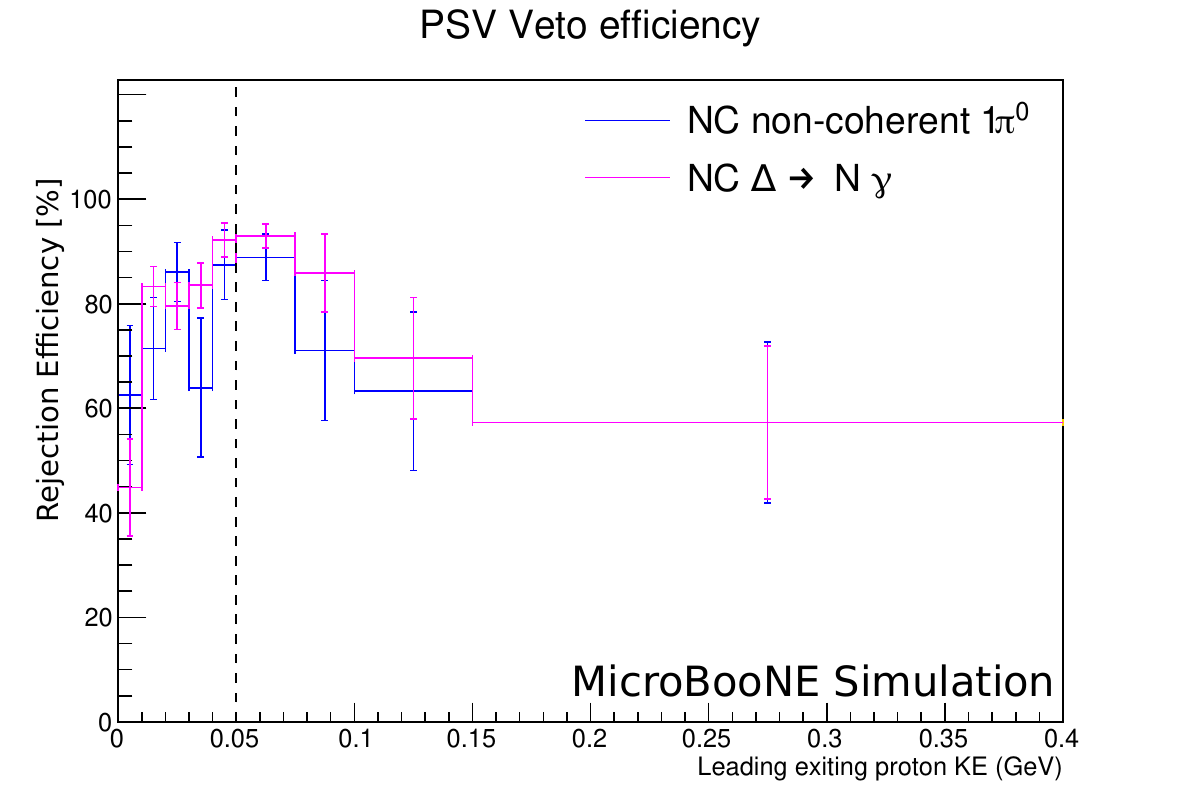}
    \caption{The rejection efficiency of the PSV BDT as a function of the KE of the leading proton exiting the nucleus for both  non-coherent NC $\pi^0$ as well as NC $\Delta \rightarrow N\gamma$ events. The error bar represent the uncertainty arising from finite MC statistics. The vertical line at 50 MeV KE is the nominal threshold for Pandora to reconstruct a proton track for comparison.}
    \label{fig:psv_veto_eff}
\end{figure}

Figure~\ref{fig:psv_veto_eff} shows the rejection efficiency of the PSV BDT on simulated NC non-coherent $1\pi^0$, NC $\Delta \rightarrow N\gamma$ (1+p), and  NC $\Delta \rightarrow N\gamma$ (0p) events. The rejection efficiency is evaluated by calculating the fraction of events that fail the PSV BDT requirement in each bin, relative to selected events passing the topological, preselection, and cosmic, CC $\nu_{e}$ and CC $\nu_{\mu}$ BDT requirements. We observe that the PSV yields a high rejection efficiency on protons at low energies on all three samples, with an average of approximately 70\% rejection efficiency for protons with KE less than 50 MeV. Due to the fact that each event has many candidate proton clusters coming from noise, cosmic induced activity, as well as EM-induced clusters associated with the reconstructed photon shower itself, a certain number of events will be vetoed due to random coincidence. To estimate this frequency we study our true coherent signal, which contains no protons, and note that only 14.8\% of these events are vetoed, indicating a dominant portion of the $\approx70\%$ veto efficiency seen in Fig.~\ref{fig:psv_veto_eff} is truly due to observing protons and not due to clusters in random coincidence. 

\subsection{Coherent-Rich Subset}
\begin{figure}[h!]
       \centering 
    \begin{subfigure}{0.48\textwidth}
        \includegraphics[width = \textwidth,trim=0cm 1.5cm 0cm 0cm, clip]{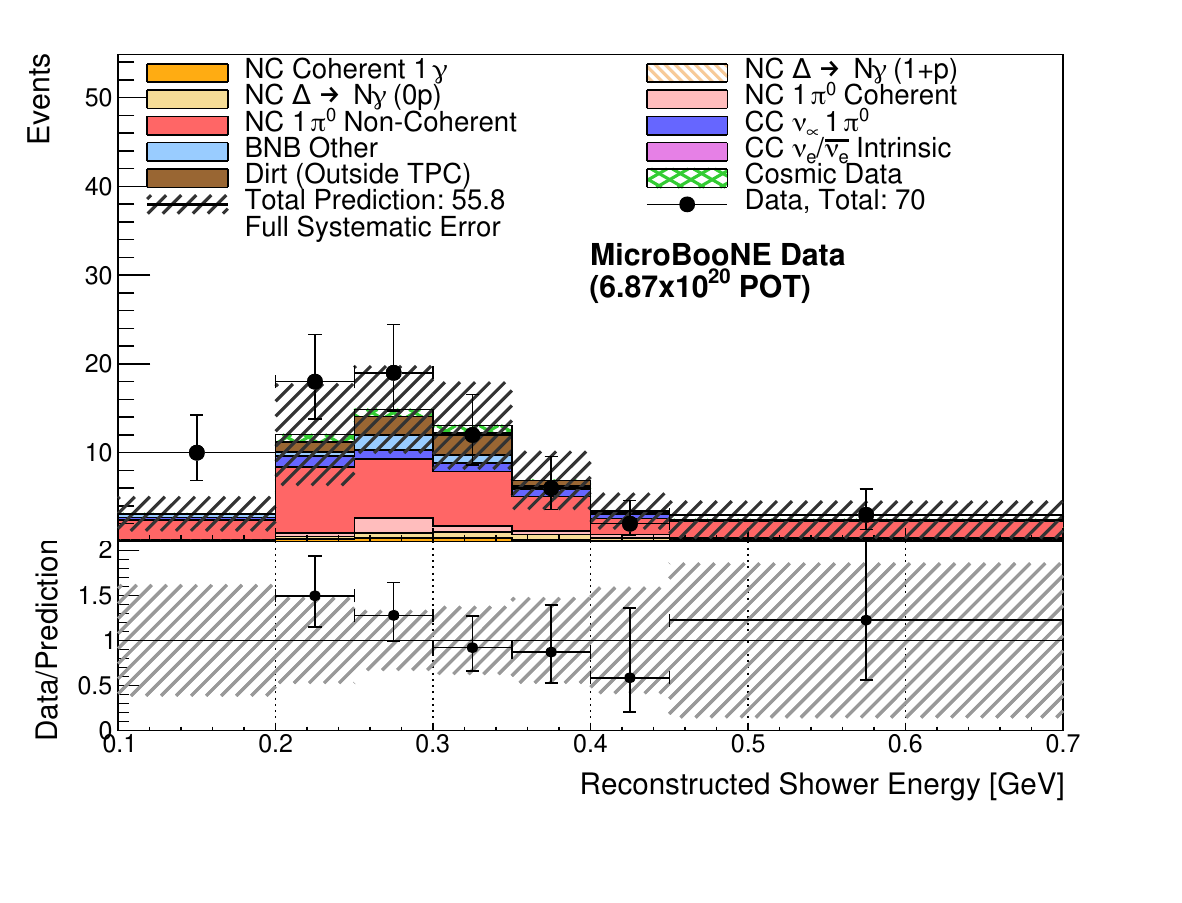}
        \caption{Reconstructed Shower Energy}
        \label{fig:cosmic_bdt}
  \end{subfigure} 
  \begin{subfigure}{0.48\textwidth}
        \includegraphics[width = \textwidth,trim=0cm 1.5cm 0cm 0cm, clip]{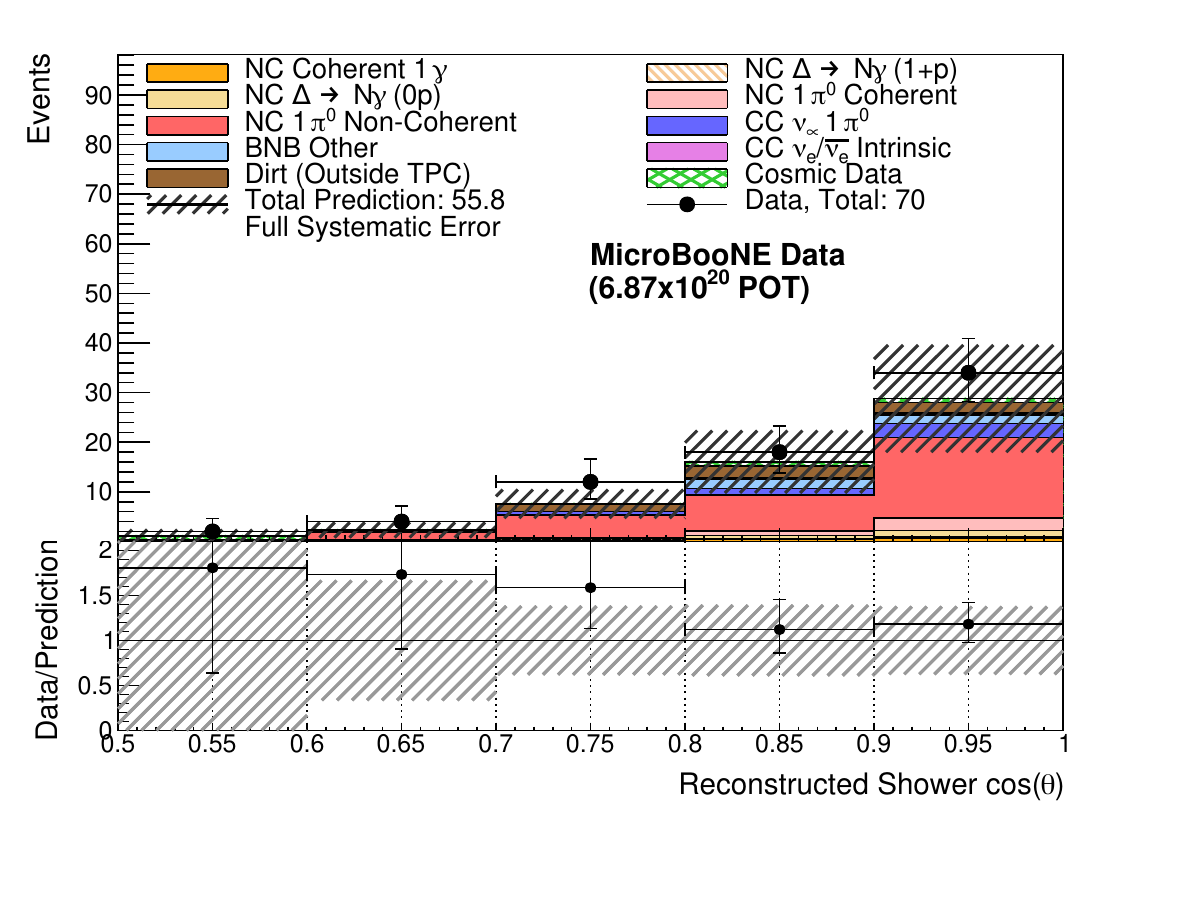}
        \caption{Reconstructed Shower cos($\theta$)}
  \end{subfigure} 
  \begin{subfigure}{0.48\textwidth}
        \includegraphics[width = \textwidth,trim=0cm 1.5cm 0cm 0cm, clip]{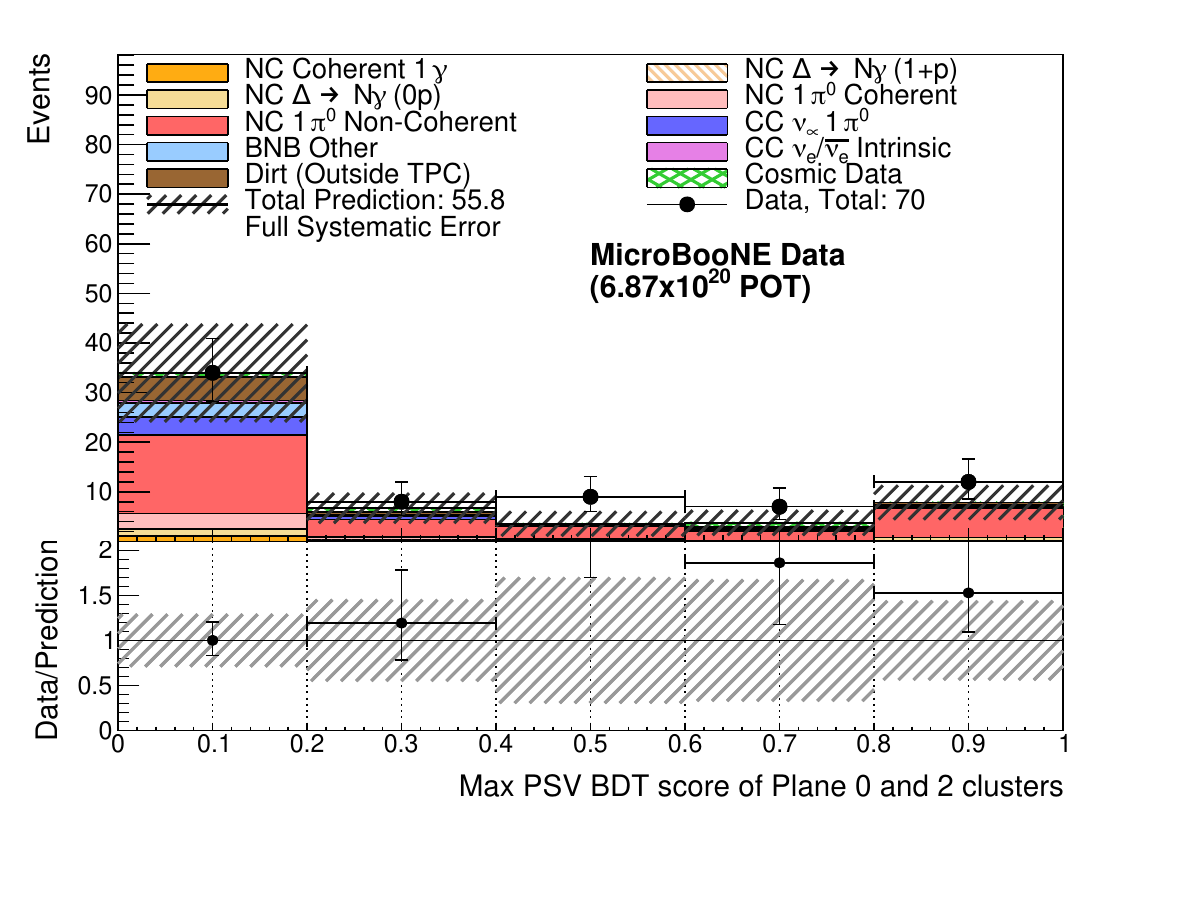}
        \caption{Maximum PSV score between Plane 0 and Plane 2}
  \end{subfigure} 
  \caption{Distributions of the single-photon selection in: (a) reconstructed shower energy; (b) reconstructed shower cos($\theta$); and (c) the maximum PSV score on planes 0 and 2. The selected events have a reconstructed shower with an energy peak at $\sim300$~MeV in a very forward direction, which is expected for the NC coherent $1\gamma$ signal.}
  \label{fig:semi_final}
\end{figure}

We first show the results at the single-photon selection stage in three variables that highlight the reconstructed shower energy and angle, as well as the PSV features of signal and backgrounds in Fig.~\ref{fig:semi_final}. The selection has a reconstructed shower distribution with an energy peak of around 300 MeV and a very forward angle, which is expected for the NC coherent $1\gamma$ signals. As expected, NC non-coherent $1\pi^0$ events are the most dominant background, and the PSV distribution at this stage motivates the additional requirement on the PSV variable to be less than 0.2. 

\begin{figure}[h!]
\centering
    \begin{subfigure}{0.49\textwidth}
        \includegraphics[width = \textwidth]{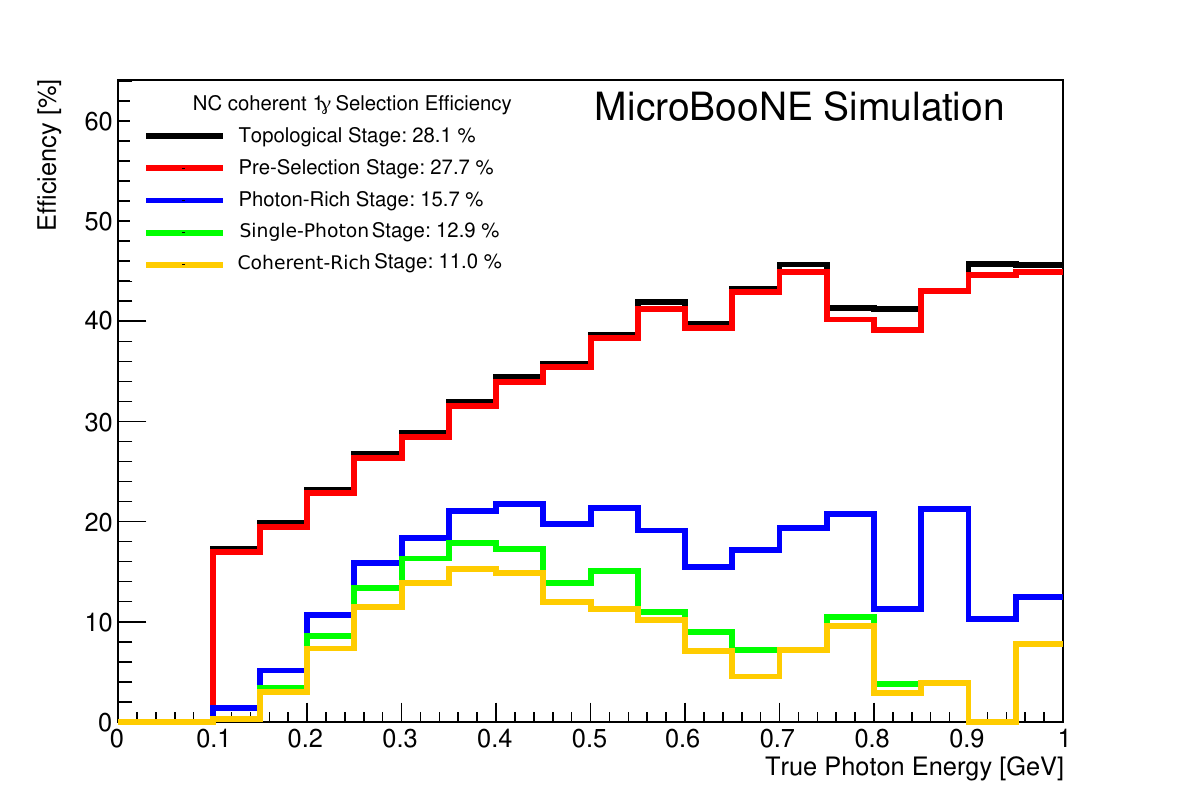}
        \caption{Selection Efficiency v.s. True Photon Energy}
        \label{}
  \end{subfigure} 
  \begin{subfigure}{0.49\textwidth}
        \includegraphics[width = \textwidth]{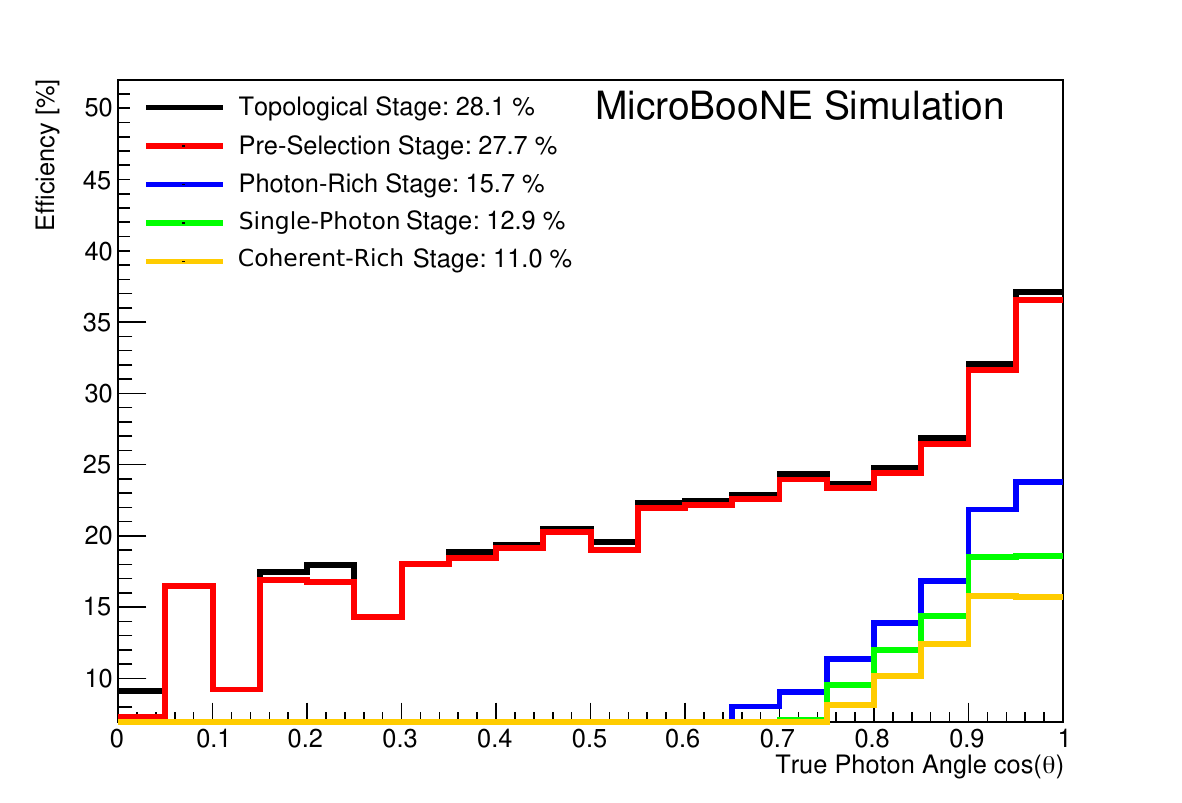}
        \caption{Selection Efficiency v.s. True Photon $\cos{\theta}$}
        \label{}
  \end{subfigure} 
    \caption{Efficiencies at various stages of the selection for the NC coherent 1$\gamma$ signal as a function of (a) true photon energy and (b) true photon angle with respect to the neutrino beam, highlighting the photon phase space targeted by this analysis. The ``photon-rich stage'' refers to the stage where the three event-level BDTs (cosmic, CC $\nu_{e}$ and CC $\nu_{\mu}$ focused BDTs) are applied. The efficiencies are calculated as 
    the ratio of the selected NC coherent $1\gamma$ events within the range plotted for each variable to the total prediction 
    of NC coherent $1\gamma$.}
    \label{fig:sig_eff_final}
\end{figure}

Table~\ref{tab:final_prediction} shows all background and signal predictions for the single-photon selection and the coherent-rich subset selection, normalized to $6.87\times 10^{20}$~POT. Prior to the PSV cut there are a total of 54.5 predicted background events and 1.3 signal events, leading to a signal-to-background ratio of 1:42. The PSV requirement further improves the signal-to-background ratio to 1:30 by filtering out almost half of the NC non-coherent $1\pi^0$ background. Figure~\ref{fig:sig_eff_final} shows the selection efficiency for the simulated NC coherent $1\gamma$ signal in the active TPC at different stages of the analysis, as a function of the true energy and angle of the outgoing photon. 

\begin{table}[h!]
    \centering
    \begin{tabular}{lrr}
    \hline\hline
        Stage & Single-Photon & Coherent-Rich  \\ 
          &   Selection & Subset \\ \hline

        NC coherent $1\gamma$ (Signal) & 1.3 & 1.1 \\
        NC $\Delta\rightarrow N\gamma$ (1+p) & 0.3 & 0.1 \\
        NC $\Delta\rightarrow N\gamma$ (0p) & 2.5 & 1.3 \\
        NC $1\pi^0$ Non-Coherent & 29.9 & 15.8 \\
        NC $1\pi^0$ Coherent & 3.8 & 3.1 \\
        CC $\nu_\mu$ $1\pi^0$ & 5.0 & 3.6\\
        CC $\nu_e$ and $\overline{\nu}_e$ & 0.6 & 0.4\\
        BNB Other & 3.5 &  2.9 \\
        Dirt (outside TPC) & 6.4 & 4.8 \\
        Cosmic Ray Data& 2.4 & 0.8 \\
        \hline
        Total Prediction (Unconstr.) & 55.8 & 34.0 \\ 
        Total Prediction (Constr.) & 45.8 & 29.0 \\ 
    \hline\hline
    \end{tabular}
    \caption{The expected event rates in the single-photon and coherent-rich subset selections for all event types, normalized to the Run~1-3 data POT ($6.87 \times 10^{20}$). The total, constrained prediction is evaluated with the conditional constraint from the $2\gamma1p$ and $2\gamma0p$ samples.}
    \label{tab:final_prediction}
\end{table}

The selection efficiencies for all signal and background categories at different analysis stages are included in Table ~\ref{tab:final_eff}, calculated relative to predictions after the topological selection. We note again that this initial topological efficiency in which a single shower is reconstructed is 28\% for our signal. The event-level BDTs successfully reject all types of backgrounds by a large fraction, especially the intrinsic CC $\nu_{e}$, BNB Other, dirt and cosmic backgrounds. Transitioning from from the single-photon selection stage to the coherent-rich subset, the cut on the PSV BDT acts mostly on the NC $\Delta \rightarrow N\gamma$ (1+p), NC $\Delta \rightarrow N \gamma$ (0p), and NC non-coherent 1$\pi^{0}$ backgrounds, with corresponding rejection efficiencies of 68.8\%, 47.9\%, and 47.4\%, respectively. Note that NC $\Delta \rightarrow N \gamma$ (0p) consists of NC $\Delta$ radiative decay events with neutrons exiting the nucleus as well as NC $\Delta$ radiative decay events with exiting protons with $\text{KE} < 50$ MeV, the latter of which is specifically targeted by the PSV. Overall, relative to the topological selection, the coherent-rich subset selection achieves a 97.5\% rejection efficiency on NC non-coherent 1$\pi^0$ and a 99.99\% rejection efficiency on the cosmic background while keeping the signal efficiency at 39\%.

\begin{table*}[hbt!]
\centering
\begin{tabular}{|c|c|c|c|}
\hline
Category & Preselection Eff. [\%] & Single-Photon Eff. [\%] & Coherent-Rich Subset Eff. [\%] \\
\hline\hline
NC Coherent 1 $\gamma$ & 98.59 & 45.85 & 39.09\% \\ \hline \hline
NC $\Delta \rightarrow N \gamma$ (0p) & 98.30 & 19.18 & 9.99 \\  
NC $\Delta \rightarrow N\gamma$ (1+p) & 97.13 & 7.99 & 2.49 \\  
NC 1 $\pi^{0}$ Coherent & 96.91 &  8.38 & 6.83 \\ 
NC 1 $\pi^{0}$ Non-Coherent & 96.39 & 4.75 & 2.50 \\ 
CC $\nu_{\mu} 1 \pi^{0}$ & 90.21 & 3.07 & 2.24\\ 
BNB Other & 82.17 &  0.30 & 0.25 \\ 
CC $\nu_{e}/\overline{\nu}_{e}$ Intrinsic & 94.59 & 0.69 & 0.47 \\ 
Dirt (Outside TPC) & 61.67 &  0.25 & 0.18 \\ 
Cosmic Data & 61.01 & 0.03 & 0.01 \\ \hline
\end{tabular}
\caption{Selection efficiencies for our simulated signal and background categories at different stages of the analysis, calculated with respect to the number of selected events after the topological requirements. }
\label{tab:final_eff}
\end{table*}

\section{Systematic Uncertainties}\label{sec:sys_error}
The sources of systematic uncertainty considered in this analysis are broken down into contributions from 1) neutrino flux, 2) neutrino interaction cross sections, 3) hadron-argon reinteractions, 4) detector response, and 5) finite MC statistics. 

The flux uncertainty includes proton delivery, hadron production, hadronic interactions, and the modeling of the horn magnetic field~\cite{miniboone_BNB_flux}. Both the intensity and position of the proton beam when it reaches the BNB target is monitored, and the uncertainty on our expected neutrino rate due to proton delivery is estimated to be a 2\% normalization uncertainty. The hadronic production concerns the production of secondary particles ($\pi^{\pm}$, $K^{\pm}$, $K^0_{L}$) which decay to produce neutrinos. Hadronic interaction cross sections impact the hadron production rate in the primary proton-Be interaction, and the rate of pion absorption in the target and the horn, which ultimately affects the rate and shape of the neutrino flux. Uncertainty on the horn magnetic field modeling comes from two sources: the intrinsic variation in the horn current, and uncertainty in the modeling of the current within the inner cylinder in the horn. Tools and techniques developed by MiniBooNE~\cite{miniboone_BNB_flux} are adopted to evaluate flux uncertainty in MicroBooNE. The ``multisim'' technique is employed for flux uncertainty evaluation where multiple MC simulated universes are generated and, in each universe, events are reweighted independently based on neutrino parentage, neutrino type, and energy. 

The neutrino interaction cross section uncertainties arise from uncertainties in the model parameters used in the \textsc{genie} prediction. This includes uncertainty in NC resonant production, the branching ratio, and the angular uncertainty on NC $\Delta$ radiative decay, uncertainties on CC quasielastic and resonance production, as well as other NC and CC interactions, and final state interactions. More details on the treatment of cross section parameter uncertainties can be found in Ref.~\cite{ub_genie_tune}. Simulated events are reweighted using \textsc{genie} reweighting tools when model parameters are varied. For the cross section uncertainty evaluation, there exist universes resulting from single parameter variations for certain model parameters as well as universes with multiple parameters varied simultaneously, which allows incorporating correlations between model parameters. Due to technical limitations, the NC and CC coherent $\pi$ production cross sections are assigned with a $\pm 100$\% normalization uncertainty.

Hadron-argon reinteractions refer to the scattering of charged hadrons off the argon nuclei through hadronic interactions during propagation, which could significantly bend the particle trajectory or lead to additional particle production. \textsc{geant4} is used to simulate hadrons propagating through the detector medium~\cite{geant4_hadron}, and events are reweighted independently for $\pi^{+}$, $\pi^{-}$, and proton interactions via the \textsc{Geant4Reweight} tool~\cite{geant4_reweight}. 

The detector systematic uncertainty stems from the difference between detector simulation and the actual detector response.  There are three main types of uncertainties: 1) uncertainty in TPC waveform from the electronic response simulation, 2) uncertainty in the light simulation, and 3) other detector effects modifying the distribution of drifting ionization electrons. MicroBooNE employs a novel data-driven technique to evaluate the uncertainty on TPC wire waveforms, by parameterizing the differences between simulated cosmic and observed cosmic data at the level of Gaussian hits after waveform deconvolution~\cite{det_err1}. This avoids the large computation required for detector simulation and signal deconvolution and allows the detector uncertainty to be evaluated in a model-agnostic way. Four variations as functions of $X$ position (drift direction), $YZ$ position (vertical and beam direction), and the angular variables $\theta_{XZ}$ and $\theta_{YZ}$ for the particle trajectory, respectively, are considered. Variations associated with the light yield consider an overall 25\% reduction in the light yield, variation in the light attenuation, and changes in Rayleigh scattering length. Other detector effects, including the space charge effect due to the buildup of argon ions~\cite{uB_Efield_measurement,SCE}, and variations in the ion recombination model~\cite{recomb} are accounted for through additional variations. Independent central value (CV) MC simulations, apart from the samples used in training the BDTs and making the final predictions, are generated together with the corresponding detector variations discussed above. The detector CV and variation samples are run through the reconstruction and analysis selection chain before being compared for detector uncertainty evaluation. 

\begin{figure}[h!]
    \centering
    \includegraphics[width=1\linewidth]{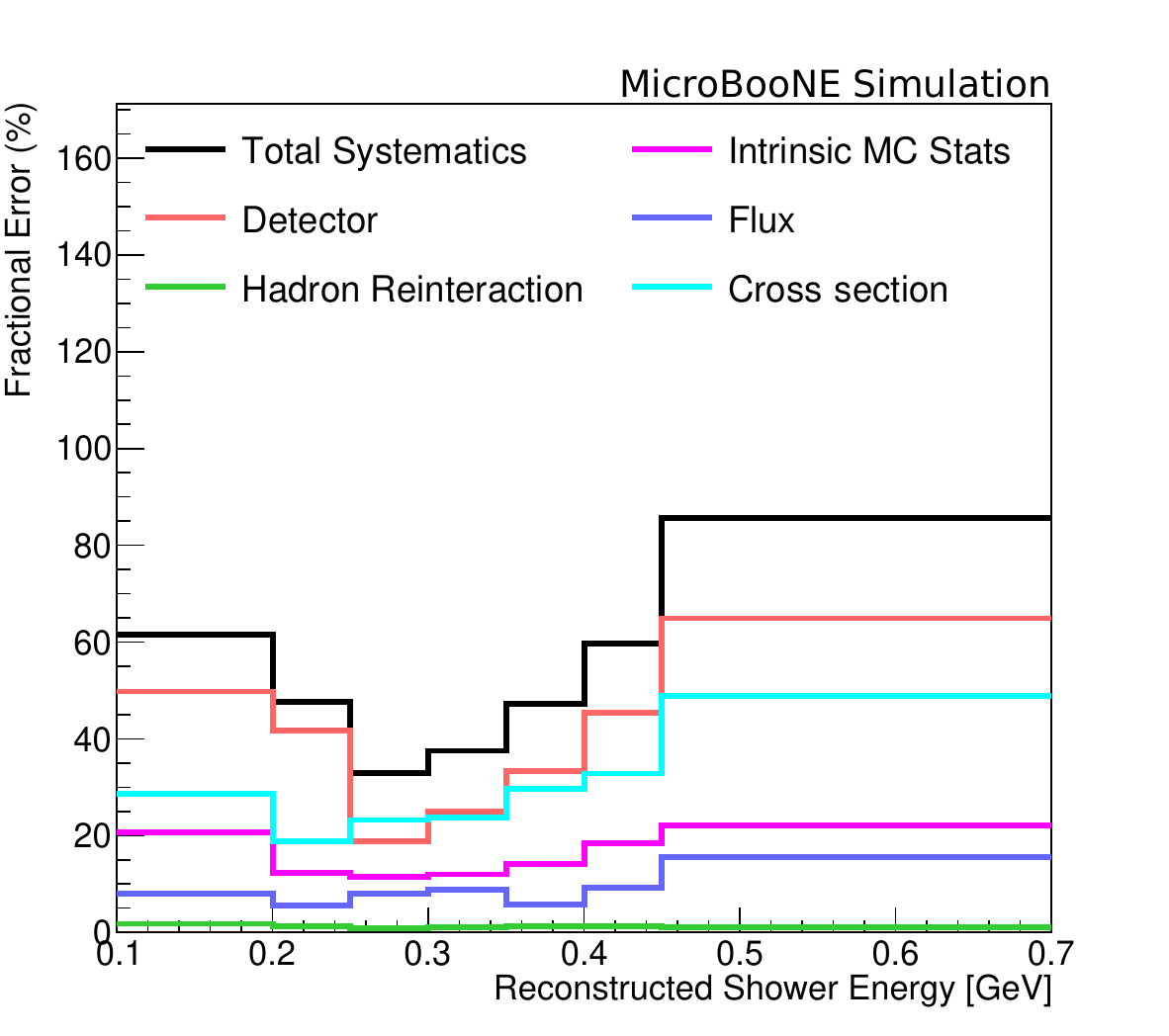}
    \caption{Overall systematic uncertainty at the single-photon selection stage, as a function of the reconstructed shower energy. The contribution from each source is highlighted in different colors and the quadrature sum of all systematics is shown in black.  }
    \label{fig:semi_final_error_budget}
\end{figure}

\begin{figure}[h!]
    \centering
    \includegraphics[width=1\linewidth]{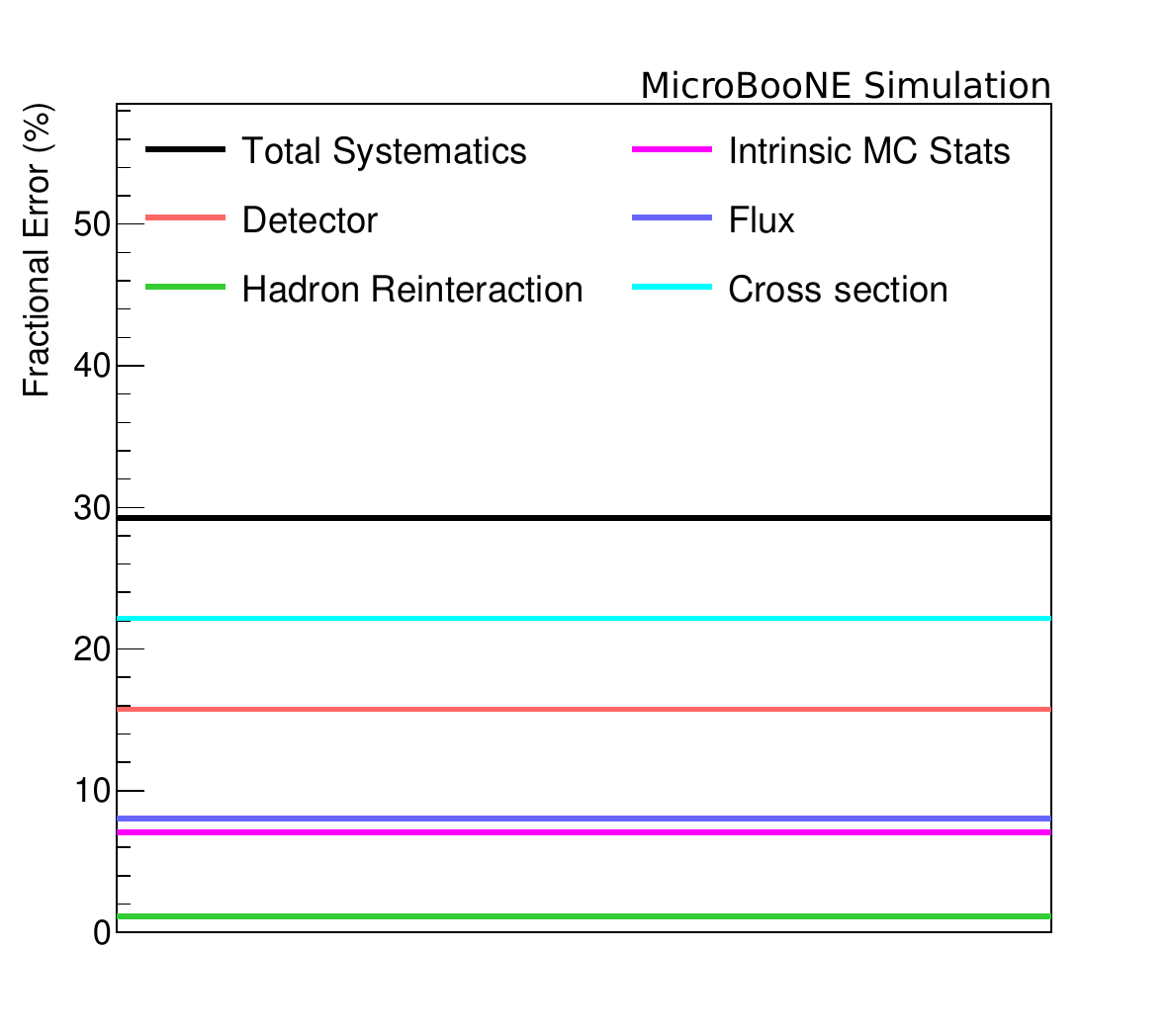}
    \caption{Overall systematic uncertainty for the coherent-rich subset selection, with contribution from each source highlighted in different colors, and in black is the sum in quadrature of all systematics. Driven by the low statistics available at the coherent-rich stage, a single bin is used. }
    \label{fig:final_error_budget}
\end{figure}

The systematic uncertainties are encapsulated in covariance matrices, which describe the covariance across bins in a predicted spectrum or across different predicted spectra. For each systematic error and associated variation(s), a covariance matrix is constructed via:
\begin{equation}
    M_{ij} = \frac{1}{N}\sum_{k = 1}^{N}(n_i^k - n_i^{CV})(n_j^k - n_{j}^{CV})
\end{equation}
where $M_{ij}$ is the resulting covariance matrix describing the covariance between bin $i$ and bin $j$ in the predicted spectrum, $N$ is the number of universes generated for a given systematic error source, $n_i^k$ is the prediction in bin $i$ in the $k$th universe and $n_i^{CV}$ is the prediction in the same bin of the CV simulation. For reweightable systematics, the $n^k$ spectrum is calculated by reweighting simulated events in the selection, and for the detector systematics, the $n^k$ spectrum is the spectrum of the variation sample after applying all selection cuts discussed in Sec.~\ref{sec:selection}. The breakdown of uncertainty contributions into the four main categories for the single-photon selection is shown in Fig.~\ref{fig:semi_final_error_budget} as a function of the reconstructed shower energy. The dominant uncertainty contribution at the single-photon selection stage is the uncertainty in detector modeling, followed by the uncertainty in cross section modeling. As this is a rare signal search, the MC statistics in the detector variation samples can get quite low in a given bin when one considers a binned spectra as in Fig.~\ref{fig:semi_final_error_budget}. Thus, for the purpose of a final fit to search for coherent single-photon events we use a single bin, integrating over the entire shower energy range. The result of this can be found in Fig.~\ref{fig:final_error_budget}, where MC statistics in the detector variation samples are less of a concern, resulting in an overall lower detector systematic uncertainty.  Nonetheless, cross section and detector uncertainty remain the two largest contributions.

\section{$\pi^0$ constraint}\label{sec:pi0_constraint}
The goal of this analysis is to search for the NC coherent $1\gamma$ process in the observed data. This is quantified by a normalization scaling factor $x$ of the nominal SM predicted rate for this process. Extraction of the parameter $x$ and its range is achieved by fitting the MC prediction to the observed data while allowing the scaling factor $x$ to vary. In order to yield higher sensitivity to the parameter $x$, external high-statistics measurements are used in the fit to constrain the predicted background and systematic uncertainty in the signal region, in a manner similar to MicroBooNE's LEE analyses~\cite{glee_delta, WC_lee, PELEE, DLEE, CombELEE}.

Given its dominant contribution, NC $1\pi^0$ selections, developed for MicroBooNE's Pandora-based NC $\Delta$ radiative decay search (detailed in Ref.~\cite{gLEE_pi0}), are chosen as the constraining sample for this analysis, with $\approx$20\% more data than was used before in Ref.~\cite{gLEE_pi0}. There are two $1\pi^0$  selections: $2\gamma1p$ and $2\gamma0p$. While both of these selections aim to reconstruct NC $1\pi^0$ events, $2\gamma1p$ specifically targets $\pi^0$ events originating from resonant $\Delta \rightarrow N \pi^0$ decays while the $2\gamma0p$ selection requires no reconstructed tracks. 
\begin{figure}[h!]
    \centering 
    \begin{subfigure}{0.49\textwidth}
        \includegraphics[width = \textwidth,trim=0cm 1.5cm 0cm 0cm, clip]{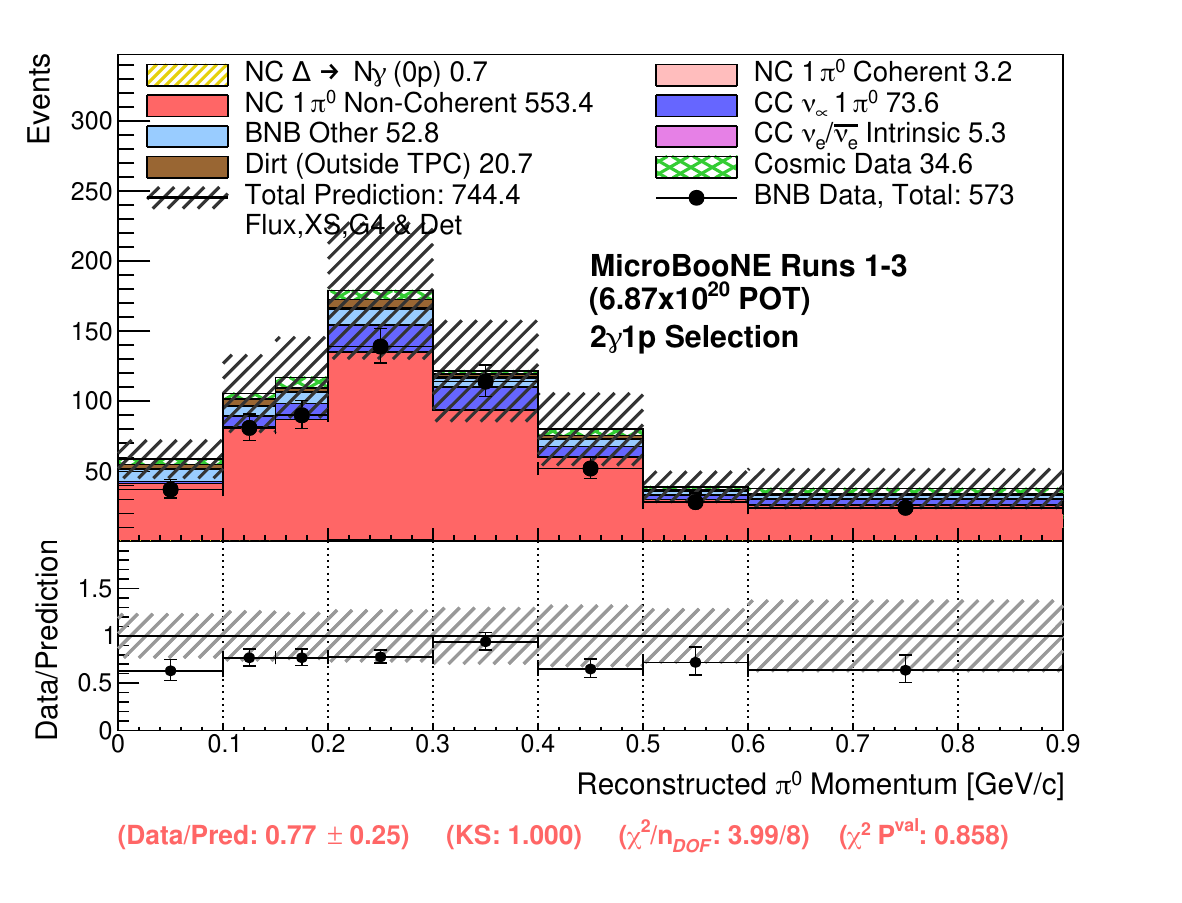}
        \caption{$2\gamma1p$}
  \end{subfigure} 
  \begin{subfigure}{0.49\textwidth}
        \includegraphics[width = \textwidth,trim=0cm 1.5cm 0cm 0cm, clip]{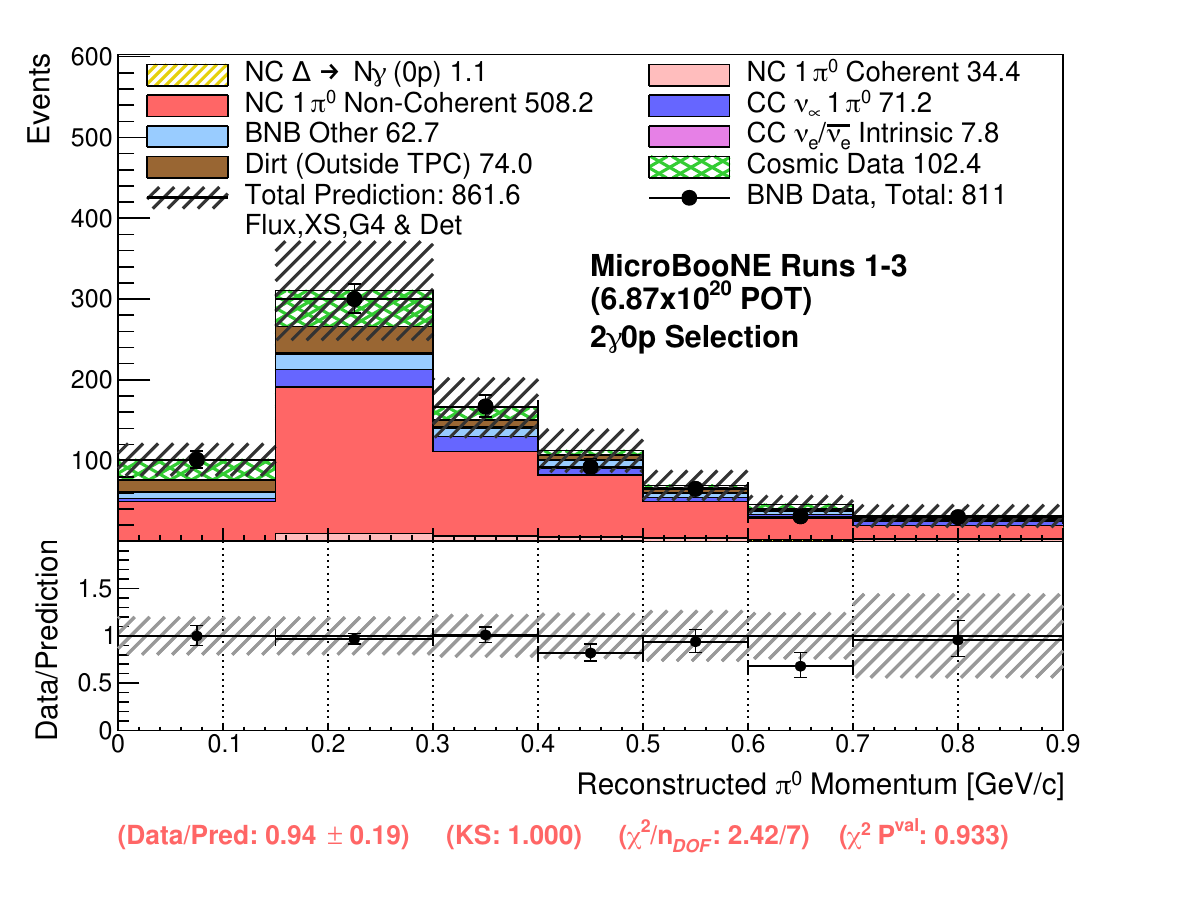}
        \caption{$2\gamma0p$}
  \end{subfigure} 
  \caption{Distributions of the $2\gamma1p$ and $2\gamma0p$ constraining samples as a function of the reconstructed $\pi^0$ momentum, shown in the binning used in the simultaneous fit. A mild deficit in data is observed in the $2\gamma1p$ sample, but in general, good data-MC agreement is seen within the systematic uncertainty.}
  \label{fig:2g}
\end{figure}

We fit three samples (signal selection, $2\gamma1p$ and $2\gamma0p$ selection) simultaneously to maximize the sensitivity to the NC coherent $1\gamma$ process. 
The distributions of the $2\gamma$ samples used in the fit are shown in Fig.~\ref{fig:2g}.  Following the conditional constraint approach described in~\cite{PELEE}, we observe that the 2$\gamma$ constraint reduces the MC background prediction from   $34.0 \pm 8.4 \text{(syst)} \pm 5.8 \text{(stats)}$  to  $29.0 \pm 5.2 \text{(syst)} \pm 5.4 \text{(stats)}$ for the coherent-rich signal selection, which corresponds to a $\sim$25\% reduction in the systematic uncertainty.

\section{Results}
\subsection{Sideband}\label{sec:sideband}
The analysis is designed as a blind analysis, where 10\% of the total data during the first run period is open and utilized to develop the analysis and validate the agreement between data and MC in variables of interest and input variables to all BDTs involved. The analysis selections presented and the backgrounds involved are similar to those of Ref.~\cite{glee_delta}, and the NC $\pi^0$ and BNB Other backgrounds with a single-shower topology are validated in Ref.~\cite{glee_delta}. The performance of the shower energy reconstruction and validation of NC $\pi^0$ modeling are demonstrated in the $2\gamma1p$ and $2\gamma0p$ NC $\pi^0$ measurements with the full dataset~\cite{gLEE_pi0}. 

\begin{figure}[h!]
    \centering 
    \begin{subfigure}{0.49\textwidth}
        \includegraphics[width = \textwidth,trim=0cm 1.5cm 0cm 0cm, clip]{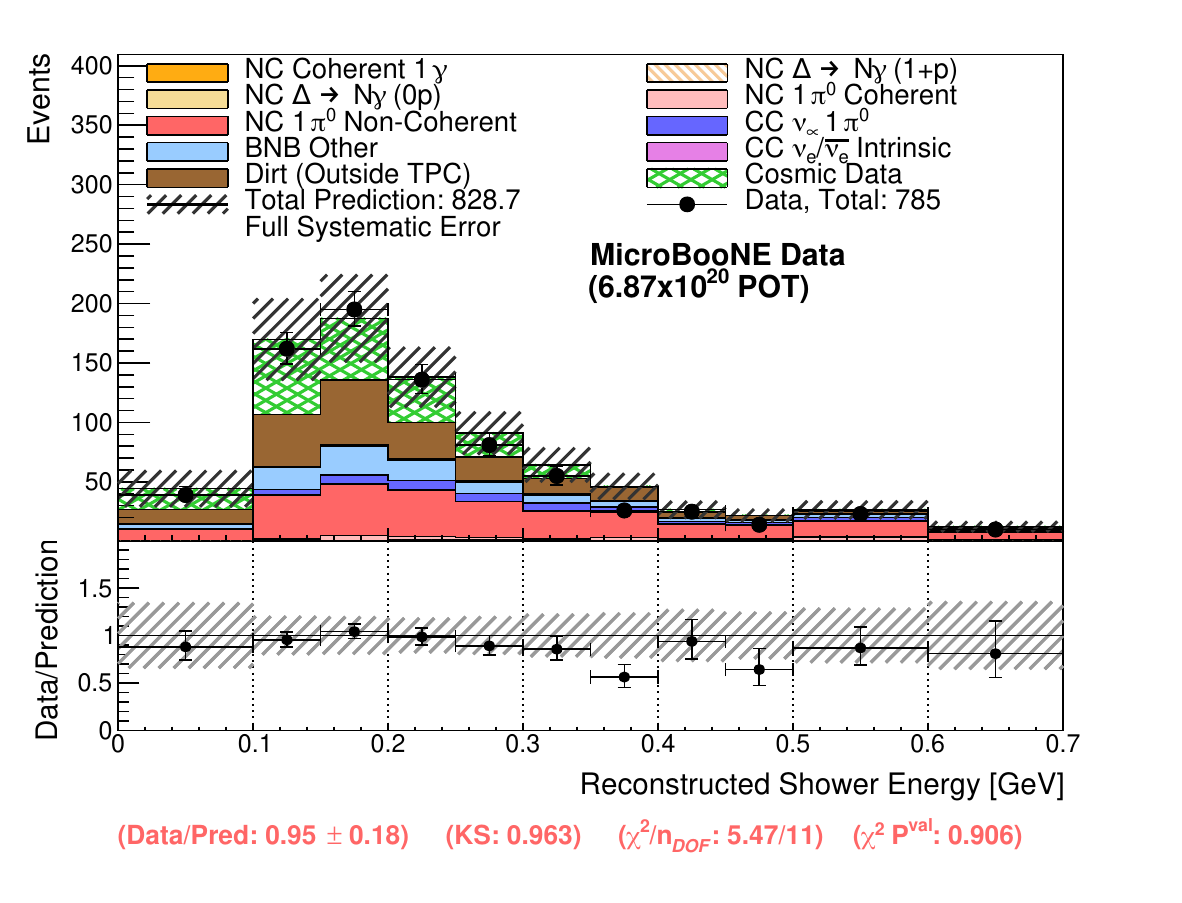}
        \caption{Reconstructed Shower Energy}
        \label{fig:cosmic_bdt}
  \end{subfigure} 
  \begin{subfigure}{0.49\textwidth}
        \includegraphics[width = \textwidth,trim=0cm 1.5cm 0cm 0cm, clip]{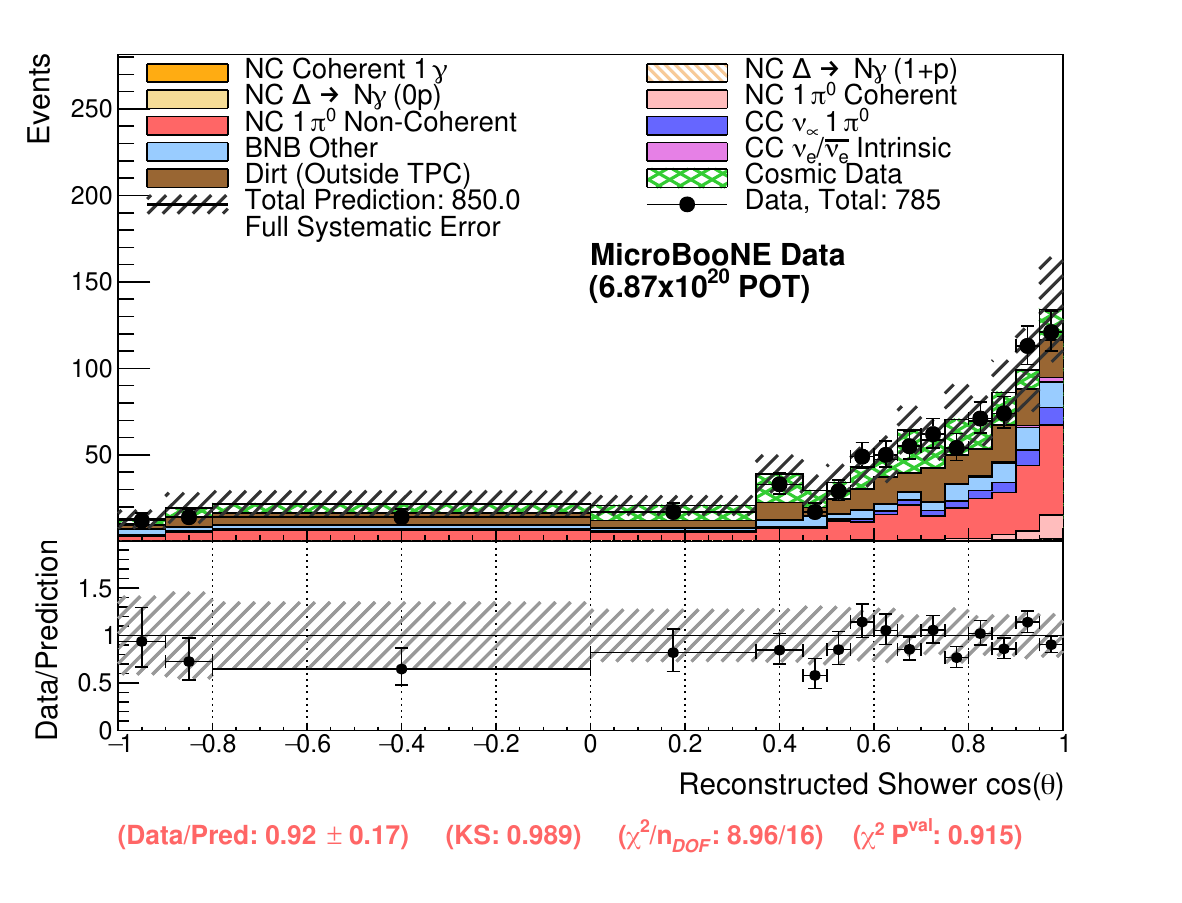}
        \caption{Reconstructed Shower cos($\theta$)}
  \end{subfigure} 
  \begin{subfigure}{0.49\textwidth}
        \includegraphics[width = \textwidth,trim=0cm 1.5cm 0cm 0cm, clip]{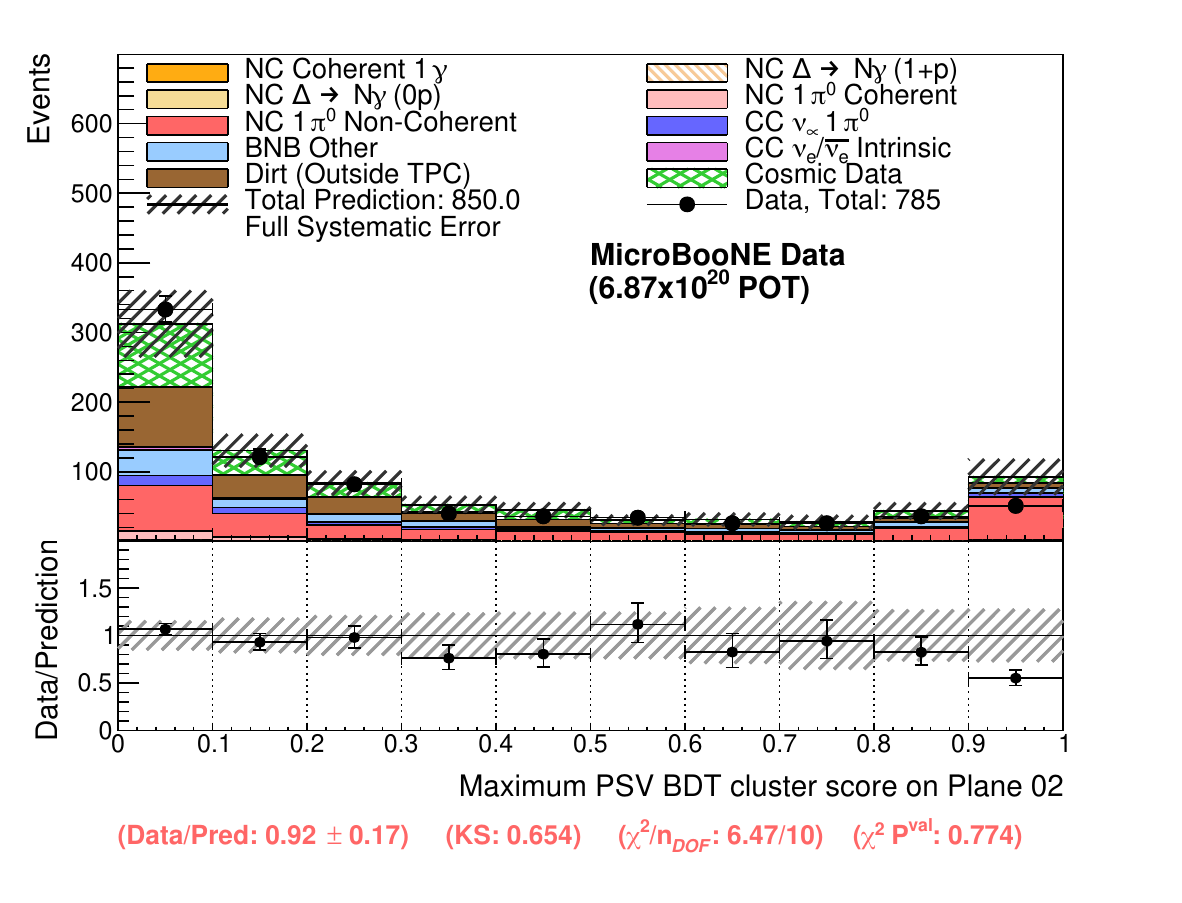}
        \caption{Maximum PSV score between Plane 0 and Plane 2}
  \end{subfigure} 
  \caption{Distributions of the sideband sample in: (a) reconstructed shower energy; (b) reconstructed shower cos($\theta$); and (c) the maximum PSV score between Plane 0 and Plane 2. A mild data deficit is observed in the highest PSV score bin, but data is found to be consistent with prediction within systematic uncertainty over all three variables.}
  \label{fig:sideband}
\end{figure}

To validate the performance of the PSV BDT, after topological and preselection cuts, we also isolate a sideband using the full dataset that is proton-rich by reversing the cut on the NC $\pi^0$ BDT ($<$0.891) while relaxing cuts on other event-level BDTs: Cosmic BDT $>$0.8, CC $\nu_{\mu}$ BDT $>$0.8, CC $\nu_{e}$  BDT $>$0.5. A total of 850.0 events are predicted, of which 54.0\% of non-cosmic MC events are predicted to have protons exiting the nucleus (regardless of proton KE). 
Consistency between data and MC is inspected through the goodness-of-fit test that incorporates systematic uncertainties and the Combined-Neyman-Pearson statistical uncertainty~\cite{CNP}. Figure~\ref{fig:sideband} shows the distributions of the reconstructed shower energy, angle, and the maximum PSV score on planes 0 and 2 for the sideband sample. A mild deficit in data is observed in the region of high PSV score. The subsample of events with high PSV score (above 0.8) was isolated and found to be uniformly distributed in shower kinematic variables. Overall, the data-MC agreement for the PSV variable is good, with a corresponding $p$-value of 0.78 when all systematic uncertainties are taken into account. Event-level and cluster-level variables including all BDT input variables were also inspected, and found to give good agreement within systematic uncertainty.

\clearpage
\subsection{Coherent-Rich Selection Results}\label{sec:final_results}
For the single-photon selection, 70 events are observed in data compared to an MC prediction of 55.8, leading to a data-MC ratio of $1.25\pm0.32 (\text{sys})$. Distributions of the reconstructed shower energy, angle (cos$\theta$) and the maximum PSV score on planes 0 and 2 were previously shown in Fig.~\ref{fig:semi_final}. Overall, the agreement between data-MC is good according to the goodness-of-fit tests with $p$-values of 0.251, 0.830, and 0.476, respectively.

\begin{figure}[h!]
    \centering
    \includegraphics[width=1\linewidth, trim=0cm 0cm 0cm 1cm,clip]{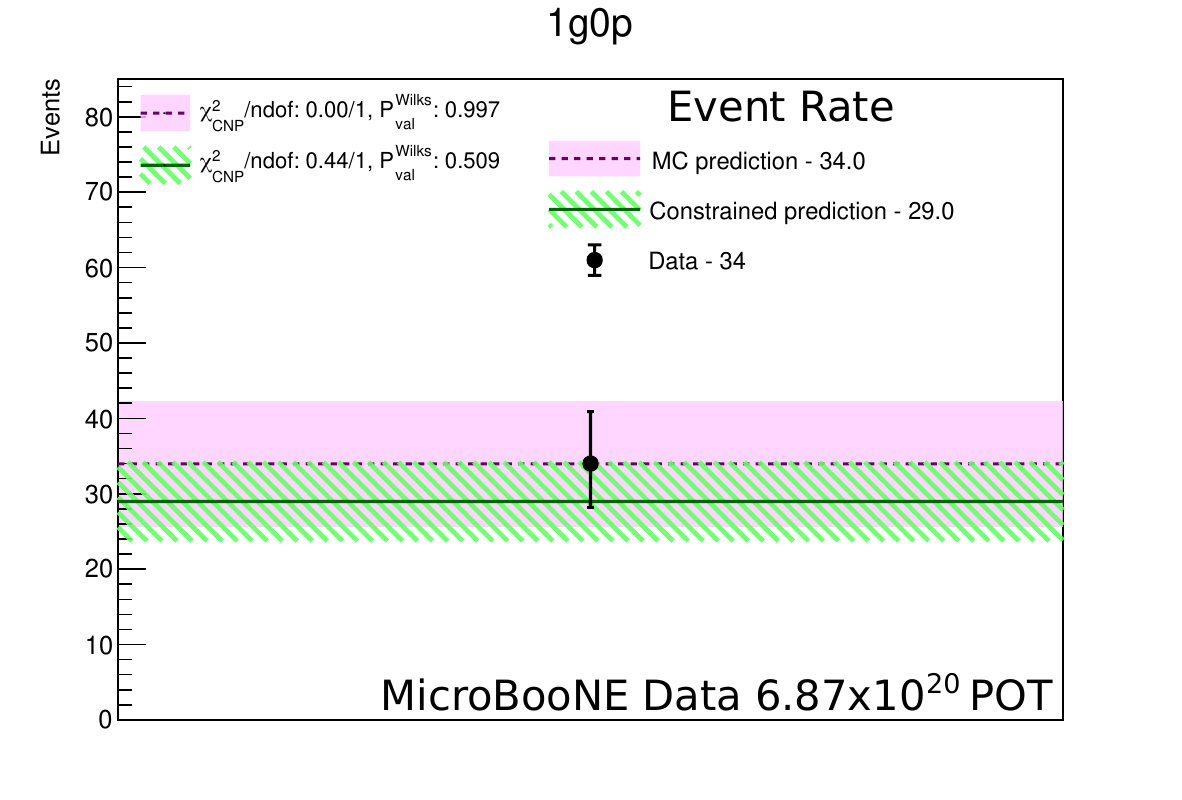}
    \caption{Comparison of observed data and MC CV predictions for the coherent-rich selection. The red (green) line and surrounding band represent the MC prediction and associated systematic uncertainty before (after) the $2\gamma$ constraint is applied. The breakdown of MC prediction in different categories can be found in Tab.~\ref{tab:final_prediction}.}
    \label{fig:final_data}
\end{figure}

\begin{figure}[h!]
    \centering 
    \begin{subfigure}{0.49\textwidth}
        \includegraphics[width = \textwidth,trim=0cm 1.5cm 0cm 0cm, clip]{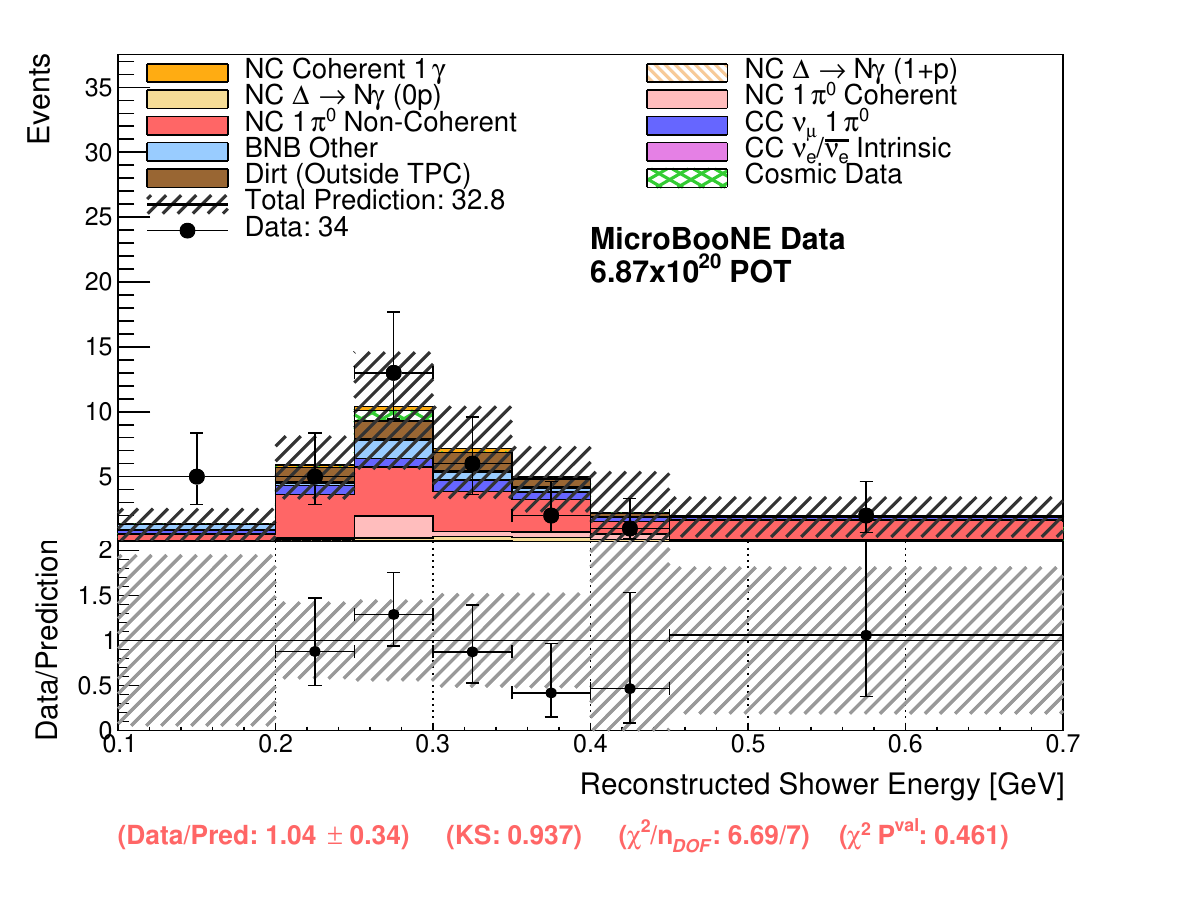}
        \caption{Reconstructed Shower Energy}
        \label{fig:cosmic_bdt}
  \end{subfigure} 
  \begin{subfigure}{0.49\textwidth}
        \includegraphics[width = \textwidth,trim=0cm 1.5cm 0cm 0cm, clip]{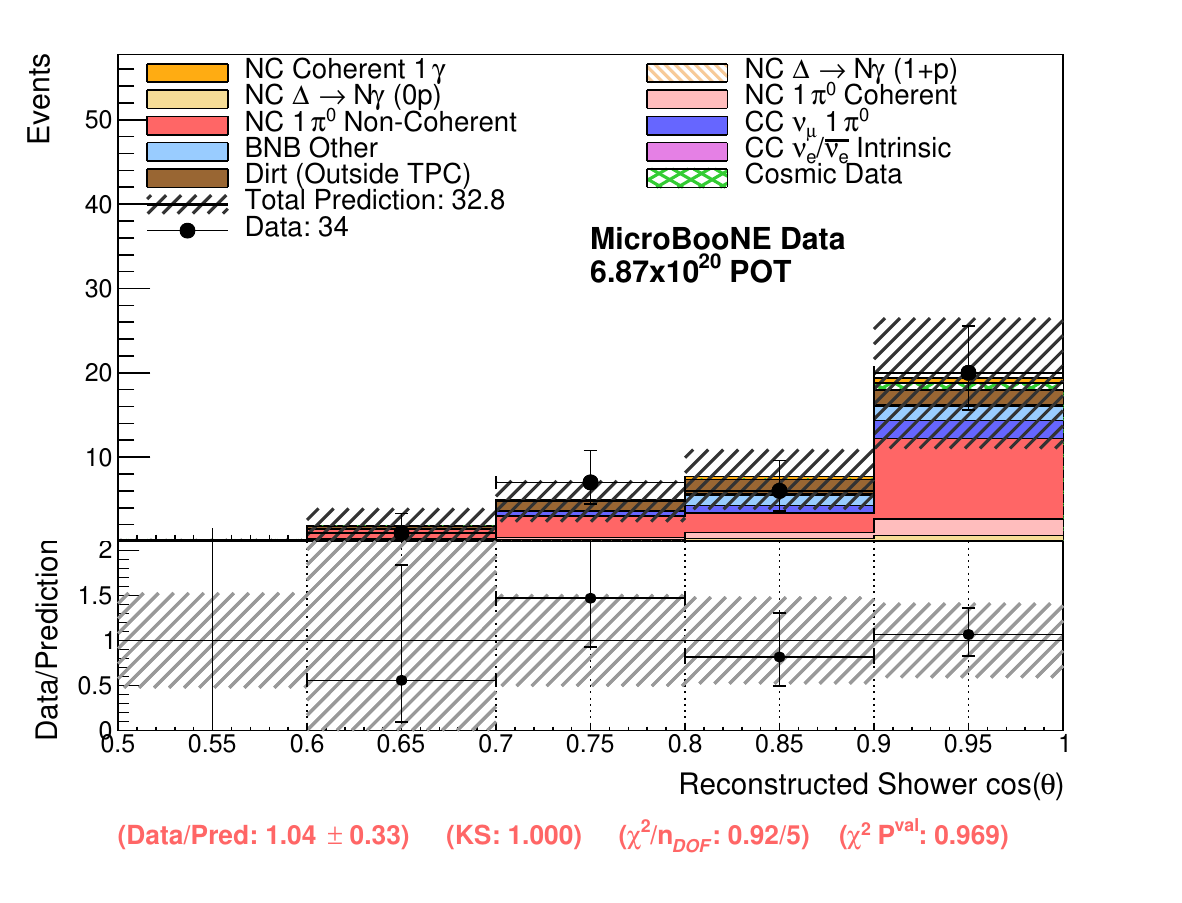}
        \caption{Reconstructed Shower cos($\theta$)}
  \end{subfigure} 
  \caption{Distributions of (a) reconstructed shower energy; (b) reconstructed shower cos($\theta$) for all events in the coherent-rich subset selection. We note that, due to low statistics, the fit only considers the total number of events in these distributions; the distributions are shown as additional information. }
  \label{fig:semi_final_data}
\end{figure}

For the coherent-rich subset selection, 34 data events are observed, compared to a constrained prediction of $29.0 \pm 5.2 \text{(syst)} \pm 5.4 \text{(stats)}$ and an unconstrained prediction of $34.0 \pm 8.4 \text{(syst)} \pm 5.8 \text{(stats)}$. The constrained background prediction is lower than unconstrained due to the mild deficit observed in data in the NC $\pi^0$ high statistics two-shower constraining channels, as can be seen in Fig.~\ref{fig:2g}. The comparison between data and MC together with results from goodness-of-fit tests are shown in Fig~\ref{fig:final_data}. Data-MC agreement slightly worsens after the constraint from the $2\gamma$ channels is applied, from a $p$-value of 1.00 to a $p$-value of 0.51, but nonetheless remains well within expectations. Distributions of the 34 data events and our simulation prediction can be found in Fig.~\ref{fig:semi_final_data}.

\begin{figure}[h!]
    \centering
    \includegraphics[width=\linewidth,trim=9.7cm 0cm 0cm 0cm, clip]{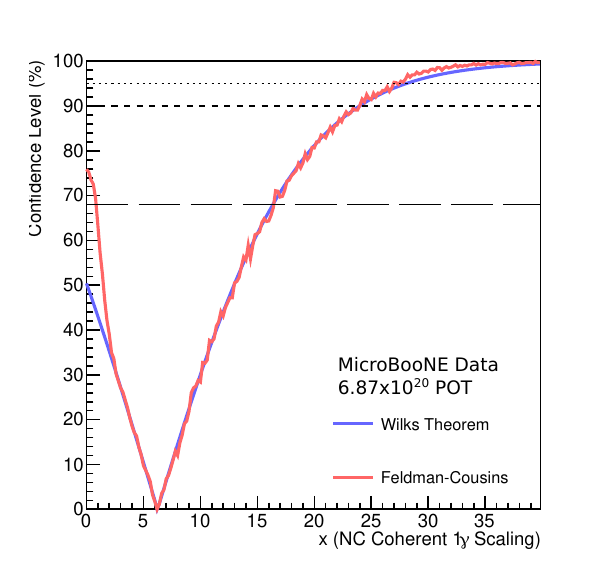}
    \caption{The confidence level for the NC coherent $1\gamma$ normalization scaling factor $x$ derived from observed data in the coherent-rich selection. The C.L.~evaluated with the Feldman-Counsins method is shown in pink with C.L.~assuming the validity of Wilks' theorem overlaid in blue. Dotted, dashed and large-dashed horizontal lines correspond to $2\sigma$, $90\%$ and $1\sigma$ respectively. }
    \label{fig:FC_corrected_fit}
\end{figure}

The observed data in the coherent-rich selection is fit to extract a normalization scaling factor $x$ for the NC coherent $1\gamma$ process, with full systematic uncertainties included except for the \textsc{genie} cross section uncertainty for the signal process. The best-fit for $x$ is found to be $x = 6.20$ with a $\chi^2/ndf = 9.7/15$. Figure~\ref{fig:FC_corrected_fit} shows the confidence level (C.L.) for different values of $x$ as evaluated using the Feldman-Cousins (FC) frequentist method~\cite{Feldman:1997qc}, with the C.L. under the assumption of Wilks' theorem overlaid. We see that curves from FC and Wilks' theorem agree with each other in the majority of the parameter space and start to differ in the parameter region near the boundary, as expected. Given the observed data, the bound at $90\%$~C.L.~is placed at $x=24.0$ and the $95\%$~C.L.~bound is at $x=28.0$ for the NC coherent $1\gamma$, corresponding to a limit on the flux-averaged cross section of $<1.49 \times 10^{-41}\text{~cm}^2$ at 90\% C.L.. Note that the slight fluctuations in the confidence level curve for the FC method are due to the finite number of pseudo-experiments used. 

\newpage
\section{Summary}
In this paper, we present the world's first search for the neutrino-induced NC coherent single-photon process with a single shower topology and no other visible vertex activity. The search is performed with the first three years of data from the MicroBooNE detector, with a neutrino beam of $\langle E_{\nu} \rangle \sim 0.8$~GeV from the Fermilab Booster Neutrino Beam. Differences in the shower kinematic properties between the shower originating from coherent photons and that from other backgrounds are leveraged to yield effective background removal, leading to 99.97\% and $>95$\% rejection efficiency on the cosmic and NC non-coherent 1$\pi^0$ background relative to Pandora's topological selection, respectively. New tools utilizing low-level TPC charge information are developed to identify and reject low-energy proton tracks near the vertex, which further reduces the dominant NC non-coherent 1$\pi^0$ background by 48\%. This search yields a signal-to-background ratio of $\approx1:30$ in the final coherent-rich selection, and good agreement between data and prediction is observed. This leads to the world's first experimental upper limit on the cross section of neutrino-induced NC coherent single-photon production on an argon nucleus below 1 GeV, of $\sigma<1.49 \times 10^{-41} \text{cm}^2$ at 90\% C.L., corresponding to 24.0 times the prediction in Ref.~\cite{coh_gamma_model}. 

The proton veto tool developed in this analysis has shown great potential in rejecting low-energy protons and represents the first high-level use of MeV scale energy deposits in MicroBooNE. This veto can be easily adapted to other coherent interaction searches. Furthermore, it indicates the way forward for enhancing the reconstruction of low-energy particles, including searching for evidence of activity due to neutron reinteractions. While this search has limited sensitivity, the selections developed can be applied to future LArTPC experiments for a search for this process with higher statistics. An example of this is the upcoming SBND experiment which is expected to collect $\mathcal{O}(20)$ times more data~\cite{SBND} than MicroBooNE, potentially enabling a limit on this process  much closer to the SM-predicted rate.

\begin{acknowledgments}
This document was prepared by the MicroBooNE collaboration using the
resources of the Fermi National Accelerator Laboratory (Fermilab), a
U.S. Department of Energy, Office of Science, Office of High Energy Physics HEP User Facility.
Fermilab is managed by Fermi Forward Discovery Group, LLC, acting
under Contract No. 89243024CSC000002.  MicroBooNE is supported by the
following: 
the U.S. Department of Energy, Office of Science, Offices of High Energy Physics and Nuclear Physics; 
the U.S. National Science Foundation; 
the Swiss National Science Foundation; 
the Science and Technology Facilities Council (STFC), part of the United Kingdom Research and Innovation; 
the Royal Society (United Kingdom); 
the UK Research and Innovation (UKRI) Future Leaders Fellowship; 
and the NSF AI Institute for Artificial Intelligence and Fundamental Interactions. 
Additional support for 
the laser calibration system and cosmic ray tagger was provided by the 
Albert Einstein Center for Fundamental Physics, Bern, Switzerland. We 
also acknowledge the contributions of technical and scientific staff 
to the design, construction, and operation of the MicroBooNE detector 
as well as the contributions of past collaborators to the development 
of MicroBooNE analyses, without whom this work would not have been 
possible. 
For the purpose of open access, the authors have applied 
a Creative Commons Attribution (CC BY) public copyright license to 
any Author Accepted Manuscript version arising from this submission.

\end{acknowledgments}

\appendix

\bibliography{references}

\begin{thebibliography}{64}%
\makeatletter
\providecommand \@ifxundefined [1]{%
 \@ifx{#1\undefined}
}%
\providecommand \@ifnum [1]{%
 \ifnum #1\expandafter \@firstoftwo
 \else \expandafter \@secondoftwo
 \fi
}%
\providecommand \@ifx [1]{%
 \ifx #1\expandafter \@firstoftwo
 \else \expandafter \@secondoftwo
 \fi
}%
\providecommand \natexlab [1]{#1}%
\providecommand \enquote  [1]{``#1''}%
\providecommand \bibnamefont  [1]{#1}%
\providecommand \bibfnamefont [1]{#1}%
\providecommand \citenamefont [1]{#1}%
\providecommand \href@noop [0]{\@secondoftwo}%
\providecommand \href [0]{\begingroup \@sanitize@url \@href}%
\providecommand \@href[1]{\@@startlink{#1}\@@href}%
\providecommand \@@href[1]{\endgroup#1\@@endlink}%
\providecommand \@sanitize@url [0]{\catcode `\\12\catcode `\$12\catcode `\&12\catcode `\#12\catcode `\^12\catcode `\_12\catcode `\%12\relax}%
\providecommand \@@startlink[1]{}%
\providecommand \@@endlink[0]{}%
\providecommand \url  [0]{\begingroup\@sanitize@url \@url }%
\providecommand \@url [1]{\endgroup\@href {#1}{\urlprefix }}%
\providecommand \urlprefix  [0]{URL }%
\providecommand \Eprint [0]{\href }%
\providecommand \doibase [0]{http://dx.doi.org/}%
\providecommand \selectlanguage [0]{\@gobble}%
\providecommand \bibinfo  [0]{\@secondoftwo}%
\providecommand \bibfield  [0]{\@secondoftwo}%
\providecommand \translation [1]{[#1]}%
\providecommand \BibitemOpen [0]{}%
\providecommand \bibitemStop [0]{}%
\providecommand \bibitemNoStop [0]{.\EOS\space}%
\providecommand \EOS [0]{\spacefactor3000\relax}%
\providecommand \BibitemShut  [1]{\csname bibitem#1\endcsname}%
\let\auto@bib@innerbib\@empty
\bibitem [{\citenamefont {Workman}\ \emph {et~al.}(2022)\citenamefont {Workman} \emph {et~al.}}]{pdg_review}%
  \BibitemOpen
  \bibfield  {author} {\bibinfo {author} {\bibfnamefont {R.~L.}\ \bibnamefont {Workman}} \emph {et~al.} (\bibinfo {collaboration} {Particle Data Group}),\ }\href {\doibase 10.1093/ptep/ptac097} {\bibfield  {journal} {\bibinfo  {journal} {PTEP}\ }\textbf {\bibinfo {volume} {2022}},\ \bibinfo {pages} {083C01} (\bibinfo {year} {2022})}\BibitemShut {NoStop}%
\bibitem [{\citenamefont {Abratenko}\ \emph {et~al.}(2022{\natexlab{a}})\citenamefont {Abratenko} \emph {et~al.}}]{PhysRevD.106.L051102}%
  \BibitemOpen
  \bibfield  {author} {\bibinfo {author} {\bibfnamefont {P.}~\bibnamefont {Abratenko}} \emph {et~al.} (\bibinfo {collaboration} {MicroBooNE Collaboration}),\ }\href {\doibase 10.1103/PhysRevD.106.L051102} {\bibfield  {journal} {\bibinfo  {journal} {Phys. Rev. D}\ }\textbf {\bibinfo {volume} {106}},\ \bibinfo {pages} {L051102} (\bibinfo {year} {2022}{\natexlab{a}})}\BibitemShut {NoStop}%
\bibitem [{\citenamefont {Abratenko}\ \emph {et~al.}(2023{\natexlab{a}})\citenamefont {Abratenko} \emph {et~al.}}]{PhysRevD.107.012004}%
  \BibitemOpen
  \bibfield  {author} {\bibinfo {author} {\bibfnamefont {P.}~\bibnamefont {Abratenko}} \emph {et~al.} (\bibinfo {collaboration} {MicroBooNE Collaboration}),\ }\href {\doibase 10.1103/PhysRevD.107.012004} {\bibfield  {journal} {\bibinfo  {journal} {Phys. Rev. D}\ }\textbf {\bibinfo {volume} {107}},\ \bibinfo {pages} {012004} (\bibinfo {year} {2023}{\natexlab{a}})}\BibitemShut {NoStop}%
\bibitem [{\citenamefont {Abratenko}\ \emph {et~al.}(2021{\natexlab{a}})\citenamefont {Abratenko} \emph {et~al.}}]{PhysRevD.104.052002}%
  \BibitemOpen
  \bibfield  {author} {\bibinfo {author} {\bibfnamefont {P.}~\bibnamefont {Abratenko}} \emph {et~al.} (\bibinfo {collaboration} {MicroBooNE Collaboration}),\ }\href {\doibase 10.1103/PhysRevD.104.052002} {\bibfield  {journal} {\bibinfo  {journal} {Phys. Rev. D}\ }\textbf {\bibinfo {volume} {104}},\ \bibinfo {pages} {052002} (\bibinfo {year} {2021}{\natexlab{a}})}\BibitemShut {NoStop}%
\bibitem [{\citenamefont {Abratenko}\ \emph {et~al.}(2022{\natexlab{b}})\citenamefont {Abratenko} \emph {et~al.}}]{PhysRevD.105.L051102}%
  \BibitemOpen
  \bibfield  {author} {\bibinfo {author} {\bibfnamefont {P.}~\bibnamefont {Abratenko}} \emph {et~al.} (\bibinfo {collaboration} {MicroBooNE Collaboration}),\ }\href {\doibase 10.1103/PhysRevD.105.L051102} {\bibfield  {journal} {\bibinfo  {journal} {Phys. Rev. D}\ }\textbf {\bibinfo {volume} {105}},\ \bibinfo {pages} {L051102} (\bibinfo {year} {2022}{\natexlab{b}})}\BibitemShut {NoStop}%
\bibitem [{\citenamefont {Abratenko}\ \emph {et~al.}(2020{\natexlab{a}})\citenamefont {Abratenko} \emph {et~al.}}]{PhysRevLett.125.201803}%
  \BibitemOpen
  \bibfield  {author} {\bibinfo {author} {\bibfnamefont {P.}~\bibnamefont {Abratenko}} \emph {et~al.} (\bibinfo {collaboration} {MicroBooNE Collaboration}),\ }\href {\doibase 10.1103/PhysRevLett.125.201803} {\bibfield  {journal} {\bibinfo  {journal} {Phys. Rev. Lett.}\ }\textbf {\bibinfo {volume} {125}},\ \bibinfo {pages} {201803} (\bibinfo {year} {2020}{\natexlab{a}})}\BibitemShut {NoStop}%
\bibitem [{\citenamefont {Abratenko}\ \emph {et~al.}(2020{\natexlab{b}})\citenamefont {Abratenko} \emph {et~al.}}]{PhysRevD.102.112013}%
  \BibitemOpen
  \bibfield  {author} {\bibinfo {author} {\bibfnamefont {P.}~\bibnamefont {Abratenko}} \emph {et~al.} (\bibinfo {collaboration} {MicroBooNE Collaboration}),\ }\href {\doibase 10.1103/PhysRevD.102.112013} {\bibfield  {journal} {\bibinfo  {journal} {Phys. Rev. D}\ }\textbf {\bibinfo {volume} {102}},\ \bibinfo {pages} {112013} (\bibinfo {year} {2020}{\natexlab{b}})}\BibitemShut {NoStop}%
\bibitem [{\citenamefont {Abratenko}\ \emph {et~al.}(2019)\citenamefont {Abratenko} \emph {et~al.}}]{PhysRevLett.123.131801}%
  \BibitemOpen
  \bibfield  {author} {\bibinfo {author} {\bibfnamefont {P.}~\bibnamefont {Abratenko}} \emph {et~al.} (\bibinfo {collaboration} {MicroBooNE Collaboration}),\ }\href {\doibase 10.1103/PhysRevLett.123.131801} {\bibfield  {journal} {\bibinfo  {journal} {Phys. Rev. Lett.}\ }\textbf {\bibinfo {volume} {123}},\ \bibinfo {pages} {131801} (\bibinfo {year} {2019})}\BibitemShut {NoStop}%
\bibitem [{\citenamefont {Aguilar-Arevalo}\ \emph {et~al.}(2021)\citenamefont {Aguilar-Arevalo} \emph {et~al.}}]{MiniBooNE_combine_osc}%
  \BibitemOpen
  \bibfield  {author} {\bibinfo {author} {\bibfnamefont {A.~A.}\ \bibnamefont {Aguilar-Arevalo}} \emph {et~al.} (\bibinfo {collaboration} {MiniBooNE Collaboration}),\ }\href {\doibase 10.1103/PhysRevD.103.052002} {\bibfield  {journal} {\bibinfo  {journal} {Phys. Rev. D}\ }\textbf {\bibinfo {volume} {103}},\ \bibinfo {pages} {052002} (\bibinfo {year} {2021})}\BibitemShut {NoStop}%
\bibitem [{\citenamefont {Abratenko}\ \emph {et~al.}(2023{\natexlab{b}})\citenamefont {Abratenko} \emph {et~al.}}]{MicroBooNE:2022sdp}%
  \BibitemOpen
  \bibfield  {author} {\bibinfo {author} {\bibfnamefont {P.}~\bibnamefont {Abratenko}} \emph {et~al.} (\bibinfo {collaboration} {{MicroBooNE Collaboration}}),\ }\href {\doibase 10.1103/PhysRevLett.130.011801} {\bibfield  {journal} {\bibinfo  {journal} {Phys. Rev. Lett.}\ }\textbf {\bibinfo {volume} {130}},\ \bibinfo {pages} {011801} (\bibinfo {year} {2023}{\natexlab{b}})},\ \Eprint {http://arxiv.org/abs/2210.10216} {arXiv:2210.10216 [hep-ex]} \BibitemShut {NoStop}%
\bibitem [{\citenamefont {Machado}\ \emph {et~al.}(2019)\citenamefont {Machado}, \citenamefont {Palamara},\ and\ \citenamefont {Schmitz}}]{SBN}%
  \BibitemOpen
  \bibfield  {author} {\bibinfo {author} {\bibfnamefont {P.~A.}\ \bibnamefont {Machado}}, \bibinfo {author} {\bibfnamefont {O.}~\bibnamefont {Palamara}}, \ and\ \bibinfo {author} {\bibfnamefont {D.~W.}\ \bibnamefont {Schmitz}},\ }\href {\doibase 10.1146/annurev-nucl-101917-020949} {\bibfield  {journal} {\bibinfo  {journal} {Annu. Rev. Nucl. Part. Sci.}\ }\textbf {\bibinfo {volume} {69}},\ \bibinfo {pages} {363} (\bibinfo {year} {2019})},\ \Eprint {http://arxiv.org/abs/https://doi.org/10.1146/annurev-nucl-101917-020949} {https://doi.org/10.1146/annurev-nucl-101917-020949} \BibitemShut {NoStop}%
\bibitem [{\citenamefont {Abud~Abed}\ \emph {et~al.}(2022)\citenamefont {Abud~Abed} \emph {et~al.}}]{DUNE:2021mtg}%
  \BibitemOpen
  \bibfield  {author} {\bibinfo {author} {\bibfnamefont {A.}~\bibnamefont {Abud~Abed}} \emph {et~al.} (\bibinfo {collaboration} {DUNE}),\ }\href {\doibase 10.1103/PhysRevD.105.072006} {\bibfield  {journal} {\bibinfo  {journal} {Phys. Rev. D}\ }\textbf {\bibinfo {volume} {105}},\ \bibinfo {pages} {072006} (\bibinfo {year} {2022})},\ \Eprint {http://arxiv.org/abs/2109.01304} {arXiv:2109.01304 [hep-ex]} \BibitemShut {NoStop}%
\bibitem [{\citenamefont {Wang}\ \emph {et~al.}(2014)\citenamefont {Wang}, \citenamefont {Alvarez-Ruso},\ and\ \citenamefont {Nieves}}]{coh_gamma_model}%
  \BibitemOpen
  \bibfield  {author} {\bibinfo {author} {\bibfnamefont {E.}~\bibnamefont {Wang}}, \bibinfo {author} {\bibfnamefont {L.}~\bibnamefont {Alvarez-Ruso}}, \ and\ \bibinfo {author} {\bibfnamefont {J.}~\bibnamefont {Nieves}},\ }\href {\doibase 10.1103/PhysRevC.89.015503} {\bibfield  {journal} {\bibinfo  {journal} {Phys. Rev. C}\ }\textbf {\bibinfo {volume} {89}},\ \bibinfo {pages} {015503} (\bibinfo {year} {2014})}\BibitemShut {NoStop}%
\bibitem [{\citenamefont {Wang}\ \emph {et~al.}(2015)\citenamefont {Wang}, \citenamefont {Alvarez-Ruso},\ and\ \citenamefont {Nieves}}]{coh_gamma_pred}%
  \BibitemOpen
  \bibfield  {author} {\bibinfo {author} {\bibfnamefont {E.}~\bibnamefont {Wang}}, \bibinfo {author} {\bibfnamefont {L.}~\bibnamefont {Alvarez-Ruso}}, \ and\ \bibinfo {author} {\bibfnamefont {J.}~\bibnamefont {Nieves}},\ }\href {\doibase 10.1016/j.physletb.2014.11.025} {\bibfield  {journal} {\bibinfo  {journal} {Phys. Lett. B}\ }\textbf {\bibinfo {volume} {740}},\ \bibinfo {pages} {16} (\bibinfo {year} {2015})},\ \Eprint {http://arxiv.org/abs/1407.6060} {arXiv:1407.6060 [hep-ph]} \BibitemShut {NoStop}%
\bibitem [{\citenamefont {Abratenko}\ \emph {et~al.}(2022{\natexlab{c}})\citenamefont {Abratenko} \emph {et~al.}}]{glee_delta}%
  \BibitemOpen
  \bibfield  {author} {\bibinfo {author} {\bibfnamefont {P.}~\bibnamefont {Abratenko}} \emph {et~al.} (\bibinfo {collaboration} {MicroBooNE Collaboration}),\ }\href {\doibase 10.1103/PhysRevLett.128.111801} {\bibfield  {journal} {\bibinfo  {journal} {Phys. Rev. Lett.}\ }\textbf {\bibinfo {volume} {128}},\ \bibinfo {pages} {111801} (\bibinfo {year} {2022}{\natexlab{c}})}\BibitemShut {NoStop}%
\bibitem [{\citenamefont {Gninenko}(2009)}]{Gninenko:2009ks}%
  \BibitemOpen
  \bibfield  {author} {\bibinfo {author} {\bibfnamefont {S.~N.}\ \bibnamefont {Gninenko}},\ }\href {\doibase 10.1103/PhysRevLett.103.241802} {\bibfield  {journal} {\bibinfo  {journal} {Phys. Rev. Lett.}\ }\textbf {\bibinfo {volume} {103}},\ \bibinfo {pages} {241802} (\bibinfo {year} {2009})},\ \Eprint {http://arxiv.org/abs/0902.3802} {arXiv:0902.3802 [hep-ph]} \BibitemShut {NoStop}%
\bibitem [{\citenamefont {Fischer}\ \emph {et~al.}(2020)\citenamefont {Fischer}, \citenamefont {Hern\'andez-Cabezudo},\ and\ \citenamefont {Schwetz}}]{Fischer:2019fbw}%
  \BibitemOpen
  \bibfield  {author} {\bibinfo {author} {\bibfnamefont {O.}~\bibnamefont {Fischer}}, \bibinfo {author} {\bibfnamefont {A.}~\bibnamefont {Hern\'andez-Cabezudo}}, \ and\ \bibinfo {author} {\bibfnamefont {T.}~\bibnamefont {Schwetz}},\ }\href {\doibase 10.1103/PhysRevD.101.075045} {\bibfield  {journal} {\bibinfo  {journal} {Phys. Rev. D}\ }\textbf {\bibinfo {volume} {101}},\ \bibinfo {pages} {075045} (\bibinfo {year} {2020})},\ \Eprint {http://arxiv.org/abs/1909.09561} {arXiv:1909.09561 [hep-ph]} \BibitemShut {NoStop}%
\bibitem [{\citenamefont {Bertuzzo}\ \emph {et~al.}(2018)\citenamefont {Bertuzzo}, \citenamefont {Jana}, \citenamefont {Machado},\ and\ \citenamefont {Zukanovich~Funchal}}]{Bertuzzo:2018itn}%
  \BibitemOpen
  \bibfield  {author} {\bibinfo {author} {\bibfnamefont {E.}~\bibnamefont {Bertuzzo}}, \bibinfo {author} {\bibfnamefont {S.}~\bibnamefont {Jana}}, \bibinfo {author} {\bibfnamefont {P.~A.~N.}\ \bibnamefont {Machado}}, \ and\ \bibinfo {author} {\bibfnamefont {R.}~\bibnamefont {Zukanovich~Funchal}},\ }\href {\doibase 10.1103/PhysRevLett.121.241801} {\bibfield  {journal} {\bibinfo  {journal} {Phys. Rev. Lett.}\ }\textbf {\bibinfo {volume} {121}},\ \bibinfo {pages} {241801} (\bibinfo {year} {2018})},\ \Eprint {http://arxiv.org/abs/1807.09877} {arXiv:1807.09877 [hep-ph]} \BibitemShut {NoStop}%
\bibitem [{\citenamefont {Ballett}\ \emph {et~al.}(2019)\citenamefont {Ballett}, \citenamefont {Pascoli},\ and\ \citenamefont {Ross-Lonergan}}]{Ballett:2018ynz}%
  \BibitemOpen
  \bibfield  {author} {\bibinfo {author} {\bibfnamefont {P.}~\bibnamefont {Ballett}}, \bibinfo {author} {\bibfnamefont {S.}~\bibnamefont {Pascoli}}, \ and\ \bibinfo {author} {\bibfnamefont {M.}~\bibnamefont {Ross-Lonergan}},\ }\href {\doibase 10.1103/PhysRevD.99.071701} {\bibfield  {journal} {\bibinfo  {journal} {Phys. Rev. D}\ }\textbf {\bibinfo {volume} {99}},\ \bibinfo {pages} {071701} (\bibinfo {year} {2019})},\ \Eprint {http://arxiv.org/abs/1808.02915} {arXiv:1808.02915 [hep-ph]} \BibitemShut {NoStop}%
\bibitem [{\citenamefont {Adams}\ \emph {et~al.}(2013)\citenamefont {Adams} \emph {et~al.}}]{SBND}%
  \BibitemOpen
  \bibfield  {author} {\bibinfo {author} {\bibfnamefont {C.}~\bibnamefont {Adams}} \emph {et~al.} (\bibinfo {collaboration} {LArTPC}),\ }in\ \href@noop {} {\emph {\bibinfo {booktitle} {{Snowmass 2013}: {Snowmass on the Mississippi}}}}\ (\bibinfo {year} {2013})\ \Eprint {http://arxiv.org/abs/1309.7987} {arXiv:1309.7987 [physics.ins-det]} \BibitemShut {NoStop}%
\bibitem [{\citenamefont {Abratenko}\ \emph {et~al.}(2025{\natexlab{a}})\citenamefont {Abratenko} \emph {et~al.}}]{uboone_enhanced_nc_delta}%
  \BibitemOpen
  \bibfield  {author} {\bibinfo {author} {\bibfnamefont {P.}~\bibnamefont {Abratenko}} \emph {et~al.} (\bibinfo {collaboration} {MicroBooNE Collaboration}),\ }\href@noop {} {\enquote {\bibinfo {title} {{Enhanced Search for Neutral Current $\Delta$ Radiative Single-Photon Production in MicroBooNE}},}\ } (\bibinfo {year} {2025}{\natexlab{a}}),\ \bibinfo {note} {\url{https://microboone.fnal.gov/nc-delta-2025/}}\BibitemShut {NoStop}%
\bibitem [{\citenamefont {Abratenko}\ \emph {et~al.}(2025{\natexlab{b}})\citenamefont {Abratenko} \emph {et~al.}}]{uboone_inclusive_gamma}%
  \BibitemOpen
  \bibfield  {author} {\bibinfo {author} {\bibfnamefont {P.}~\bibnamefont {Abratenko}} \emph {et~al.} (\bibinfo {collaboration} {MicroBooNE Collaboration}),\ }\href@noop {} {\enquote {\bibinfo {title} {{Inclusive Search for Anomalous Single-Photon Production in MicroBooNE}},}\ } (\bibinfo {year} {2025}{\natexlab{b}}),\ \bibinfo {note} {\url{https://microboone.fnal.gov/inclusive-single-photon-2025/}}\BibitemShut {NoStop}%
\bibitem [{\citenamefont {Nygren}(1974)}]{Nygren:1974nfi}%
  \BibitemOpen
  \bibfield  {author} {\bibinfo {author} {\bibfnamefont {D.~R.}\ \bibnamefont {Nygren}},\ }\href@noop {} {\bibfield  {journal} {\bibinfo  {journal} {1974 PEP summer study, eConf}\ }\textbf {\bibinfo {volume} {C740805}},\ \bibinfo {pages} {58} (\bibinfo {year} {1974})}\BibitemShut {NoStop}%
\bibitem [{\citenamefont {Rubbia}()}]{Rubbia:117852}%
  \BibitemOpen
  \bibfield  {author} {\bibinfo {author} {\bibfnamefont {C.}~\bibnamefont {Rubbia}},\ }\href {https://cds.cern.ch/record/117852/files/CERN-EP-INT-77-8.pdf} {\ }\bibinfo {note} {{CERN-EP/77-08 (1977)}}\BibitemShut {NoStop}%
\bibitem [{\citenamefont {Acciarri}\ \emph {et~al.}(2019)\citenamefont {Acciarri} \emph {et~al.}}]{PhysRevD.99.012002}%
  \BibitemOpen
  \bibfield  {author} {\bibinfo {author} {\bibfnamefont {R.}~\bibnamefont {Acciarri}} \emph {et~al.} (\bibinfo {collaboration} {ArgoNeuT Collaboration}),\ }\href {\doibase 10.1103/PhysRevD.99.012002} {\bibfield  {journal} {\bibinfo  {journal} {Phys. Rev. D}\ }\textbf {\bibinfo {volume} {99}},\ \bibinfo {pages} {012002} (\bibinfo {year} {2019})}\BibitemShut {NoStop}%
\bibitem [{\citenamefont {Andringa}\ \emph {et~al.}(2023)\citenamefont {Andringa} \emph {et~al.}}]{Andringa:2023aax}%
  \BibitemOpen
  \bibfield  {author} {\bibinfo {author} {\bibfnamefont {S.}~\bibnamefont {Andringa}} \emph {et~al.},\ }\href {\doibase 10.1088/1361-6471/acad17} {\bibfield  {journal} {\bibinfo  {journal} {J. Phys. G}\ }\textbf {\bibinfo {volume} {50}},\ \bibinfo {pages} {033001} (\bibinfo {year} {2023})}\BibitemShut {NoStop}%
\bibitem [{\citenamefont {Castiglioni}\ \emph {et~al.}(2020)\citenamefont {Castiglioni}, \citenamefont {Foreman}, \citenamefont {Lepetic}, \citenamefont {Littlejohn}, \citenamefont {Malaker},\ and\ \citenamefont {Mastbaum}}]{Castiglioni:2020tsu}%
  \BibitemOpen
  \bibfield  {author} {\bibinfo {author} {\bibfnamefont {W.}~\bibnamefont {Castiglioni}}, \bibinfo {author} {\bibfnamefont {W.}~\bibnamefont {Foreman}}, \bibinfo {author} {\bibfnamefont {I.}~\bibnamefont {Lepetic}}, \bibinfo {author} {\bibfnamefont {B.~R.}\ \bibnamefont {Littlejohn}}, \bibinfo {author} {\bibfnamefont {M.}~\bibnamefont {Malaker}}, \ and\ \bibinfo {author} {\bibfnamefont {A.}~\bibnamefont {Mastbaum}},\ }\href {\doibase 10.1103/PhysRevD.102.092010} {\bibfield  {journal} {\bibinfo  {journal} {Phys. Rev. D}\ }\textbf {\bibinfo {volume} {102}},\ \bibinfo {pages} {092010} (\bibinfo {year} {2020})},\ \Eprint {http://arxiv.org/abs/2006.14675} {arXiv:2006.14675 [physics.ins-det]} \BibitemShut {NoStop}%
\bibitem [{\citenamefont {Abratenko}\ \emph {et~al.}(2022{\natexlab{d}})\citenamefont {Abratenko} \emph {et~al.}}]{Abratenko_2022}%
  \BibitemOpen
  \bibfield  {author} {\bibinfo {author} {\bibfnamefont {P.}~\bibnamefont {Abratenko}} \emph {et~al.} (\bibinfo {collaboration} {{MicroBooNE Collaboration}}),\ }\href {\doibase 10.1088/1748-0221/17/11/P11022} {\bibfield  {journal} {\bibinfo  {journal} {JINST}\ }\textbf {\bibinfo {volume} {17}},\ \bibinfo {pages} {P11022} (\bibinfo {year} {2022}{\natexlab{d}})},\ \Eprint {http://arxiv.org/abs/2203.10147} {arXiv:2203.10147 [physics.ins-det]} \BibitemShut {NoStop}%
\bibitem [{\citenamefont {Abratenko}\ \emph {et~al.}(2024)\citenamefont {Abratenko} \emph {et~al.}}]{PhysRevD.109.052007}%
  \BibitemOpen
  \bibfield  {author} {\bibinfo {author} {\bibfnamefont {P.}~\bibnamefont {Abratenko}} \emph {et~al.} (\bibinfo {collaboration} {MicroBooNE}),\ }\href {\doibase 10.1103/PhysRevD.109.052007} {\bibfield  {journal} {\bibinfo  {journal} {Phys. Rev. D}\ }\textbf {\bibinfo {volume} {.109}},\ \bibinfo {pages} {052007} (\bibinfo {year} {2024})},\ \Eprint {http://arxiv.org/abs/2307.03102} {arXiv:2307.03102 [hep-ex]} \BibitemShut {NoStop}%
\bibitem [{\citenamefont {Acciarri}\ \emph {et~al.}(2017{\natexlab{a}})\citenamefont {Acciarri} \emph {et~al.}}]{ub_TPC_design}%
  \BibitemOpen
  \bibfield  {author} {\bibinfo {author} {\bibfnamefont {R.}~\bibnamefont {Acciarri}} \emph {et~al.} (\bibinfo {collaboration} {{MicroBooNE Collaboration}}),\ }\href {\doibase 10.1088/1748-0221/12/02/P02017} {\bibfield  {journal} {\bibinfo  {journal} {JINST}\ }\textbf {\bibinfo {volume} {12}},\ \bibinfo {pages} {P02017} (\bibinfo {year} {2017}{\natexlab{a}})},\ \Eprint {http://arxiv.org/abs/1612.05824} {arXiv:1612.05824 [physics.ins-det]} \BibitemShut {NoStop}%
\bibitem [{\citenamefont {Aguilar-Arevalo}\ \emph {et~al.}(2009)\citenamefont {Aguilar-Arevalo} \emph {et~al.}}]{miniboone_BNB_flux}%
  \BibitemOpen
  \bibfield  {author} {\bibinfo {author} {\bibfnamefont {A.~A.}\ \bibnamefont {Aguilar-Arevalo}} \emph {et~al.} (\bibinfo {collaboration} {MiniBooNE Collaboration}),\ }\href {\doibase 10.1103/PhysRevD.79.072002} {\bibfield  {journal} {\bibinfo  {journal} {Phys. Rev. D}\ }\textbf {\bibinfo {volume} {79}},\ \bibinfo {pages} {072002} (\bibinfo {year} {2009})}\BibitemShut {NoStop}%
\bibitem [{\citenamefont {Andreopoulos}\ \emph {et~al.}(2010)\citenamefont {Andreopoulos} \emph {et~al.}}]{GENIE}%
  \BibitemOpen
  \bibfield  {author} {\bibinfo {author} {\bibfnamefont {C.}~\bibnamefont {Andreopoulos}} \emph {et~al.},\ }\href {\doibase https://doi.org/10.1016/j.nima.2009.12.009} {\bibfield  {journal} {\bibinfo  {journal} {Nucl. Instrum. Meth. A}\ }\textbf {\bibinfo {volume} {614}},\ \bibinfo {pages} {87} (\bibinfo {year} {2010})}\BibitemShut {NoStop}%
\bibitem [{\citenamefont {Abratenko}\ \emph {et~al.}(2022{\natexlab{e}})\citenamefont {Abratenko} \emph {et~al.}}]{ub_genie_tune}%
  \BibitemOpen
  \bibfield  {author} {\bibinfo {author} {\bibfnamefont {P.}~\bibnamefont {Abratenko}} \emph {et~al.} (\bibinfo {collaboration} {MicroBooNE Collaboration}),\ }\href {\doibase 10.1103/PhysRevD.105.072001} {\bibfield  {journal} {\bibinfo  {journal} {Phys. Rev. D}\ }\textbf {\bibinfo {volume} {105}},\ \bibinfo {pages} {072001} (\bibinfo {year} {2022}{\natexlab{e}})}\BibitemShut {NoStop}%
\bibitem [{\citenamefont {Abe}\ \emph {et~al.}(2016)\citenamefont {Abe} \emph {et~al.}}]{T2K_cc_data}%
  \BibitemOpen
  \bibfield  {author} {\bibinfo {author} {\bibfnamefont {K.}~\bibnamefont {Abe}} \emph {et~al.} (\bibinfo {collaboration} {T2K Collaboration}),\ }\href {\doibase 10.1103/PhysRevD.93.112012} {\bibfield  {journal} {\bibinfo  {journal} {Phys. Rev. D}\ }\textbf {\bibinfo {volume} {93}},\ \bibinfo {pages} {112012} (\bibinfo {year} {2016})}\BibitemShut {NoStop}%
\bibitem [{\citenamefont {Berger}\ and\ \citenamefont {Sehgal}(2009)}]{PhysRevD.79.053003}%
  \BibitemOpen
  \bibfield  {author} {\bibinfo {author} {\bibfnamefont {C.}~\bibnamefont {Berger}}\ and\ \bibinfo {author} {\bibfnamefont {L.~M.}\ \bibnamefont {Sehgal}},\ }\href {\doibase 10.1103/PhysRevD.79.053003} {\bibfield  {journal} {\bibinfo  {journal} {Phys. Rev. D}\ }\textbf {\bibinfo {volume} {79}},\ \bibinfo {pages} {053003} (\bibinfo {year} {2009})}\BibitemShut {NoStop}%
\bibitem [{\citenamefont {Bodek}\ \emph {et~al.}(2008)\citenamefont {Bodek}, \citenamefont {Avvakumov}, \citenamefont {Bradford},\ and\ \citenamefont {Budd}}]{Bodek:2007ym}%
  \BibitemOpen
  \bibfield  {author} {\bibinfo {author} {\bibfnamefont {A.}~\bibnamefont {Bodek}}, \bibinfo {author} {\bibfnamefont {S.}~\bibnamefont {Avvakumov}}, \bibinfo {author} {\bibfnamefont {R.}~\bibnamefont {Bradford}}, \ and\ \bibinfo {author} {\bibfnamefont {H.~S.}\ \bibnamefont {Budd}},\ }\href {\doibase 10.1140/epjc/s10052-007-0491-4} {\bibfield  {journal} {\bibinfo  {journal} {Eur. Phys. J. C}\ }\textbf {\bibinfo {volume} {53}},\ \bibinfo {pages} {349} (\bibinfo {year} {2008})},\ \Eprint {http://arxiv.org/abs/0708.1946} {arXiv:0708.1946 [hep-ex]} \BibitemShut {NoStop}%
\bibitem [{\citenamefont {Merenyi}\ \emph {et~al.}(1992)\citenamefont {Merenyi}, \citenamefont {Mann}, \citenamefont {Kafka}, \citenamefont {Leeson}, \citenamefont {Saitta}, \citenamefont {Schneps}, \citenamefont {Derrick},\ and\ \citenamefont {Musgrave}}]{PhysRevD.45.743}%
  \BibitemOpen
  \bibfield  {author} {\bibinfo {author} {\bibfnamefont {R.}~\bibnamefont {Merenyi}}, \bibinfo {author} {\bibfnamefont {W.~A.}\ \bibnamefont {Mann}}, \bibinfo {author} {\bibfnamefont {T.}~\bibnamefont {Kafka}}, \bibinfo {author} {\bibfnamefont {W.}~\bibnamefont {Leeson}}, \bibinfo {author} {\bibfnamefont {B.}~\bibnamefont {Saitta}}, \bibinfo {author} {\bibfnamefont {J.}~\bibnamefont {Schneps}}, \bibinfo {author} {\bibfnamefont {M.}~\bibnamefont {Derrick}}, \ and\ \bibinfo {author} {\bibfnamefont {B.}~\bibnamefont {Musgrave}},\ }\href {\doibase 10.1103/PhysRevD.45.743} {\bibfield  {journal} {\bibinfo  {journal} {Phys. Rev. D}\ }\textbf {\bibinfo {volume} {45}},\ \bibinfo {pages} {743} (\bibinfo {year} {1992})}\BibitemShut {NoStop}%
\bibitem [{\citenamefont {Snider}\ and\ \citenamefont {Petrillo}(2017)}]{larsoft}%
  \BibitemOpen
  \bibfield  {author} {\bibinfo {author} {\bibfnamefont {E.}~\bibnamefont {Snider}}\ and\ \bibinfo {author} {\bibfnamefont {G.}~\bibnamefont {Petrillo}},\ }\href {\doibase 10.1088/1742-6596/898/4/042057} {\bibfield  {journal} {\bibinfo  {journal} {Journal of Physics: Conference Series}\ }\textbf {\bibinfo {volume} {898}},\ \bibinfo {pages} {042057} (\bibinfo {year} {2017})}\BibitemShut {NoStop}%
\bibitem [{\citenamefont {Agostinelli}\ \emph {et~al.}(2003)\citenamefont {Agostinelli} \emph {et~al.}}]{geant4}%
  \BibitemOpen
  \bibfield  {author} {\bibinfo {author} {\bibfnamefont {S.}~\bibnamefont {Agostinelli}} \emph {et~al.},\ }\href {\doibase https://doi.org/10.1016/S0168-9002(03)01368-8} {\bibfield  {journal} {\bibinfo  {journal} {Nuclear Instruments and Methods in Physics Research Section A: Accelerators, Spectrometers, Detectors and Associated Equipment}\ }\textbf {\bibinfo {volume} {506}},\ \bibinfo {pages} {250} (\bibinfo {year} {2003})}\BibitemShut {NoStop}%
\bibitem [{\citenamefont {Adams}\ \emph {et~al.}(2018{\natexlab{a}})\citenamefont {Adams} \emph {et~al.}}]{ub_signal_process_1}%
  \BibitemOpen
  \bibfield  {author} {\bibinfo {author} {\bibfnamefont {C.}~\bibnamefont {Adams}} \emph {et~al.} (\bibinfo {collaboration} {{MicroBooNE Collaboration}}),\ }\href {\doibase 10.1088/1748-0221/13/07/P07006} {\bibfield  {journal} {\bibinfo  {journal} {JINST}\ }\textbf {\bibinfo {volume} {13}},\ \bibinfo {pages} {P07006} (\bibinfo {year} {2018}{\natexlab{a}})},\ \Eprint {http://arxiv.org/abs/1802.08709} {arXiv:1802.08709 [physics.ins-det]} \BibitemShut {NoStop}%
\bibitem [{\citenamefont {Adams}\ \emph {et~al.}(2018{\natexlab{b}})\citenamefont {Adams} \emph {et~al.}}]{ub_signal_process_2}%
  \BibitemOpen
  \bibfield  {author} {\bibinfo {author} {\bibfnamefont {C.}~\bibnamefont {Adams}} \emph {et~al.} (\bibinfo {collaboration} {{MicroBooNE Collaboration}}),\ }\href {\doibase 10.1088/1748-0221/13/07/P07007} {\bibfield  {journal} {\bibinfo  {journal} {JINST}\ }\textbf {\bibinfo {volume} {13}},\ \bibinfo {pages} {P07007} (\bibinfo {year} {2018}{\natexlab{b}})},\ \Eprint {http://arxiv.org/abs/1804.02583} {arXiv:1804.02583 [physics.ins-det]} \BibitemShut {NoStop}%
\bibitem [{\citenamefont {Jaskolski}\ and\ \citenamefont {Wojcik}(2011)}]{electron_recomb}%
  \BibitemOpen
  \bibfield  {author} {\bibinfo {author} {\bibfnamefont {M.}~\bibnamefont {Jaskolski}}\ and\ \bibinfo {author} {\bibfnamefont {M.}~\bibnamefont {Wojcik}},\ }\href {\doibase 10.1021/jp201149w} {\bibfield  {journal} {\bibinfo  {journal} {J. Phys. Chem. A}\ }\textbf {\bibinfo {volume} {115}},\ \bibinfo {pages} {4317} (\bibinfo {year} {2011})},\ \bibinfo {note} {pMID: 21473614},\ \Eprint {http://arxiv.org/abs/https://doi.org/10.1021/jp201149w} {https://doi.org/10.1021/jp201149w} \BibitemShut {NoStop}%
\bibitem [{\citenamefont {Adams}\ \emph {et~al.}(2020{\natexlab{a}})\citenamefont {Adams} \emph {et~al.}}]{ub_SCE_effect}%
  \BibitemOpen
  \bibfield  {author} {\bibinfo {author} {\bibfnamefont {C.}~\bibnamefont {Adams}} \emph {et~al.} (\bibinfo {collaboration} {{MicroBooNE Collaboration}}),\ }\href {\doibase 10.1088/1748-0221/15/03/P03022} {\bibfield  {journal} {\bibinfo  {journal} {JINST}\ }\textbf {\bibinfo {volume} {15}},\ \bibinfo {pages} {P03022} (\bibinfo {year} {2020}{\natexlab{a}})},\ \Eprint {http://arxiv.org/abs/1907.11736} {arXiv:1907.11736 [physics.ins-det]} \BibitemShut {NoStop}%
\bibitem [{\citenamefont {Adams}\ \emph {et~al.}(2020{\natexlab{b}})\citenamefont {Adams} \emph {et~al.}}]{uB_Efield_measurement}%
  \BibitemOpen
  \bibfield  {author} {\bibinfo {author} {\bibfnamefont {C.}~\bibnamefont {Adams}} \emph {et~al.} (\bibinfo {collaboration} {{MicroBooNE Collaboration}}),\ }\href {\doibase 10.1088/1748-0221/15/07/P07010} {\bibfield  {journal} {\bibinfo  {journal} {JINST}\ }\textbf {\bibinfo {volume} {15}},\ \bibinfo {pages} {P07010} (\bibinfo {year} {2020}{\natexlab{b}})},\ \Eprint {http://arxiv.org/abs/1910.01430} {arXiv:1910.01430 [physics.ins-det]} \BibitemShut {NoStop}%
\bibitem [{\citenamefont {Adams}\ \emph {et~al.}(2021)\citenamefont {Adams} \emph {et~al.}}]{cosmic_rate}%
  \BibitemOpen
  \bibfield  {author} {\bibinfo {author} {\bibfnamefont {C.}~\bibnamefont {Adams}} \emph {et~al.} (\bibinfo {collaboration} {{MicroBooNE Collaboration}}),\ }\href {\doibase 10.1088/1748-0221/16/04/P04004} {\bibfield  {journal} {\bibinfo  {journal} {JINST}\ }\textbf {\bibinfo {volume} {16}},\ \bibinfo {pages} {P04004} (\bibinfo {year} {2021})},\ \Eprint {http://arxiv.org/abs/2012.14324} {arXiv:2012.14324 [physics.ins-det]} \BibitemShut {NoStop}%
\bibitem [{\citenamefont {Acciarri}\ \emph {et~al.}(2018)\citenamefont {Acciarri} \emph {et~al.}}]{pandora_reco}%
  \BibitemOpen
  \bibfield  {author} {\bibinfo {author} {\bibfnamefont {R.}~\bibnamefont {Acciarri}} \emph {et~al.} (\bibinfo {collaboration} {{MicroBooNE Collaboration}}),\ }\href {\doibase 10.1140/epjc/s10052-017-5481-6} {\bibfield  {journal} {\bibinfo  {journal} {Eur. Phys. J. C}\ }\textbf {\bibinfo {volume} {78}},\ \bibinfo {pages} {82} (\bibinfo {year} {2018})},\ \Eprint {http://arxiv.org/abs/1708.03135} {arXiv:1708.03135 [hep-ex]} \BibitemShut {NoStop}%
\bibitem [{\citenamefont {Frühwirth}(1987)}]{kalman_filter}%
  \BibitemOpen
  \bibfield  {author} {\bibinfo {author} {\bibfnamefont {R.}~\bibnamefont {Frühwirth}},\ }\href {\doibase https://doi.org/10.1016/0168-9002(87)90887-4} {\bibfield  {journal} {\bibinfo  {journal} {Nucl. Instrum. Meth. A}\ }\textbf {\bibinfo {volume} {262}},\ \bibinfo {pages} {444} (\bibinfo {year} {1987})}\BibitemShut {NoStop}%
\bibitem [{\citenamefont {Acciarri}\ \emph {et~al.}(2017{\natexlab{b}})\citenamefont {Acciarri} \emph {et~al.}}]{MicroBooNE:2017kvv}%
  \BibitemOpen
  \bibfield  {author} {\bibinfo {author} {\bibfnamefont {R.}~\bibnamefont {Acciarri}} \emph {et~al.} (\bibinfo {collaboration} {MicroBooNE}),\ }\href {\doibase 10.1088/1748-0221/12/09/P09014} {\bibfield  {journal} {\bibinfo  {journal} {JINST}\ }\textbf {\bibinfo {volume} {12}},\ \bibinfo {pages} {P09014} (\bibinfo {year} {2017}{\natexlab{b}})},\ \Eprint {http://arxiv.org/abs/1704.02927} {arXiv:1704.02927 [physics.ins-det]} \BibitemShut {NoStop}%
\bibitem [{\citenamefont {Abratenko}\ \emph {et~al.}(2023{\natexlab{c}})\citenamefont {Abratenko} \emph {et~al.}}]{gLEE_pi0}%
  \BibitemOpen
  \bibfield  {author} {\bibinfo {author} {\bibfnamefont {P.}~\bibnamefont {Abratenko}} \emph {et~al.} (\bibinfo {collaboration} {MicroBooNE Collaboration}),\ }\href {\doibase 10.1103/PhysRevD.107.012004} {\bibfield  {journal} {\bibinfo  {journal} {Phys. Rev. D}\ }\textbf {\bibinfo {volume} {107}},\ \bibinfo {pages} {012004} (\bibinfo {year} {2023}{\natexlab{c}})}\BibitemShut {NoStop}%
\bibitem [{\citenamefont {Alvarez-Ruso}\ \emph {et~al.}(2021)\citenamefont {Alvarez-Ruso} \emph {et~al.}}]{genie_v3.2}%
  \BibitemOpen
  \bibfield  {author} {\bibinfo {author} {\bibfnamefont {L.}~\bibnamefont {Alvarez-Ruso}} \emph {et~al.},\ }\href {\doibase 10.1140/epjs/s11734-021-00295-7} {\bibfield  {journal} {\bibinfo  {journal} {Eur. Phys. J. ST}\ }\textbf {\bibinfo {volume} {230}},\ \bibinfo {pages} {4449} (\bibinfo {year} {2021})}\BibitemShut {NoStop}%
\bibitem [{\citenamefont {Abratenko}\ \emph {et~al.}(2021{\natexlab{b}})\citenamefont {Abratenko} \emph {et~al.}}]{spacecharge}%
  \BibitemOpen
  \bibfield  {author} {\bibinfo {author} {\bibfnamefont {P.}~\bibnamefont {Abratenko}} \emph {et~al.} (\bibinfo {collaboration} {MicroBooNE}),\ }\href {\doibase 10.1103/PhysRevApplied.15.064071} {\bibfield  {journal} {\bibinfo  {journal} {Phys. Rev. Applied}\ }\textbf {\bibinfo {volume} {15}},\ \bibinfo {pages} {064071} (\bibinfo {year} {2021}{\natexlab{b}})},\ \Eprint {http://arxiv.org/abs/2101.05076} {arXiv:2101.05076 [physics.ins-det]} \BibitemShut {NoStop}%
\bibitem [{\citenamefont {Chen}\ and\ \citenamefont {Guestrin}(2016)}]{xgboost}%
  \BibitemOpen
  \bibfield  {author} {\bibinfo {author} {\bibfnamefont {T.}~\bibnamefont {Chen}}\ and\ \bibinfo {author} {\bibfnamefont {C.}~\bibnamefont {Guestrin}},\ }in\ \href {\doibase 10.1145/2939672.2939785} {\emph {\bibinfo {booktitle} {Proceedings of the 22nd ACM SIGKDD International Conference on Knowledge Discovery and Data Mining}}},\ \bibinfo {series and number} {KDD ’16}\ (\bibinfo  {publisher} {ACM},\ \bibinfo {year} {2016})\BibitemShut {NoStop}%
\bibitem [{\citenamefont {Ester}\ \emph {et~al.}(1996)\citenamefont {Ester}, \citenamefont {Kriegel}, \citenamefont {Sander},\ and\ \citenamefont {Xiaowei}}]{dbscan}%
  \BibitemOpen
  \bibfield  {author} {\bibinfo {author} {\bibfnamefont {M.}~\bibnamefont {Ester}}, \bibinfo {author} {\bibfnamefont {H.~P.}\ \bibnamefont {Kriegel}}, \bibinfo {author} {\bibfnamefont {J.}~\bibnamefont {Sander}}, \ and\ \bibinfo {author} {\bibfnamefont {X.}~\bibnamefont {Xiaowei}},\ }\href {\doibase 10.1023/A:1009745219419} {\bibfield  {journal} {\bibinfo  {journal} {Data Min. Knowl. Discov}\ ,\ \bibinfo {pages} {169–194}} (\bibinfo {year} {1996})}\BibitemShut {NoStop}%
\bibitem [{\citenamefont {Wright}\ and\ \citenamefont {Kelsey}(2015)}]{geant4_hadron}%
  \BibitemOpen
  \bibfield  {author} {\bibinfo {author} {\bibfnamefont {D.~H.}\ \bibnamefont {Wright}}\ and\ \bibinfo {author} {\bibfnamefont {M.~H.}\ \bibnamefont {Kelsey}},\ }\href {https://www.sciencedirect.com/science/article/pii/S0168900215011134} {\bibfield  {journal} {\bibinfo  {journal} {Nucl. Instrum. Meth. A}\ }\textbf {\bibinfo {volume} {804}},\ \bibinfo {pages} {175} (\bibinfo {year} {2015})}\BibitemShut {NoStop}%
\bibitem [{\citenamefont {Calcutt}\ \emph {et~al.}(2021)\citenamefont {Calcutt}, \citenamefont {Thorpe}, \citenamefont {Mahn},\ and\ \citenamefont {Fields}}]{geant4_reweight}%
  \BibitemOpen
  \bibfield  {author} {\bibinfo {author} {\bibfnamefont {J.}~\bibnamefont {Calcutt}}, \bibinfo {author} {\bibfnamefont {C.}~\bibnamefont {Thorpe}}, \bibinfo {author} {\bibfnamefont {K.}~\bibnamefont {Mahn}}, \ and\ \bibinfo {author} {\bibfnamefont {L.}~\bibnamefont {Fields}},\ }\href {\doibase 10.1088/1748-0221/16/08/P08042} {\bibfield  {journal} {\bibinfo  {journal} {JINST}\ }\textbf {\bibinfo {volume} {16}},\ \bibinfo {pages} {P08042} (\bibinfo {year} {2021})}\BibitemShut {NoStop}%
\bibitem [{\citenamefont {Abratenko}\ \emph {et~al.}(2022{\natexlab{f}})\citenamefont {Abratenko} \emph {et~al.}}]{det_err1}%
  \BibitemOpen
  \bibfield  {author} {\bibinfo {author} {\bibfnamefont {P.}~\bibnamefont {Abratenko}} \emph {et~al.},\ }\href {\doibase 10.1140/epjc/s10052-022-10270-8} {\bibfield  {journal} {\bibinfo  {journal} {Eur. Phys. J. C}\ }\textbf {\bibinfo {volume} {82}},\ \bibinfo {pages} {454} (\bibinfo {year} {2022}{\natexlab{f}})}\BibitemShut {NoStop}%
\bibitem [{\citenamefont {Abratenko}\ \emph {et~al.}(2020{\natexlab{c}})\citenamefont {Abratenko} \emph {et~al.}}]{SCE}%
  \BibitemOpen
  \bibfield  {author} {\bibinfo {author} {\bibfnamefont {P.}~\bibnamefont {Abratenko}} \emph {et~al.} (\bibinfo {collaboration} {{MicroBooNE Collaboration}}),\ }\href {\doibase 10.1088/1748-0221/15/12/P12037} {\bibfield  {journal} {\bibinfo  {journal} {JINST}\ }\textbf {\bibinfo {volume} {15}},\ \bibinfo {pages} {P12037} (\bibinfo {year} {2020}{\natexlab{c}})},\ \Eprint {http://arxiv.org/abs/2008.09765} {arXiv:2008.09765 [physics.ins-det]} \BibitemShut {NoStop}%
\bibitem [{\citenamefont {Acciarri}\ \emph {et~al.}(2013)\citenamefont {Acciarri} \emph {et~al.}}]{recomb}%
  \BibitemOpen
  \bibfield  {author} {\bibinfo {author} {\bibfnamefont {R.}~\bibnamefont {Acciarri}} \emph {et~al.} (\bibinfo {collaboration} {ArgoNeuT}),\ }\href {\doibase 10.1088/1748-0221/8/08/P08005} {\bibfield  {journal} {\bibinfo  {journal} {JINST}\ }\textbf {\bibinfo {volume} {8}},\ \bibinfo {pages} {P08005} (\bibinfo {year} {2013})},\ \Eprint {http://arxiv.org/abs/1306.1712} {arXiv:1306.1712 [physics.ins-det]} \BibitemShut {NoStop}%
\bibitem [{\citenamefont {Abratenko}\ \emph {et~al.}(2022{\natexlab{g}})\citenamefont {Abratenko} \emph {et~al.}}]{WC_lee}%
  \BibitemOpen
  \bibfield  {author} {\bibinfo {author} {\bibfnamefont {P.}~\bibnamefont {Abratenko}} \emph {et~al.} (\bibinfo {collaboration} {MicroBooNE Collaboration}),\ }\href {\doibase 10.1103/PhysRevD.105.112005} {\bibfield  {journal} {\bibinfo  {journal} {Phys. Rev. D}\ }\textbf {\bibinfo {volume} {105}},\ \bibinfo {pages} {112005} (\bibinfo {year} {2022}{\natexlab{g}})}\BibitemShut {NoStop}%
\bibitem [{\citenamefont {Abratenko}\ \emph {et~al.}(2022{\natexlab{h}})\citenamefont {Abratenko} \emph {et~al.}}]{PELEE}%
  \BibitemOpen
  \bibfield  {author} {\bibinfo {author} {\bibfnamefont {P.}~\bibnamefont {Abratenko}} \emph {et~al.} (\bibinfo {collaboration} {MicroBooNE Collaboration}),\ }\href {\doibase 10.1103/PhysRevD.105.112004} {\bibfield  {journal} {\bibinfo  {journal} {Phys. Rev. D}\ }\textbf {\bibinfo {volume} {105}},\ \bibinfo {pages} {112004} (\bibinfo {year} {2022}{\natexlab{h}})}\BibitemShut {NoStop}%
\bibitem [{\citenamefont {Abratenko}\ \emph {et~al.}(2022{\natexlab{i}})\citenamefont {Abratenko} \emph {et~al.}}]{DLEE}%
  \BibitemOpen
  \bibfield  {author} {\bibinfo {author} {\bibfnamefont {P.}~\bibnamefont {Abratenko}} \emph {et~al.} (\bibinfo {collaboration} {MicroBooNE Collaboration}),\ }\href {\doibase 10.1103/PhysRevD.105.112003} {\bibfield  {journal} {\bibinfo  {journal} {Phys. Rev. D}\ }\textbf {\bibinfo {volume} {105}},\ \bibinfo {pages} {112003} (\bibinfo {year} {2022}{\natexlab{i}})}\BibitemShut {NoStop}%
\bibitem [{\citenamefont {Abratenko}\ \emph {et~al.}(2022{\natexlab{j}})\citenamefont {Abratenko} \emph {et~al.}}]{CombELEE}%
  \BibitemOpen
  \bibfield  {author} {\bibinfo {author} {\bibfnamefont {P.}~\bibnamefont {Abratenko}} \emph {et~al.} (\bibinfo {collaboration} {MicroBooNE Collaboration}),\ }\href {\doibase 10.1103/PhysRevLett.128.241801} {\bibfield  {journal} {\bibinfo  {journal} {Phys. Rev. Lett.}\ }\textbf {\bibinfo {volume} {128}},\ \bibinfo {pages} {241801} (\bibinfo {year} {2022}{\natexlab{j}})}\BibitemShut {NoStop}%
\bibitem [{\citenamefont {Ji}\ \emph {et~al.}(2020)\citenamefont {Ji}, \citenamefont {Gu}, \citenamefont {Qian}, \citenamefont {Wei},\ and\ \citenamefont {Zhang}}]{CNP}%
  \BibitemOpen
  \bibfield  {author} {\bibinfo {author} {\bibfnamefont {X.}~\bibnamefont {Ji}}, \bibinfo {author} {\bibfnamefont {W.}~\bibnamefont {Gu}}, \bibinfo {author} {\bibfnamefont {X.}~\bibnamefont {Qian}}, \bibinfo {author} {\bibfnamefont {H.}~\bibnamefont {Wei}}, \ and\ \bibinfo {author} {\bibfnamefont {C.}~\bibnamefont {Zhang}},\ }\href {\doibase 10.1016/j.nima.2020.163677} {\bibfield  {journal} {\bibinfo  {journal} {Nucl. Instrum. Meth. A}\ }\textbf {\bibinfo {volume} {961}},\ \bibinfo {pages} {163677} (\bibinfo {year} {2020})}\BibitemShut {NoStop}%
\bibitem [{\citenamefont {Feldman}\ and\ \citenamefont {Cousins}(1998)}]{Feldman:1997qc}%
  \BibitemOpen
  \bibfield  {author} {\bibinfo {author} {\bibfnamefont {G.~J.}\ \bibnamefont {Feldman}}\ and\ \bibinfo {author} {\bibfnamefont {R.~D.}\ \bibnamefont {Cousins}},\ }\href {\doibase 10.1103/PhysRevD.57.3873} {\bibfield  {journal} {\bibinfo  {journal} {Phys. Rev. D}\ }\textbf {\bibinfo {volume} {57}},\ \bibinfo {pages} {3873} (\bibinfo {year} {1998})},\ \Eprint {http://arxiv.org/abs/physics/9711021} {arXiv:physics/9711021} \BibitemShut {NoStop}%
\end{thebibliography}%

\end{document}